\documentclass[nofootinbib,twocolumn,preprintnumbers]{revtex4-1}
\pdfoutput=1
\usepackage{amsmath,amsthm,amssymb,multirow,psfrag}
\usepackage{epsfig}

\graphicspath{{./Figures/}}

\begin{document}

\def\lsim{\mathrel{\rlap{\lower4pt\hbox{\hskip1pt$\sim$}}
    \raise1pt\hbox{$<$}}}
\def\gsim{\mathrel{\rlap{\lower4pt\hbox{\hskip1pt$\sim$}}
    \raise1pt\hbox{$>$}}}
\newcommand{\vev}[1]{ \left\langle {#1} \right\rangle }
\newcommand{\bra}[1]{ \langle {#1} | }
\newcommand{\ket}[1]{ | {#1} \rangle }
\newcommand{\ev}{ {\rm eV} }
\newcommand{\kev}{{\rm keV}}
\newcommand{\mev}{{\rm MeV}}
\newcommand{\gev}{{\rm GeV}}
\newcommand{\tev}{{\rm TeV}}
\newcommand{\mpl}{$M_{Pl}$}
\newcommand{\mw}{$M_{W}$}
\newcommand{\Ft}{F_{T}}
\newcommand{\Zparity}{\mathbb{Z}_2}
\newcommand{\BLambda}{\boldsymbol{\lambda}}
\newcommand{\be}{\begin{eqnarray}}
\newcommand{\ee}{\end{eqnarray}}
\newcommand{\met}{\;\not\!\!\!{E}_T}

\title{Scrutinizing the Higgs Signal and Background in the $2e2\mu$ Golden Channel}
\author{\bf{Yi Chen$\, ^{a}$, Nhan Tran$\, ^{b}$,\, Roberto Vega-Morales$^{b,c}$}}
\email{Corresponding author:\\ robertovegamorales2010@u.northwestern.edu}

\affiliation{
$^a$ Physics Department, California Institute of Technology, Pasadena, CA, USA\\
$^b$ Fermi National Accelerator Laboratory (FNAL), Batavia, IL, USA,\\
$^c$ Department of Physics and Astronomy, Northwestern University, Evanston, IL, USA}
\date{\today}

\begin{abstract}

Kinematic distributions in the decays of the newly discovered resonance to four leptons are a powerful probe of the tensor structure of its couplings to electroweak gauge bosons.  We present analytic calculations for both signal and background of the fully differential cross section for the `Golden Channel' $e^+e^-\mu^+\mu^-$ final state. We include all interference effects between intermediate gauge bosons and allow them to be on- or off-shell. For the signal we compute the fully differential decay width for general scalar couplings to $ZZ$,~$\gamma\gamma$, and $Z\gamma$.  For the background we compute the leading order fully differential cross section for $q\bar{q}$ annihilation into $Z$ and $\gamma$ gauge bosons, including the contribution from the resonant $Z\rightarrow 2e2\mu$ process. We also present singly and doubly differential projections and study the interference effects on the differential spectra. These expressions can be used in a variety of ways to uncover the nature of the newly discovered resonance or any new scalars decaying to neutral gauge bosons which might be discovered in the future.

\end{abstract}
\preprint{FERMILAB-PUB-12-588-E-PPD-T, nuhep-th/12-13, CALT-68-2894}

\maketitle


\section{Introduction}
\label{sec:Intro}
With the recent discovery of a new resonance at the LHC~\cite{:2012gk,:2012gu} the focus now shifts to the determination of its detailed properties including its spin, CP, and electroweak (EW) quantum numbers.  It has been shown in recent studies~\cite{Gao:2010qx,DeRujula:2010ys,Gainer:2011xz,Bolognesi:2012mm,Boughezal:2012tz,Stolarski:2012ps} and also emphasized for quite some time~\cite{Gunion:1985mc,Soni:1993jc,Choi:2002jk,Matsuura:1991pj,Buszello:2002uu}, that the decay to four charged leptons is a powerful channel in accomplishing this goal. Because of the experimental precision with which this channel is measured, it offers one of the few opportunities to use analytic methods to analyze the data. We thus seek to extend previous analytic calculations of both the signal and the standard model (SM) background and present completely general, leading order (LO) fully differential cross sections for the $2e2\mu$ final state mediated by intermediate $Z$ and $\gamma$ gauge bosons. In addition to performing discovery/exclusion analysis and signal hypothesis testing one could, with enough data, experimentally determine all possible couplings of a spin-0 scalar to pairs of neutral electroweak (EW) gauge bosons in one multi-parameter fit using these expressions.

Analytic expressions are ideal for use in the matrix element method (MEM) taking full advantage of all of the kinematic information in the event.  One can then use the fully differential cross section to build a likelihood function~\cite{Fiedler:2010sg,Volobouev:2011vb} to be used as a discriminant.   For a recent study of the golden channel comparing existing leading order MEM-based approaches and software~\cite{Belyaev:2012qa}, along with providing code which calculates kinematic discriminants based on the Madgraph~\cite{Alwall:2011uj} matrix element squared see~\cite{Avery:2012um}.  We view this `analytic approach' as equivalent and complementary to these other approaches.  These analytic expressions also allow for more flexibility in performing multidimensional fits to determine coupling values which will be useful when performing parameter extraction.  Our parametrization allows for easy implementation of various hypothesis tests as well as the addition of NLO effects which can also be implemented into an MEM framework~\cite{Campbell:2012ct,Campbell:2012cz}.

For the signal we compute the fully differential decay width for the process $\varphi \rightarrow ZZ + Z\gamma + \gamma\gamma \rightarrow 2e2\mu$ where $\varphi$ is a spin-0 scalar.  We allow for the most general CP odd/even mixtures and include all interference between intermediate vector bosons. While these expressions are applicable to the newly discovered boson at 125 GeV, they are also applicable for any new scalar decaying to neutral gauge bosons.  This allows one to consider a variety of hypotheses which can be tested against one another.  It should be emphasized however that for optimal performance, even when testing between two different signal CP and spin hypothesis, one should also include the background in the discriminant since in any given sample it is not known with full certainty which are background and which are signal events.  Thus we seek to provide both signal and background distributions which can be used together to build a complete likelihood.  

For the background we compute the fully differential cross section for the $q\bar{q}\rightarrow 2e2\mu$ process. This includes the contributions from all the intermediate vector bosons through both t-channel pair production and the singly resonant four-lepton production s-channel process $q\bar{q}\rightarrow Z\rightarrow 2e2\mu$.  We include all interference effects between the intermediate vector bosons as well the interference between the s-channel and t-channel diagrams which can affect the differential distributions as we will see below.  Also, unlike the analytic calculations in~\cite{Hagiwara:1986vm,Gainer:2011xz} of the golden channel background differential cross section, these expressions are valid for a much larger energy range for the four lepton invariant mass as well as the invariant mass of each lepton pair.  In particular, since these also include the $\gamma\gamma$ contribution one can probe lower values in the differential mass spectrums, which as we will see is a highly discriminating region. The intermediate vector bosons are allowed to be on or off-shell and in what follows we do not distinguish between the two.  We do not discuss the $4e$ and $4\mu$ final states explicitly, but in some kinematic regimes the interference effects between identical final state particles can be sizable~\cite{Avery:2012um}. We leave an inclusion of these final states to future work.

Although other channels can also probe the tensor couplings of a resonance to neutral gauge bosons, the golden channel, with a four body final state has the advantage of extra kinematic variables, such as the azimuthal angle between lepton decay planes.  This variable would be unavailable for example in the $\gamma\gamma$ final state.  In addition to offering more kinematic observables, the golden channel offers the unique opportunity to test all of the possible tensor couplings including any potential interference effects between the different operators in one direct (and very precise) fully correlated measurement without any recourse to theoretical input (other than the production cross section of course). This allows for stringent tests of the SM to be performed and perhaps allow us to uncover new physics which may be hiding in subtle effects within the golden channel.

In addition to presenting the calculation of the fully differential cross sections we examine various singly and doubly differential distributions and elucidate the subtle interference effects between the different contributions to the signal and background.  Of course a proper treatment of the golden channel requires careful study of detector resolution and acceptance effects, but we leave that to ongoing analyses.  

The organization of this paper is as follows: in Sec.~\ref{sec:events} we briefly review the kinematics of the four lepton final state.  In Sec.~\ref{sec:Signal} we describe the calculation of the signal fully differential cross section and examine the differential mass spectra for a variety of signal hypotheses. In Sec.~\ref{sec:Background} we describe the calculation of the background fully differential cross section and examine how the kinematic variables are affected by NLO and \emph{pdf} effects before concluding in Sec.~\ref{sec:conc}.  We also present in the Appendix a pair of expressions for the signal and background doubly differential mass spectrums and also show plots for a multitude of singly and doubly differential spectra.

\section{Four Lepton Events}
\label{sec:events}

The kinematics of four lepton ($4\ell$) events are described in detail in many places in the literature and here we use the convention found in~\cite{Gao:2010qx}. We comment on the kinematics briefly and point out that in the case of the background the physical interpretation of the kinematic variables is not as intuitive as in the case of previous studies which only considered the t-channel $ZZ$ contribution. Now since we include the contribution from resonant four lepton production, the lepton pairs do not necessarily reconstruct to a physical particle.  In this case, resonant production of a $Z$ (or possibly $\gamma$) is followed by decay to charged leptons one of which radiates a $Z/\gamma$, which again decays to charged leptons (see Fig.~\ref{fig:Zdiags}). The first lepton pair which radiates the second vector boson does not reconstruct to the $Z$ boson four momentum, which in this case is also equal to the invariant mass of the $4\ell$ system. The kinematics remain unchanged, but now we must interpret the angles defined in the lepton pair rest frame with respect to the direction of momentum of the \emph{lepton pair system} as opposed to that of one of the gauge bosons.  Thus, we have the following more general interpretations for the kinematic variables defined in the $4\ell$ rest frame;
\begin{itemize}
\item $M_{1,2}$ -- The invariant mass of the two lepton pair systems. 
\item $\Theta$ -- The `production angle' between the momentum vectors of the lepton pair which reconstructs to $M_1$ and the total $4\ell$ system momentum.
\item $\theta_{1,2}$ -- Polar angle of the momentum vectors of $e^-,\mu^-$ in the lepton pair rest frame.
\item $\Phi_1$ -- The angle between the plane formed by the $M_1$ lepton pair and the `production plane' formed out of the momenta of the incoming partons and the momenta of the two lepton pair systems.
\item $\Phi$ -- The angle between the decay planes of the final state lepton pairs in the rest frame of the $4\ell$ system.
\end{itemize}
These variables are all independent subject to the constraint $(M_1 + M_2) \leq \sqrt{s}$ where $s$ is the invariant mass squared of the $4\ell$ system.  We have also ignored the irrelevant azimuthal production angle.

\begin{figure}
\includegraphics[width=0.40\textwidth]{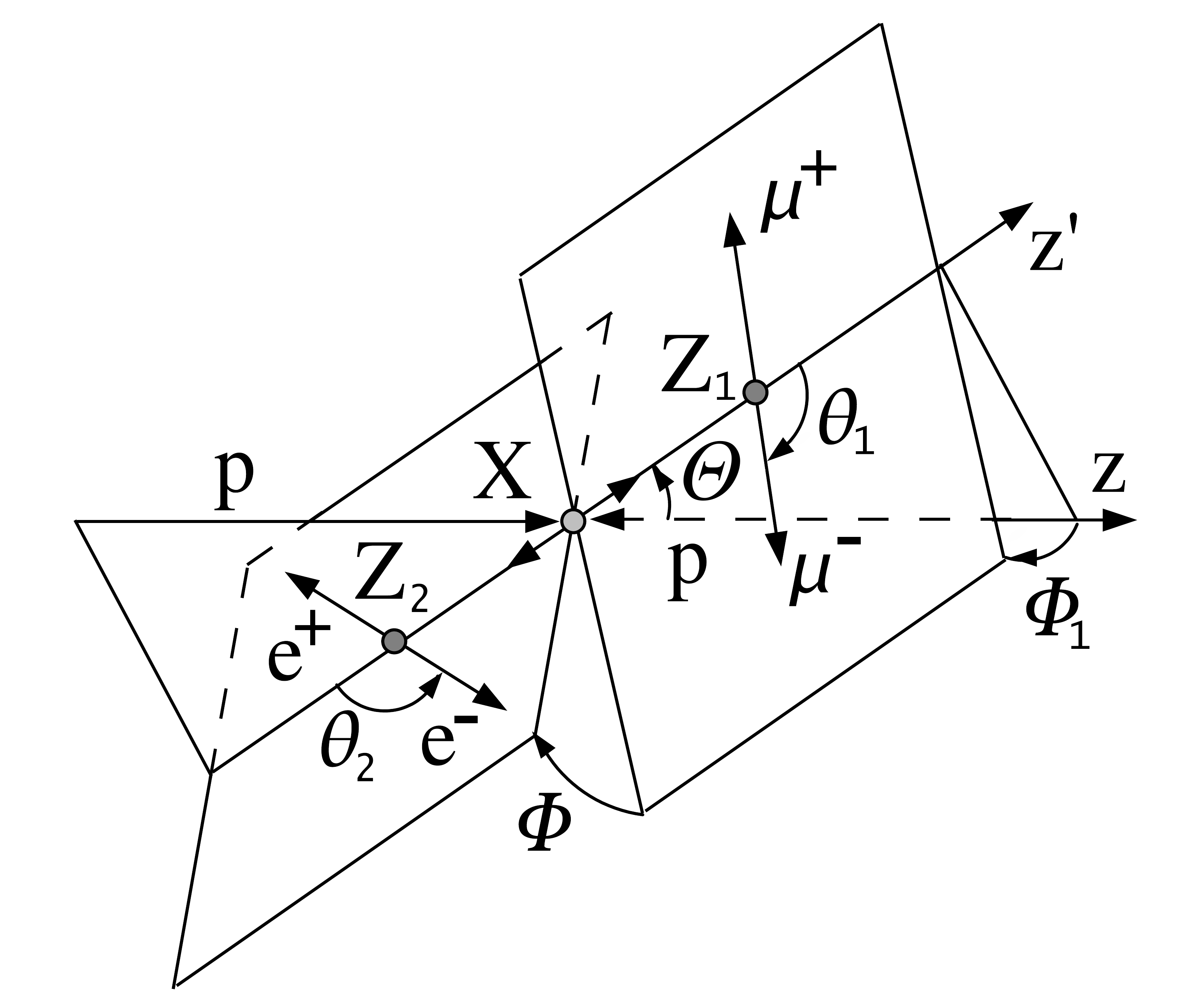}
\caption{Definition of angles in the four lepton CM frame $X$. }
\label{fig:DecayPlanes}
\end{figure}

In the case of the signal events one can replace `lepton pair' momentum with $Z$ or $\gamma$ momentum since in those cases, both lepton pairs do indeed decay from a vector boson and the intuition follows that found in Fig~\ref{fig:DecayPlanes}.  The same can be said for background events which proceed through t-channel pair production.  In these cases, the angle $\Phi_1$ defines the azimuthal angle between the di-boson production plane and the plane formed by the lepton pair which reconstructs to $M_1$ and $\Theta$ is the vector boson production angle.  Other than this more subtle interpretation of the various kinematic variable however, in practice the definitions of these variables are left unchanged from the definitions found in~\cite{Gao:2010qx} which we follow from here on.

\section{Signal}
\label{sec:Signal}

In this section we present the calculation of the signal fully differential cross section and examine the differential mass spectra for several signal hypothesis.  We take our signal to be a general spin-0 scalar and consider all possible couplings to any combination of $Z$ and $\gamma$ pairs allowing for mixtures of both CP even and odd interactions.  Previous studies have analytically computed the $ZZ$~\cite{DeRujula:2010ys,Gao:2010qx} contribution to the golden channel, but as far as we are aware, none consider the contributions from the $Z\gamma$ and $\gamma\gamma$ intermediate states. There are also interference effects between the intermediate state which are not present when $\gamma$ is not allowed to decay.  As we will see, these effects can manifest themselves in the kinematic distributions. Of course for a SM Higgs, the $Z\gamma$ and $\gamma\gamma$ contributions to the golden channel are expected to be small, but this need not be true for a general scalar or if the discovered resonance turns out to have enhanced couplings to $Z\gamma$ or to $\gamma\gamma$.  How large these effects are once one takes into account detector and acceptance effects deserves careful study, but we leave this for ongoing work.  

The most general couplings of a spinless particle to two gauge bosons with four momenta $k_1$ and $k_2$ can be expressed as,  
\begin{eqnarray}
\label{eq:sigvert}
i\Gamma^{\mu\nu}_{ij} &=& v^{-1} 
\Big( A_{1ij} m_Z^2 g^{\mu\nu} + 
A_{2ij} (k_1\cdot k_2 g^{\mu\nu} - k_1^\nu k_2^\mu)\nonumber \\
&& + A_{3ij} \epsilon_{\mu\nu\alpha\beta} k_1^\alpha k_2^\beta \Big)
\end{eqnarray}
where $ij=ZZ,Z\gamma$, or $\gamma\gamma$.  The $A_{1,2,3}$ are  dimensionless arbitrary complex form factors and $v$ is the Higgs vacuum expectation value (vev), which we have chosen as our overall normalization.  For the case of a scalar coupling to $Z\gamma$ or $\gamma\gamma$ electromagnetic gauge invariance requires $A_1 = 0$, while for $ZZ$ it can be generated at tree level as in the SM or by higher dimensional operators.  We have chosen to write the vertex in this form to make the connection with operators in the Lagrangian which may generate them more transparent. For example the following list of operators may generate a coupling as in Eq.(\ref{eq:sigvert}),
\begin{eqnarray}
\label{eq:siglag}
\mathcal{L} &\sim& \frac{1}{v} 
\varphi \Big(g_h m_Z^2Z^\mu Z_\mu + g_Z Z^{\mu\nu}Z_{\mu\nu} + \tilde{g}_Z Z^{\mu\nu} \widetilde{Z}_{\mu\nu} \nonumber \\
&& +~g_{Z\gamma} F^{\mu\nu}Z_{\mu\nu} + \tilde{g}_{Z\gamma} F^{\mu\nu} \widetilde{Z}_{\mu\nu} \nonumber \\
&& +~g_{\gamma} F^{\mu\nu}F_{\mu\nu} + \tilde{g}_{\gamma} F^{\mu\nu} \widetilde{F}_{\mu\nu} + ... \Big)
\end{eqnarray}
where $Z_\mu$ is the $Z$ field while $V_{\mu\nu} = \partial_\mu V_\nu - \partial_\nu V_\mu$ the usual bosonic field strengths.  The dual field strengths are defined as $\widetilde{V}_{\mu\nu} = \frac{1}{2} \epsilon_{\mu\nu\rho\sigma} V^{\rho \sigma}$ and the $...$ is for operators of dimension higher than five.  For a given model many of these are of course zero.  If $\varphi$ is the Standard Model Higgs, then $g_h=i$, while $g_Z$, $g_{Z\gamma}$ and $g_{\gamma\gamma}$ are $\ne 0$, but loop induced and small. Detailed studies of the $ZZ$ contribution to the golden channel mediated through the operators with coefficients $g_h$, $g_Z$ were conducted in~\cite{Cao:2009ah,DeRujula:2010ys,Bolognesi:2012mm}. The operators corresponding to $g_{Z\gamma}$ were studied in~\cite{Stolarski:2012ps} for the golden channel final state and in~\cite{Gainer:2011aa} for the $\ell^+\ell^-\gamma$ final state and both were shown to be useful discriminators. 

Other recent studies of these operators, though not only through the golden channel final state, have also been done. The pseudo scalar couplings $\tilde{g}_Z$, $\tilde{g}_{Z\gamma}$, $\tilde{g}_{\gamma}$ were studied recently in the context of the newly discovered resonance in~\cite{Coleppa:2012eh} where it was shown that a purely CP odd scalar is disfavored as the new resonance.  The analysis of~\cite{Low:2012rj} shows that with a fit of the $\gamma\gamma$, $ZZ^*$, and $WW^*$ rates, as well as the absence of a large anomaly in continuum $Z\gamma$, that the scenario of the four lepton decays being due to $g_Z$ or $g_{Z\gamma}$ is strongly disfavored. While these statements contain few assumptions, they are still model dependent and should be confirmed by direct measurements.  

Even if the newly discovered resonance appears to be `SM like', it is still possible that it can have contributions to the $2e2\mu$ channel coming from operators other than $g_h$ which are slightly enhanced relative to the SM prediction. Here we are motivated by asking what information can be extracted from this channel with out any a-priori reference to other measurements or theoretical input.  In addition, there still exists the possibility that another scalar resonance will be discovered which can also decay to EW gauge boson pairs.  In this case it may have comparable contributions from the various operators.  Thus we allow for all operators in Eq.~(\ref{eq:siglag}) to contribute simultaneously including all interference effects between the $ZZ$, $Z\gamma$, and $\gamma\gamma$ intermediate states.  Because the vertex in terms of arbitrary complex form factors is more general than the Lagrangian, for purposes of the calculation we use Eq.(\ref{eq:sigvert}) explicitly. Below we summarize the details of the calculation.

\subsection{Calculation}
\label{subsec:SigCalc}

\begin{figure}
\includegraphics[width=0.23\textwidth]{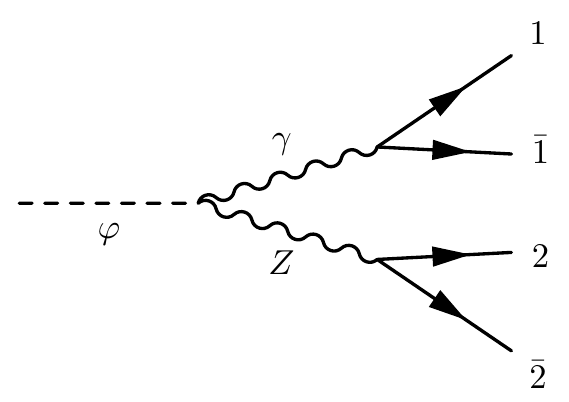}
\includegraphics[width=0.23\textwidth]{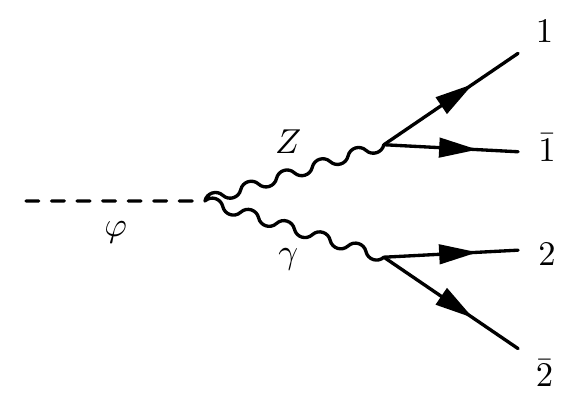}\\
\includegraphics[width=0.23\textwidth]{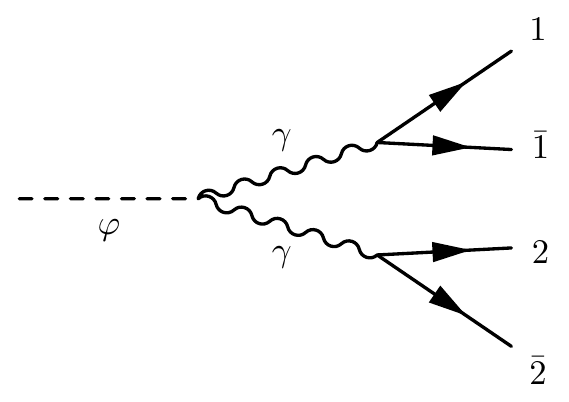}
\includegraphics[width=0.23\textwidth]{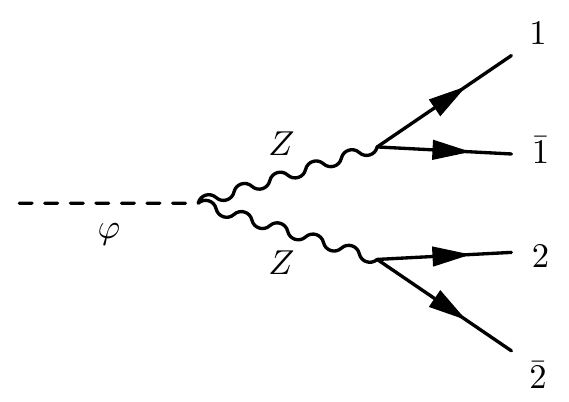}
\caption{Feynman diagrams contributing to $\varphi \rightarrow 2\ell_12\ell_2$. The arrows are to indicate the direction of momentum flow.}
\label{fig:sigdiags}
\end{figure}

To compute the process $\varphi \rightarrow ZZ + Z\gamma + \gamma\gamma \rightarrow 4\ell$ we include the diagrams shown in Fig.~\ref{fig:sigdiags} and parametrize the scalar coupling to gauge bosons as in Eq.~\ref{eq:sigvert}.  The total amplitude can be written as,
\begin{equation}
\begin{array}{ccc}
\label{eq:sigamp}
\mathcal{M} = \mathcal{M}_{ZZ} +\mathcal{M}_{Z\gamma} +\mathcal{M}_{\gamma Z} + \mathcal{M}_{\gamma\gamma} 
\end{array}
\end{equation}
which upon squaring gives,
\begin{equation}
\begin{array}{ccc}
\label{eq:sigamp2}
|\mathcal{M}|^2 = |\mathcal{M}_{ZZ}|^2 +  |\mathcal{M}_{Z\gamma}|^2 +  |\mathcal{M}_{\gamma Z}|^2 +  |\mathcal{M}_{\gamma\gamma}|^2\\
+ 2Re\Big(\mathcal{M}_{ZZ} \mathcal{M}^\ast_{Z\gamma} + \mathcal{M}_{ZZ}\mathcal{M}^\ast_{\gamma Z} + \mathcal{M}_{ZZ}\mathcal{M}^\ast_{\gamma \gamma}\\
\mathcal{M}_{\gamma\gamma} \mathcal{M}^\ast_{Z\gamma} + \mathcal{M}_{\gamma\gamma}\mathcal{M}^\ast_{\gamma Z} + \mathcal{M}_{Z\gamma}\mathcal{M}^\ast_{\gamma Z} \Big)~.
\end{array}
\end{equation}
An explicit calculation of all of these terms is overwhelming, but things can be simplified greatly by taking the final state leptons to be massless.  In this case, the momentum dependent terms in the $Z$ boson propagator numerators do not contribute.  This leads to the propagators of both $Z$ and $\gamma$ to have the same Lorentz structure, namely the Minkowski metric $g_{\mu\nu}$. This implies that all of these terms have the same general \emph{Lorentz} structure.  The only difference from these terms comes from Breit-wigner factors in the propagators as well as in the couplings of the vector bosons, some of which are zero thus `turning off' the contributions from their corresponding Lorentz structure.  To see this, let us consider the amplitude for any combination of intermediate $Z$ and $\gamma$ shown in Fig.\ref{fig:sigdiags},
\small{
\begin{eqnarray}
\label{eqn:SigAmplitude}
&&\mathcal{M}_{ij} = 
\bar{u}_2 \left( i\gamma^\gamma (g_{2R}^j P_R + g_{2L}^j P_L) \right) v_2
\left( \frac{-ig_{\nu\gamma}}{M_2^2 - m_j^2 + im_j\Gamma_j} \right)\nonumber \\
&&\Gamma^{\mu\nu}_{ij}
\left( \frac{-ig_{\mu\sigma}}{M_1^2 - m_i^2 + im_i\Gamma_i} \right)
\bar{u}_1  \left( i\gamma^\sigma (g_{1R}^i P_R + g_{1L}^i P_L) \right) v_1
\end{eqnarray}}
\\
where $i,j$ label $Z$ or $\gamma$ while $1$ and $2$ label the final state leptons and can in principal be $e$ or $\mu$.  In the $4e$ and $4\mu$ case one must also include the interference between identical particles, but we do not address that issue here\footnote{We have computed it in~\cite{Chen:2013ejz}.}. Upon squaring the amplitude and summing over final state lepton polarizations we can obtain a general amplitude squared which encompasses any of terms in Eq.(\ref{eq:sigamp2}) and is given by,
\begin{eqnarray}
\label{eq:gensigamp2}
 \mathcal{M}_{ij}\mathcal{M}^\ast_{\bar{i}\bar{j}} = 
 (D_{1i} D_{2j} D^\ast_{1\bar{i}} D^\ast_{2\bar{j}})^{-1}\nonumber \\
(g_{\mu\sigma} \mathcal{T}_{1i\bar{i}}^{\sigma\bar{\sigma}} g_{\bar{\mu}\bar{\sigma}})(g_{\nu\gamma} \mathcal{T}_{2 j\bar{j}}^{\gamma\bar{\gamma}} g_{\bar{\nu}\bar{\gamma}})\Gamma_{ij}^{\mu\nu}\Gamma_{\bar{i}\bar{j}}^{\ast\bar{\mu}\bar{\nu}}
\end{eqnarray}
where
\begin{equation}
\begin{array}{ccc}
\label{eq:leptrace}
\mathcal{T}_{1i\bar{i}}^{\sigma\bar{\sigma}} = (g^i_{1R} g^{\bar{i}}_{1R} + g^i_{1L} g^{\bar{i}}_{1L})Tr(\not p_1\gamma^\sigma \not p_{\bar{1}} \gamma^{\bar{\sigma}})/2\\
+ (g^i_{1R} g^{\bar{i}}_{1R} - g^i_{1L} g^{\bar{i}}_{1L})Tr(\not p_1\gamma^\sigma \not p_{\bar{1}} \gamma^{\bar{\sigma}}\gamma^5)/2\\
\\
D_{1i} =  M_1^2 - m_i^2 + i\Gamma_i m_i 
\end{array}
\end{equation}
and $\Gamma_{ij}^{\mu\nu}$ are given in Eq.(\ref{eq:sigvert}).  The $g^i_{R,L}$ are at this point general left and right handed couplings of a `$Z$-like' spin-1 vector boson to a pair of fermions.  The bars are to indicate that the corresponding index belongs to the conjugated amplitude and are distinct indices from the un-bared ones.  We treat all couplings at every vertex encountered when tracing over the Dirac strings as distinct as well as all Breit-Wigner factors so for any amplitude squared term there can in principal be four different vector bosons as intermediate states.  In the case of the photon we have of course $g^\gamma_{R}=g^\gamma_{L}=-e_{em}$ and $m_\gamma=\Gamma_\gamma=0$. Since at this stage the various couplings and masses are completely general, Eq.(\ref{eq:gensigamp2}) applies to any process where a scalar decays to two spin-1 vector bosons which then decay to massless fermions through `Z-like' couplings.  

Expanding out the terms in Eq.(\ref{eq:gensigamp2}) we can write the amplitude squared as,
\begin{equation}
\begin{array}{ccc}
\label{eq:gensigamp2exp}
\mathcal{M}_{ij}\mathcal{M}^\ast_{\bar{i}\bar{j}} = \mathcal{C}^{++}_{ij\bar{i}\bar{j}} L^{++}_{ij\bar{i}\bar{j}} + \mathcal{C}^{+-}_{ij\bar{i}\bar{j}} L^{+-}_{ij\bar{i}\bar{j}}  \ +\\
\\
\mathcal{C}^{-+}_{ij\bar{i}\bar{j}} L^{-+}_{ij\bar{i}\bar{j}} + \mathcal{C}^{--}_{ij\bar{i}\bar{j}} L^{--}_{ij\bar{i}\bar{j}} = \sum\limits_{ab}  \mathcal{C}^{ab}_{ij\bar{i}\bar{j}} L^{ab}_{ij\bar{i}\bar{j}} 
\end{array}
\end{equation}
where $a,b = (+,-)$ with $a$ and $b$ corresponding to the fermion pairs labeled $1$ and $2$ respectively and
\begin{equation}
\begin{array}{ccc}
\label{eq:gaugestruc}
\mathcal{C}^{\pm\pm}_{ij\bar{i}\bar{j}} = \frac{(g^i_{1R} g^{\bar{i}}_{1R} \pm g^i_{1L} g^{\bar{i}}_{1L})(g^j_{2R} g^{\bar{j}}_{2R} \pm g^j_{2L} g^{\bar{j}}_{2L})}{4(D_{1i} D_{2j} D^\ast_{1\bar{i}} D^\ast_{2\bar{j}})}\\
\\
L^{\pm\pm}_{ij\bar{i}\bar{j}} = (g_{\mu\sigma} T_{1\pm}^{\sigma\bar{\sigma}} g_{\bar{\mu}\bar{\sigma}})(g_{\nu\gamma} T_{2\pm}^{\gamma\bar{\gamma}} g_{\bar{\nu}\bar{\gamma}})\Gamma_{ij}^{\mu\nu}\Gamma_{\bar{i}\bar{j}}^{\ast\bar{\mu}\bar{\nu}}~.
\end{array} 
\end{equation}
The $T^{\sigma\bar{\sigma}}_{1\pm}$ are the Dirac traces found in Eq.(\ref{eq:leptrace}) and $\pm$ indicates whether the trace ends with a $\gamma^5$~($-$) or not~($+$). The full amplitude squared can then be built out of the objects\footnote{Expressions for the various coefficients and Lorentz structure can be obtained by emailing the corresponding author.}  in Eq.(\ref{eq:gaugestruc}), 
\begin{equation}
\begin{array}{ccc}
\label{eq:simgensigamp2}
(\mathcal{M}_{ij}\mathcal{M}^\ast_{\bar{i}\bar{j}})^{ab} = \mathcal{C}^{ab}_{ij\bar{i}\bar{j}} L^{ab}_{ij\bar{i}\bar{j}}~.
\end{array}
\end{equation}
Since all of the angular information is contained in the $L^{ab}_{ij\bar{i}\bar{j}}$ we can take advantage of the simple nature of these terms to perform the desired integration \emph{before} summing over $\pm$ and the various vector boson intermediate states, after which analytic integration becomes unmanageable.  Expressions for the $L^{ab}_{ij\bar{i}\bar{j}}$ are obtained in terms of invariant dot products and CM variables. These  objects can be used to build the differential cross section of any scalar decay to four massless fermions via two spin-1 vector bosons.  From these one can also reproduce analytic results for other processes such as the semi-leptonic decay of the Higgs to $\ell\nu j j$~\cite{Dobrescu:2009zf}. 

The final fully differential decay width can now be written as,
\begin{equation}
\begin{array}{ccc}
\label{eq:sigdiffcxn}
\frac{d\Gamma_{\varphi}}{dM_1^2dM_2^2d\Omega} = \Pi_{4\ell} \sum\limits_{ab} \Big(\sum\limits_{ij\bar{i}\bar{j}} \mathcal{C}^{ab}_{ij\bar{i}\bar{j}} L^{ab}_{ij\bar{i}\bar{j}} \Big)
\end{array}
\end{equation}
where $d\Omega=dc_\Theta dc_{\theta_2} dc_{\theta_1} d\Phi d\Phi_1$ ($c_\theta = \cos{\theta}$) and $ \Pi_{4\ell}$ is the final state lepton four body phase space derived following~\cite{Nakamura:2010zzi} and given by,
\begin{eqnarray}
\label{eq:phasespace}
&&\Pi_{4\ell}=
(\frac{1}{2 \pi})^2 (\frac{1}{32 \pi^2})^2 (\frac{1}{32 \pi s}) \nonumber \\
&&\cdot\Big(1 + \frac{(M_1^2 - M_2^2)^2}{s^2} - \frac{2(M_1^2 + M_2^2)}{s}\Big)^{1/2}~.
\end{eqnarray}
We can obtain the differential mass spectrum\footnote{We give an analytic expression for a particular hypothesis in the Appendix.} via,
\begin{equation}
\begin{array}{ccc}
\label{eq:diffmassspec}
\frac{d\Gamma_{\varphi}}{dM_1^2dM_2^2} =  \Pi_{4\ell} \sum\limits_{ab} \Big(\sum\limits_{ij\bar{i}\bar{j}} \mathcal{C}^{ab}_{ij\bar{i}\bar{j}} (\int d\Omega L^{ab}_{ij\bar{i}\bar{j}}) \Big)~.   
\end{array}
\end{equation}
We note that we perform the sum over vector bosons before the sum over $\pm$ which allows for greater simplification of the expressions.  We can now go on to examine the differential mass spectrum for different signal hypothesis.  In the Appendix we show various singly and doubly differential spectra for a number of signal hypotheses. We also give in Eq.(\ref{eq:sigm1m2ZA}) of the Appendix, an explicit expression for the doubly differential mass spectrum of a scalar with SM-like $ZZ$ couplings and both CP even and CP odd $Z\gamma$ couplings including all interference effects.
\subsection{The Differential Mass Spectra}
\label{sec:SigSinglyDoubly}
In this section we examine the singly differential mass spectra for various signal hypotheses and give a feel for how $M_1$ and $M_2$ might be able to distinguish between the different operators in Eq.(\ref{eq:siglag}). Explicitly we consider the following cases\footnote{We have validated these cases with FeynRules/CalcHEP~\cite{Belyaev:2012qa,Christensen:2008py} and the Monte Carlo generator introduced in~\cite{Gao:2010qx}.},
\begin{itemize}
\item ${\bf 1}$: SM including $Z\gamma$ and $\gamma\gamma$ ($A_{1ZZ} = 2, A_{2Z\gamma} = 0.007, A_{2\gamma\gamma} = -0.008$)\footnote{Values obtained from~\cite{Low:2012rj} after translating to our parametraziation.}
\item ${\bf 2}$: SM coupling to $ZZ$ plus enhanced $Z\gamma$ and $\gamma\gamma$ ($A_{1ZZ} = 2, A_{2Z\gamma} = 6*0.007, A_{2\gamma\gamma} = -1.3*0.008$)
\item ${\bf 3}$: SM coupling to $ZZ$ plus CP odd couplings to  $\gamma\gamma$ and  $Z\gamma$ ($A_{1ZZ} = 2, A_{3Z\gamma} = 0.01, A_{3\gamma\gamma} = 0.01$)
\item ${\bf 4}$:  CP odd/even mixed coupling to $ZZ$ only ($A_{1ZZ}=2, A_{3ZZ}= 0.1$)
\item ${\bf 5}$:  General Scalar ($A_{1ZZ} = 0.1,A_{2ZZ} = 1, A_{2Z\gamma} = 0.01, A_{2\gamma\gamma} = 0.01,A_{3ZZ} = 1, A_{3Z\gamma} = 0.01, A_{3\gamma\gamma} = 0.01$)
\end{itemize}
where we also show the values for the couplings chosen in Eq.(\ref{eq:sigvert}). Couplings whose values are not shown in a given hypotheses are taken to be zero and we take all values at $\sqrt{s}=m_h=125$ GeV. Note that all of these couplings can be interpreted in terms of the couplings in Eq.(\ref{eq:siglag}) if we assume only up to dimension 5 operators contribute. 

\begin{figure}
\includegraphics[width=0.45\textwidth]{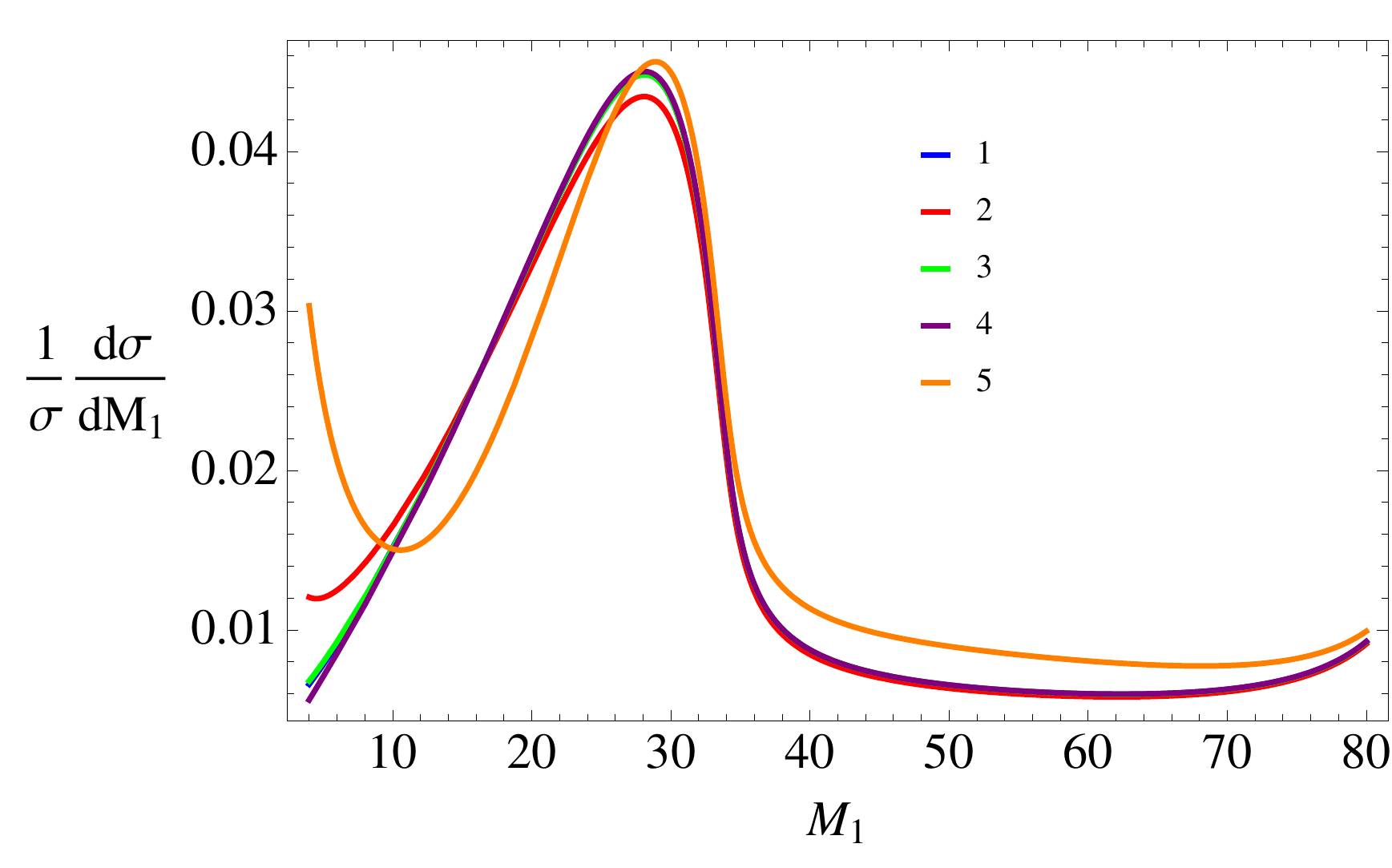}
\includegraphics[width=0.45\textwidth]{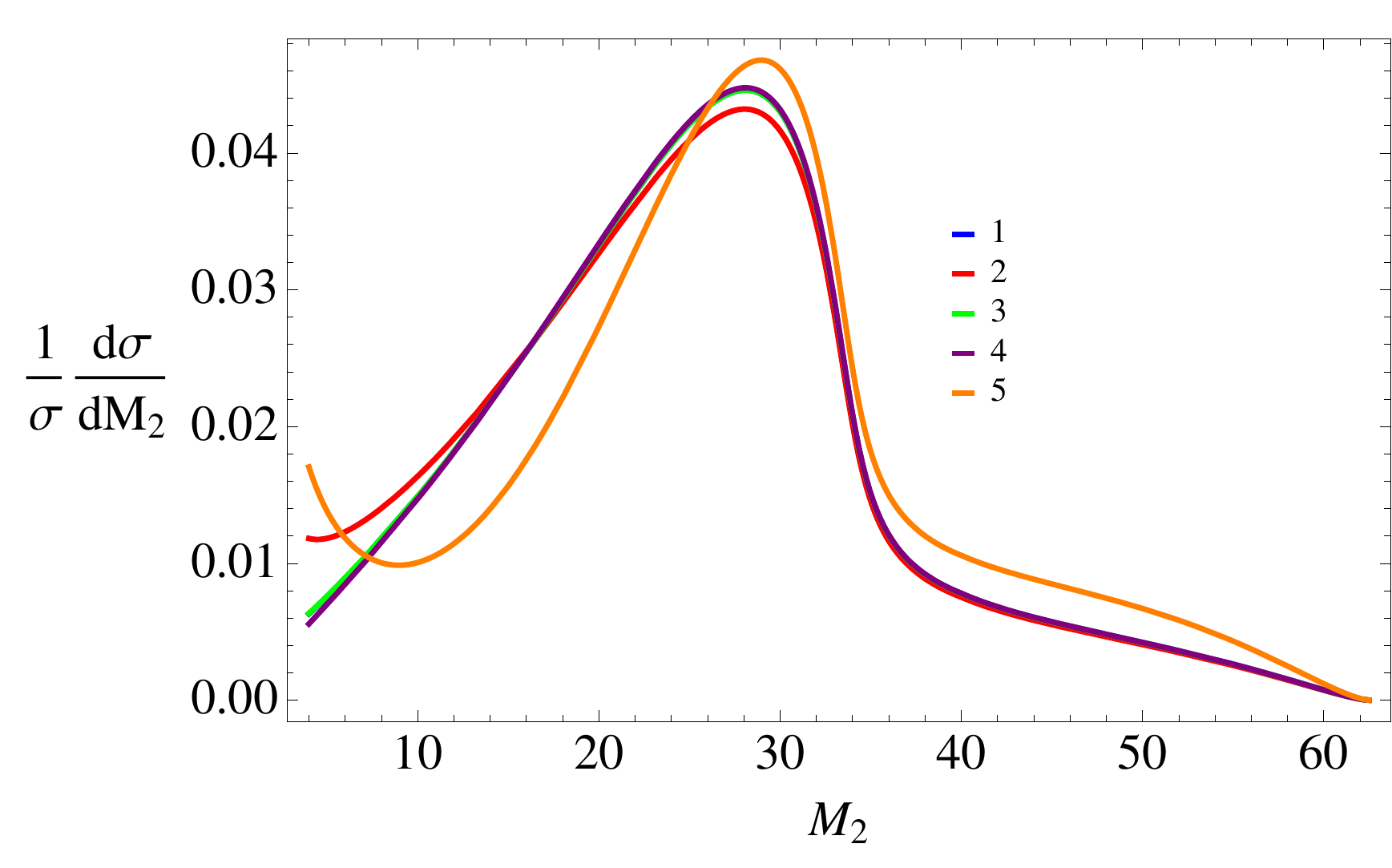}
\caption{In the top plot we take 4~GeV $< M_{1,2} < 120$ GeV while in the bottom plot we take the range 40~GeV $< M_1 < 120$~GeV,~4~GeV $< M_2 < 120$ GeV at $\sqrt{s}=m_h=125$ GeV when integrating over phase space. The SM is shown in blue, but is essentially indistinguishable from hypothesis 3 (see text).}
\label{fig:massspecs}
\end{figure}

We obtain the differential mass spectra via Eq.(\ref{eq:diffmassspec}), followed by integration over $M_1$ or $M_2$, and compare them for different hypotheses.  These are shown in Fig.~\ref{fig:massspecs} for two ranges.  The first range we take 4~GeV$ < M_{1,2}< 120$~GeV treating $M_1$ and $M_2$ symmetrically shown in the top plot.  In this case we only show the $M_1$ distribution since it is identical to the $M_2$ distribution and only show the lower mass region above which the different cases are very similar.

We also consider the more `experimental' cut requiring a wide window around the $Z$ boson mass 40~GeV $< M_1 < 120$~GeV  and 4~GeV $< M_2<$ 120~GeV for the `off-shell' vector boson. In this case the $M_1$ distribution is indistinguishable for the separate cases so we only show the $M_2$ distribution. One can see, that in particular in the low mass region, these variables can be highly discriminating between the different cases. We point out also that, if values of $M_2 \lesssim 10$~GeV can be probed, the requirement of a window around a $Z$ boson may decrease sensitivity to certain hypotheses which have a sizable $\gamma\gamma$ or $Z\gamma$ component such as hypothesis 5. 

Our lower bound on $M_2$ is chosen to be $4$ GeV since lower values of $M_2$ runs the risk of contamination from $J/\psi$ states whose mass is $\sim 3$ GeV. We emphasize that experimental analyses should be made to push down as far possible since as can be seen in Fig.~\ref{fig:massspecs}, one needs to be able to probe $M_2$ below $\sim 10$ GeV in order to discriminate between a SM scalar (hypothesis 1) and one with enhanced $Z\gamma$ and $\gamma\gamma$ couplings  (hypothesis 2) for example. Though current experimental signal searches in the golden do not yet consider such low values for $M_2$, it seems feasible to push the $M_2$ cut down further as was done in the CMS observation of the $Z\rightarrow \ell^+\ell^-\ell^+\ell^-$ process~\cite{CMS:2012bw}. We therefore include this highly interesting region here and hope that it may motivate efforts to push the $M_2$ reach lower. We leave a complete analysis including detector effects to an ongoing study.


\section{Background}
\label{sec:Background}

The dominant irreducible background to the golden channel comes from $q\bar{q}$ annihilation into gauge bosons.  At energies $\sim 125$ GeV the dominant contribution comes from t-channel $Z\gamma$ production.  However, as we will see contributions from s-channel $Z\rightarrow 4\ell$ diagrams can effect the angular distributions such as the distribution of the angle between the decay planes $\Phi$ defined in Sec.\ref{sec:events}.  Furthermore, we include the $ZZ$ and $\gamma\gamma$ contributions since in principal these are always present and may have observable interference effects due to the fact that they add at the amplitude level when decaying to charged leptons.  In addition, the inclusion of these contributions allows for considering a much larger energy range in one fully differential cross section than can be considered when including only the $t$ and $u$ channel contributions.  Of course NLO effects, including the $gg$ initiated process~\cite{Zecher:1994kb, Binoth:2008pr, Kauer:2012hd} will contribute as well, but these are expected to be small and mainly only effect the `input' invariant mass (and overall normalization) for the fully differential cross sections.  We will examine this point below.  

It should also be noted that ideally one would like to include the $4e$ and $4\mu$ final states which in some kinematic regimes can have non-negligible contributions from interference between final state particles~\cite{Avery:2012um}.  The inclusion of this channel would allow for greater sensitivity for the same amount of luminosity.  However, because of the interference between identical final states in this case, the Lorentz structure becomes severely more complicated and we thus leave this calculation for future work. 

\subsection{Calculation}
\label{subsec:BackCalc}

\begin{figure}
\includegraphics[width=0.23\textwidth]{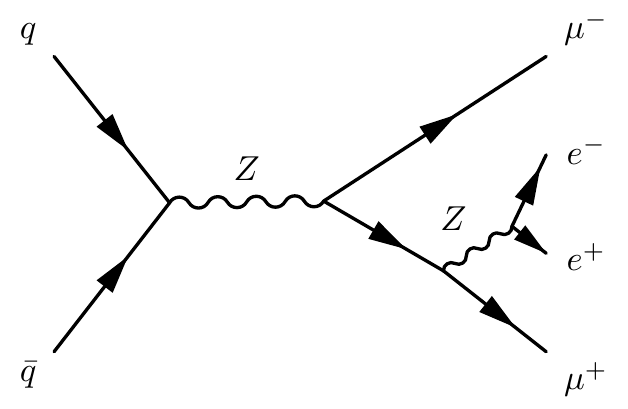}
\includegraphics[width=0.23\textwidth]{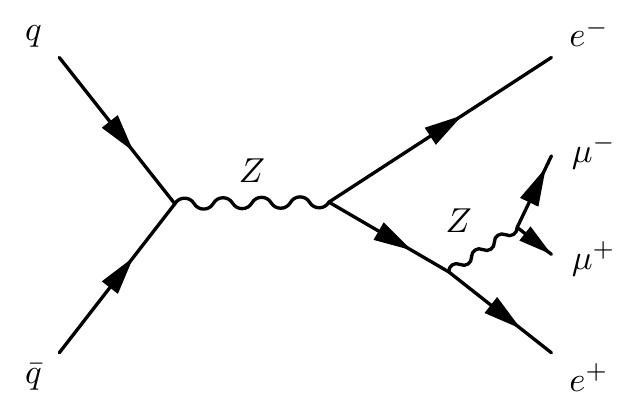}\\
\includegraphics[width=0.23\textwidth]{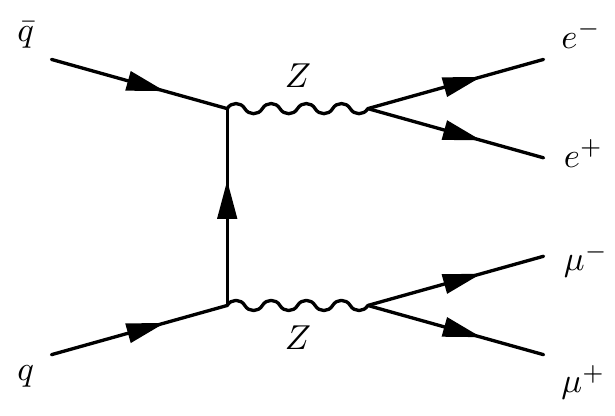}
\caption{Feynman diagrams contributing to $q\bar{q} \rightarrow ZZ \rightarrow 2e2\mu$ and $q\bar{q} \rightarrow Z \rightarrow 2e2\mu$. The arrows are to indicate the direction of momentum flow.}
\label{fig:Zdiags}
\end{figure}

The background calculation is much more involved than the signal calculation due to a higher number of Feynman diagrams in addition to a more complicated Lorentz structure.  As in the signal case the amplitude can be written as,
\begin{equation}
\begin{array}{ccc}
\label{eq:bgamp}
\mathcal{M} = \mathcal{M}_{ZZ} +\mathcal{M}_{Z\gamma} +\mathcal{M}_{\gamma Z} + \mathcal{M}_{\gamma\gamma}~.
\end{array}
\end{equation}
Now however, each of these amplitudes breaks down into six `sub-amplitudes'.  To see this, let us first consider the $ZZ$ mediated decays. There are three diagrams which contribute to the $2e2\mu$ process shown in Fig.~\ref{fig:Zdiags}.~First there is the t-channel contribution shown in the bottom diagram.  This contribution (and its u-channel counterpart) has been computed previously for both on-shell~\cite{Hagiwara:1986vm} and off-shell~\cite{Gainer:2011xz} $Z$ bosons.  The second contribution comes from resonant $2e2\mu$ production proceeding through the top two diagrams.  Each of these diagrams also has a corresponding `crossed' diagram taking into account the other possibility for attaching the vector boson lines.  This gives six diagrams for the $ZZ$ contribution to the golden channel.  Similarly, there are six more for the $\gamma\gamma$ contribution plus six for $Z\gamma$ and six for $\gamma Z$ giving a total of twenty four diagrams.  At first this many diagrams can seem intractable, but as we will see, when organized in a proper manner the calculation is relatively straightforward with the help of Tracer~\cite{Jamin:1993} to perform the Lorentz contraction.  

\begin{figure}
\includegraphics[width=0.23\textwidth]{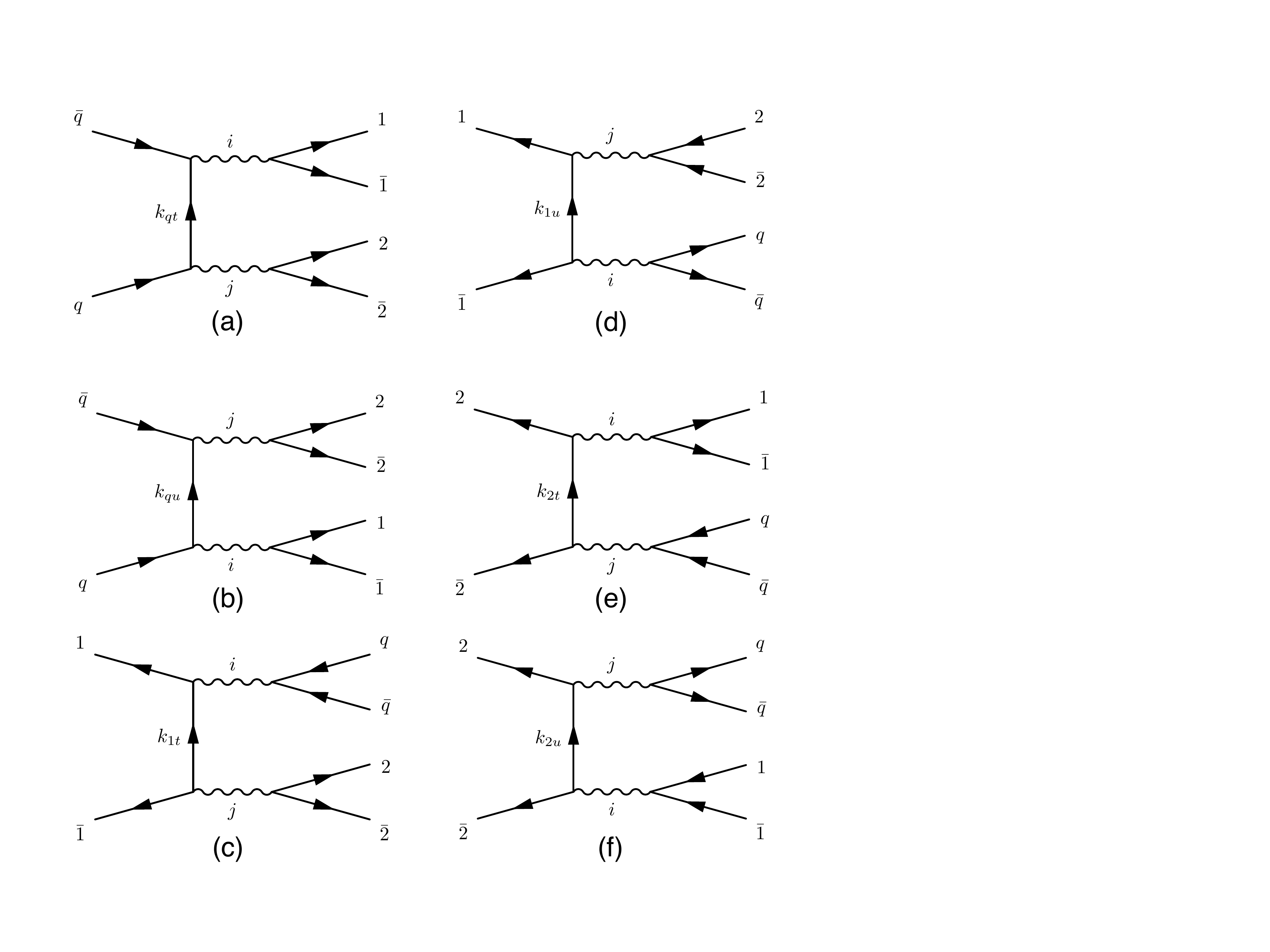}
\includegraphics[width=0.23\textwidth]{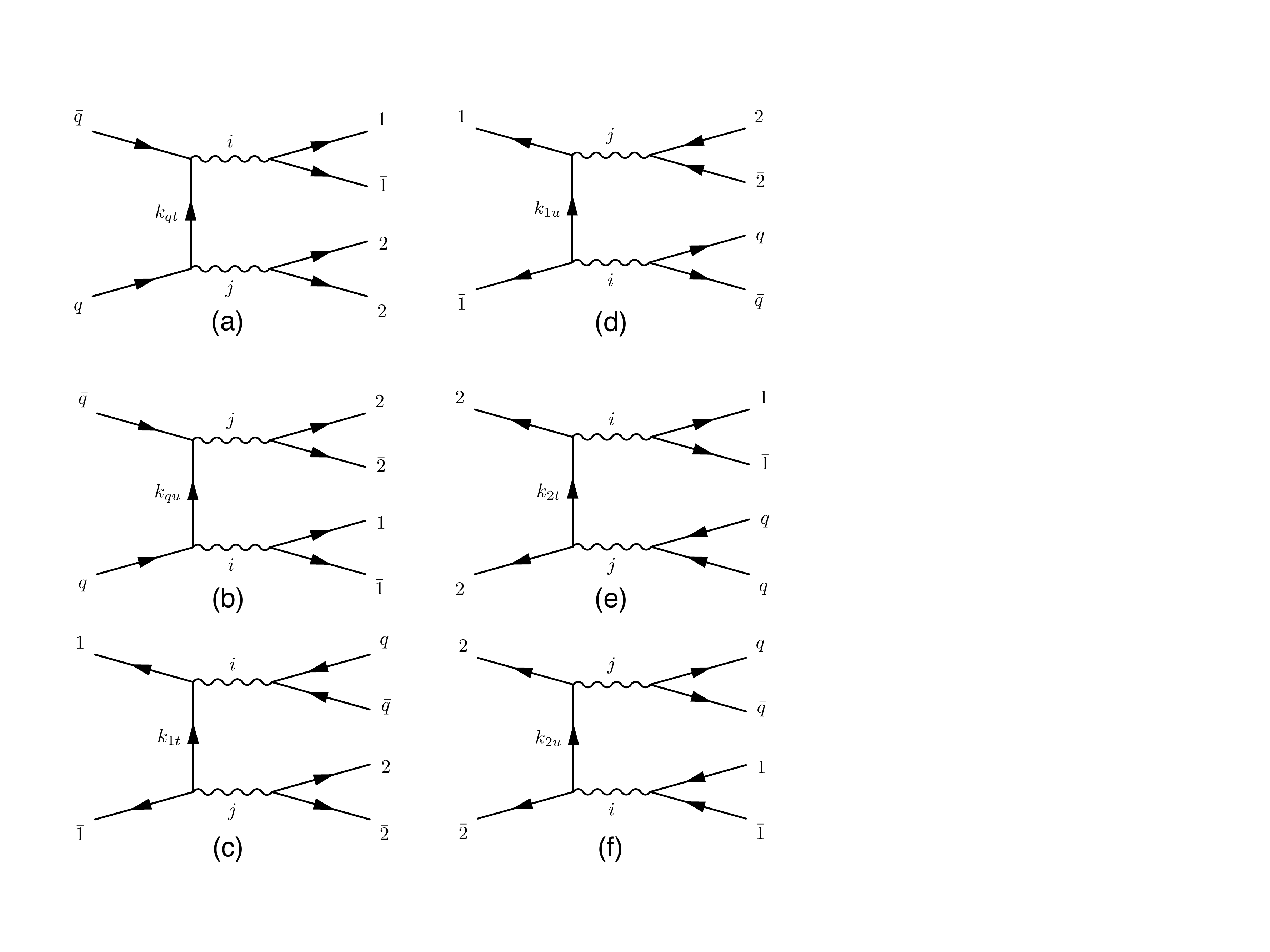}\\
\includegraphics[width=0.23\textwidth]{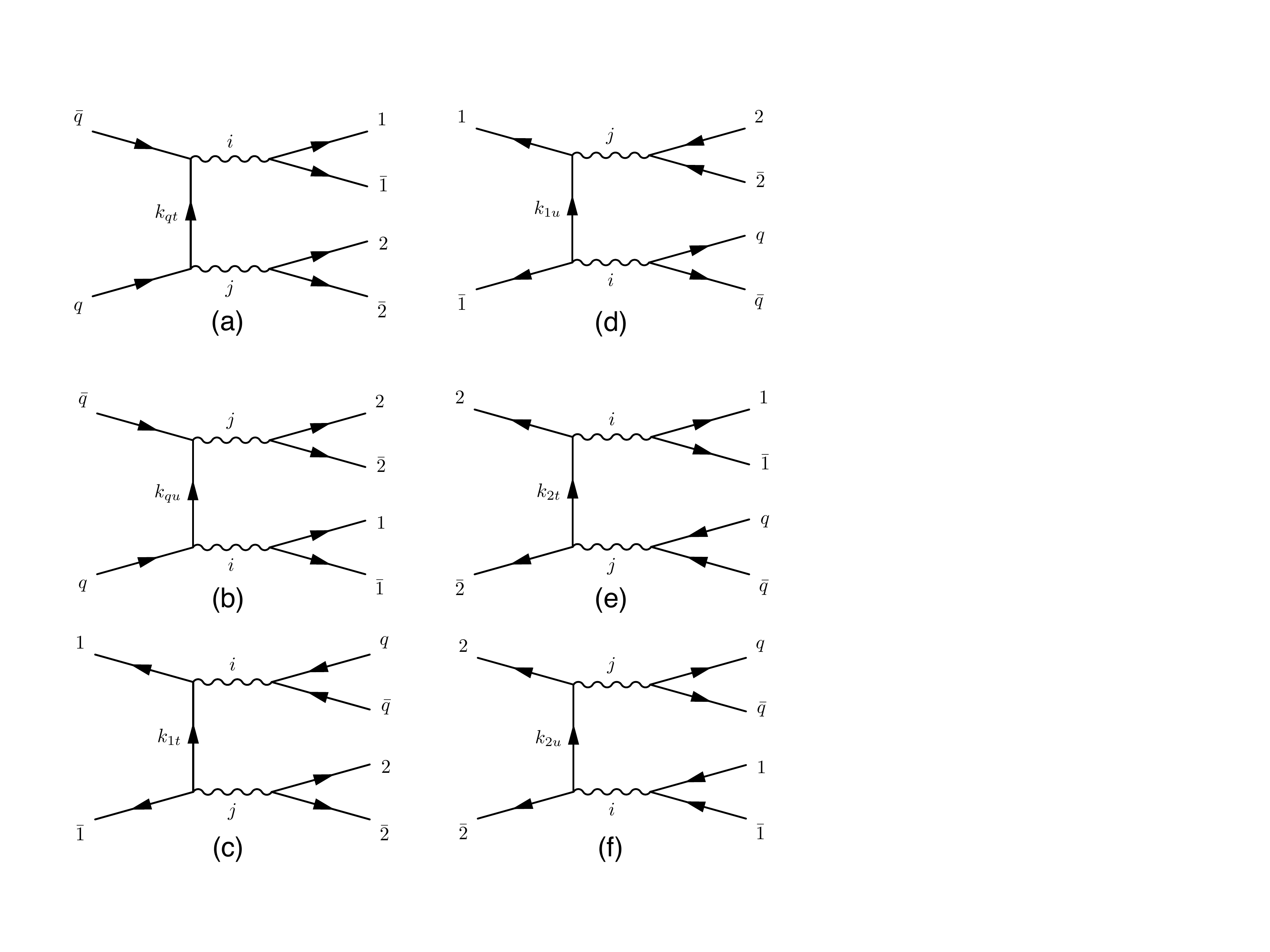}
\includegraphics[width=0.23\textwidth]{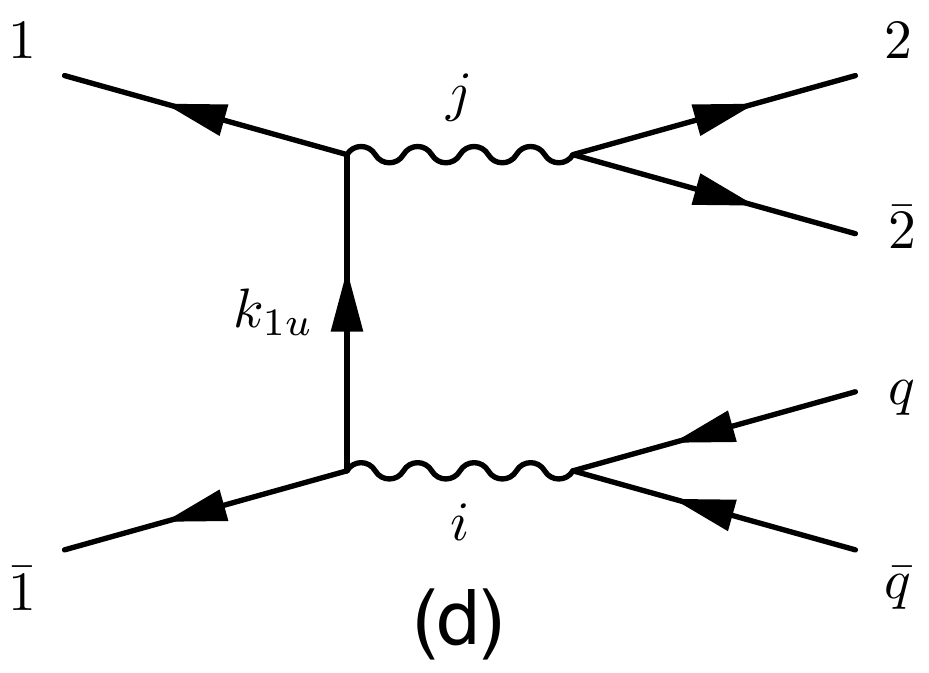}\\
\includegraphics[width=0.23\textwidth]{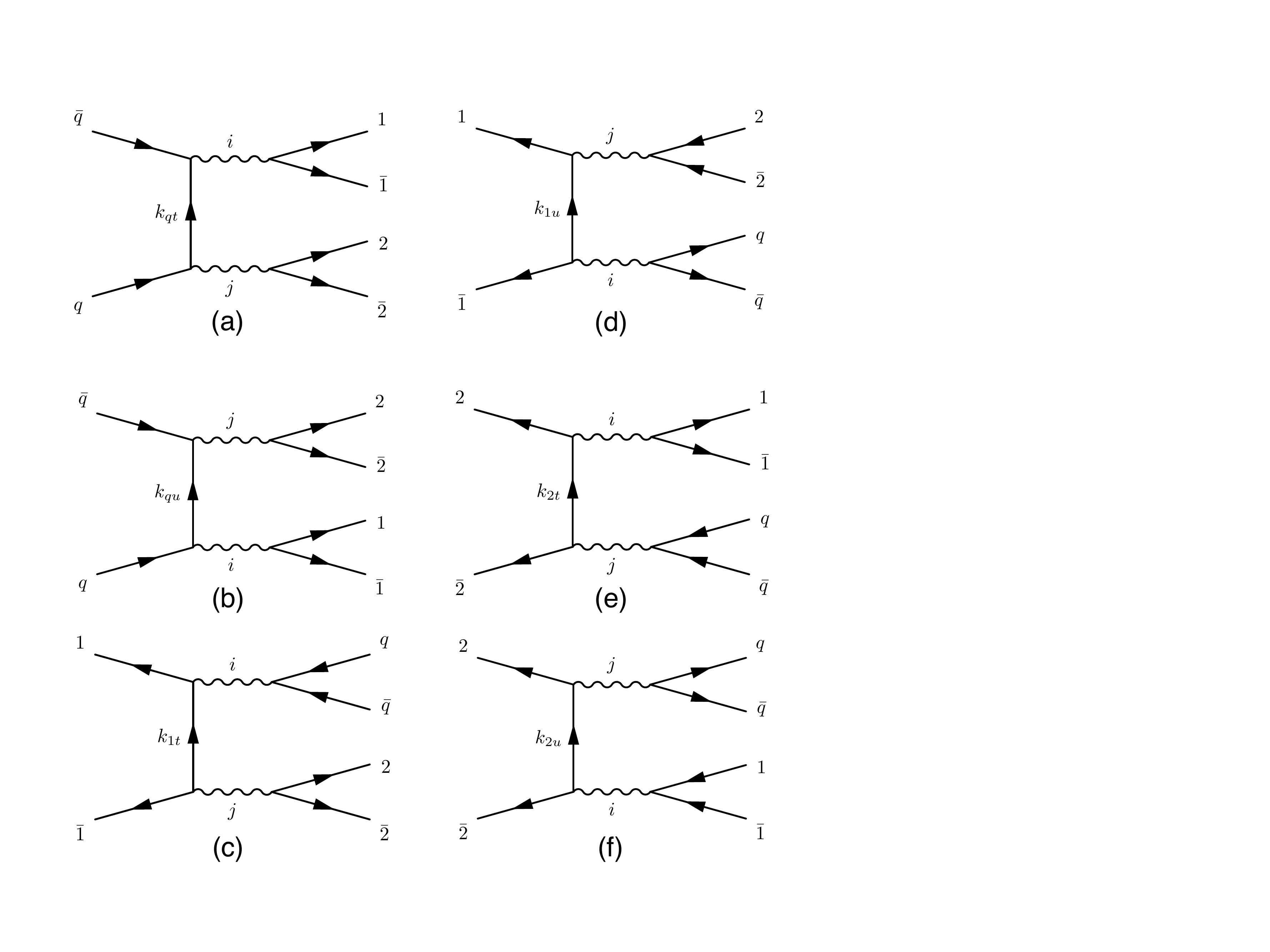}
\includegraphics[width=0.23\textwidth]{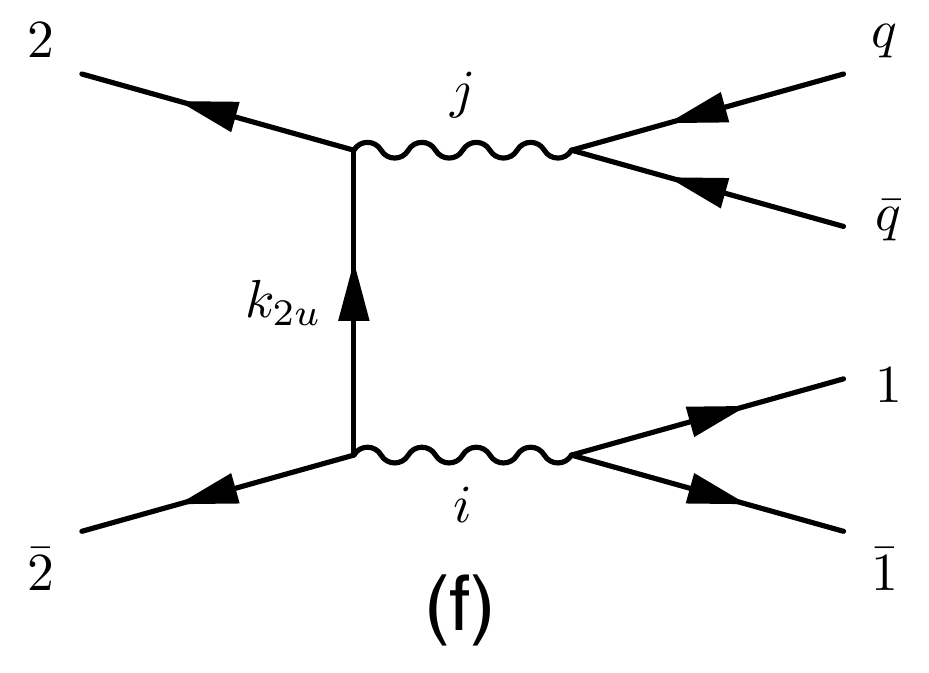}
\caption{Feynman diagrams contributing to $q\bar{q} \rightarrow V_i V_j\rightarrow 2\ell_1 2\ell_2$ and $q\bar{q} \rightarrow V_i\rightarrow 2\ell_1 2\ell_2$.  Note that diagrams $( c)-(f)$ are in fact s-channel diagrams so the fermions labeled by $1$ and $2$ are not to be confused as being in the initial state.  This is taken into account in how the various momenta are assigned as indicated by the arrows.}
\label{fig:XYZdiags}
\end{figure}

To begin we first note that the six diagrams can be `twisted' and arranged into the form found in Fig.~\ref{fig:XYZdiags} where we now allow the vector bosons to take on any $Z$ or $\gamma$, but once chosen are treated as fixed. We use the conventions indicated in the diagrams and in particular refer to the diagrams $(a)$, $( c )$, and $(e)$ as `t-channel' type diagrams and $(b)$, $(d)$, and $(f)$ as `u-channel'.  This is not to be confused with the typical vocabulary for this process which refers to diagrams $(a)$ and $(b)$ as t and u channel and diagrams $( c)-(f)$ as s-channel.  We find this re-naming convenient for organizing and reducing the many terms which need to be computed for the differential cross section. The Lorentz structure for all of these amplitudes is clearly the same.  One needs only to keep proper track of how the various momentum are routed through each diagram.  We can see this by considering the amplitude explicitly.  Using the massless initial quark and final state lepton approximation we can write any of the amplitudes in Fig.~\ref{fig:XYZdiags} as,
\small{
\begin{equation}
\label{eq:BgAmplitude}
\begin{array}{ccc}
\mathcal{M}^n_{Xij} =
\bar{u}_Z \left( i\gamma^\sigma (g_{ZR}^j P_R + g_{ZL}^j P_L) \right) v_Z
\left( \frac{-ig_{\mu\sigma}}{M_Z^2 - m_j^2 + im_j\Gamma_j} \right) \\
\bar{v}_X 
\left( i\gamma^\mu (g_{XR}^i P_R + g_{XL}^i P_L) \right)
\left( \frac{i\not k_{Xn}}{k_{Xn}^2} \right)
\left( i\gamma^\nu (g_{XR}^j P_R + g_{XL}^j P_L) \right)
u_X \\
\left( \frac{-ig_{\nu\gamma}}{M_Y^2 - m_i^2 + im_i\Gamma_i} \right)
\bar{u}_Y  \left( i\gamma^\gamma (g_{YR}^i P_R + g_{YL}^i P_L) \right) v_Y\
\end{array}
\end{equation}
}
\\
where we label the amplitude by the `long' dirac string, in this case $X$. The labels $X/Y/Z = 1,2,q$ where $1,2$ are for final state lepton pairs while $q$ is for the initial state quarks.  The $i,j = Z, \gamma$ label the vector bosons and $n = t, u$ labels the t and u-channel diagrams in our new vocabulary. The invariant masses are defined as $M_Y^2 = (p_Y + p_{\bar{Y}})^2$.  The internal fermion momentum are given in terms of external momentum by,
\begin{eqnarray}
\label{eq:fermionmom}
&&k_{qt} = p_q - (p_1 + p_{\bar{1}}),~~k_{qu} = p_q - (p_2 + p_{\bar{2}})\nonumber\\
&&k_{1t} = - p_{\bar{1}} -( p_2 + p_{\bar{2}}),~~k_{1u} = (p_q + p_{\bar{q}}) - p_{\bar{1}}\nonumber\\
&&k_{2t} = (p_q + p_{\bar{q}}) - p_{\bar{2}},~~k_{2u} = - p_{\bar{2}} - (p_1 + p_{\bar{1}})~.
\end{eqnarray}  
Note that the invariant masses $M_Y$ and $M_Z$ do not necessarily correspond to the invariant mass formed by the final state lepton pairs, as they do in the signal case and in previous analytic calculations of the golden channel background which neglect the s-channel diagrams. Now with the inclusion of the resonant four lepton processes in $( c )-(f)$ we have for these diagrams $M_{Y/Z}^2 = M_q^2  = s$ which is equal to the invariant mass of the four lepton system.  To obtain any of the physical amplitudes one simply assigns the appropriate labels to Eq.(\ref{eq:BgAmplitude}) as well as the appropriate momenta. Thus for example for diagram $( c)$ we have $X \rightarrow 1$, $Y \rightarrow q$, $Z \rightarrow 2$, and $n \rightarrow t$.  To switch from t-channel type to u-channel diagrams (staying in the same row in Fig.\ref{fig:XYZdiags}) one simply takes $t\rightarrow u$ and $\gamma^\sigma \leftrightarrow \gamma^\gamma$.  Of course at this stage all these labels are arbitrary meaning that the amplitude in Eq.(\ref{eq:BgAmplitude}) applies to any process with this topology and Lorentz structure.  Note that for the $Z$ propagators we drop the momentum dependent terms since they do not contribute in the massless lepton approximation.  As mentioned in the signal case, for the photon $g^\gamma_R=g^\gamma_L$ and $m_\gamma=\Gamma_\gamma=0$, but for now we take the couplings and propagators as general.  

As in the case of the signal, the next step is to find a generalized amplitude squared for any two of the six diagrams.  Although there are in principal thirty six terms when squaring the amplitudes, these organize themselves into only two distinct types of Lorentz structure.  The first type is found when multiplying any two diagrams in the same row of Fig.~\ref{fig:XYZdiags}.  This is the Lorentz structure found in our previous calculations of the $ZZ$ contribution in which only diagrams $\mathcal{M}_{qt}$ and $\mathcal{M}_{qu}$ are included (the top row). The square of $\mathcal{M}_{1t} +\mathcal{M}_{1u}$ and $\mathcal{M}_{2t} +\mathcal{M}_{2u}$ (second and third rows) will also exhibit this Lorentz structure.  The second type of Lorentz structure is obtained when taking the product of any two diagrams in different rows.  In the conventional language, interference between the first and second row or first and third row corresponds to interference between the t-channel di-boson production amplitudes and the s-channel diagrams.  Interference between the second and third row corresponds to interference between the two types of s-channel diagrams.  We first discuss the `squared' terms where the amplitudes are contained within the same row before examining the interference terms between rows.

Using the conventions just described, we can write the product of any two amplitudes within a row as
\begin{equation}
\begin{array}{ccc}
\label{eq:genbgampXX}
 \mathcal{M}^n_{Xij} \mathcal{M}^{m\ast}_{X\bar{i}\bar{j}} = (D_{Yi} D_{Zj} D^\ast_{Y\bar{i}} D^\ast_{Z\bar{j}})^{-1}\\
(g_{\mu\sigma} \mathcal{T}_{Yi\bar{i}}^{\sigma\bar{\sigma}} g_{\bar{\mu}\bar{\sigma}})(g_{\nu\gamma} \mathcal{T}_{Z j\bar{j}}^{\gamma\bar{\gamma}} g_{\bar{\nu}\bar{\gamma}})
\mathcal{T}_{X ij\bar{i}\bar{j}nm}^{\nu\mu\bar{\mu}\bar{\nu}}
\end{array}
\end{equation}
where the $\mathcal{T}_{Yi\bar{i}}^{\sigma\bar{\sigma}}$ and $D_{Yi}$ are defined similarly to those in Eq.(\ref{eq:leptrace}) and the long Dirac string is given by,
\begin{equation}
\begin{array}{ccc}
\label{eq:trace8}
\mathcal{T}_{X ij\bar{i}\bar{j}nm}^{\nu\mu\bar{\mu}\bar{\nu}} = (g^i_{XR}g^j_{XR}g^{\bar{i}}_{XR}g^{\bar{j}}_{XR} + g^i_{XL}g^j_{XL} g^{\bar{i}}_{XL}g^{\bar{j}}_{XL})/2\\
\cdot Tr(\not{p}_X\gamma^\nu \not k_{Xn} \gamma^\mu \not{p}_{\bar{X}} \gamma^{\bar{\mu}} \not{k}_{Xm} \gamma^{\bar{\nu}}) ~+\\ 
(g^i_{XR}g^j_{XR}g^{\bar{i}}_{XR}g^{\bar{j}}_{XR} - g^i_{XL}g^j_{XL} g^{\bar{i}}_{XL}g^{\bar{j}}_{XL})\\
\cdot Tr(\not{p}_X\gamma^\nu \not k_{Xn} \gamma^\mu \not{p}_{\bar{X}} \gamma^{\bar{\mu}} \not{k}_{Xm} \gamma^{\bar{\nu}}\gamma^5)/2. 
\end{array}
\end{equation}
Expanding out the terms in Eq.(\ref{eq:genbgampXX}) we can organize in a manner similar to Eq.(\ref{eq:gensigamp2exp}) and write the amplitude squared as,
\begin{equation}
\begin{array}{ccc}
\label{eq:genbgXXexp}
 \mathcal{M}^n_{Xij} \mathcal{M}^{m\ast}_{X\bar{i}\bar{j}} = \sum\limits_{abc} \mathcal{C}^{abc}_{XXij\bar{i}\bar{j}} L^{abc}_{XXnm} 
\end{array}
\end{equation}
where again $a,b,c = (+,-)$ in the order $X,Y,Z$ and
\begin{equation}
\begin{array}{ccc}
\label{eq:XXgaugestruc}
\mathcal{C}^{\pm\pm\pm}_{XXij\bar{i}\bar{j}} = (8D_{Yi} D_{Zj} D^\ast_{Y\bar{i}} D^\ast_{Z\bar{j}})^{-1}\\
(g^i_{XR}g^j_{XR}g^{\bar{i}}_{XR}g^{\bar{j}}_{XR} \pm g^i_{XL}g^j_{XL} g^{\bar{i}}_{XL}g^{\bar{j}}_{XL})\\
\cdot (g^i_{YR} g^{\bar{i}}_{YR} \pm g^i_{YL} g^{\bar{i}}_{YL})(g^j_{ZR} g^{\bar{j}}_{ZR} \pm g^j_{ZL} g^{\bar{j}}_{ZL})\\
\\
L^{\pm\pm\pm}_{XXnm} = (g_{\mu\sigma} T_{Y\pm}^{\sigma\bar{\sigma}} g_{\bar{\mu}\bar{\sigma}})(g_{\nu\gamma} T_{Z\pm}^{\gamma\bar{\gamma}} g_{\bar{\nu}\bar{\gamma}})T_{Xnm\pm}^{\nu\mu\bar{\mu}\nu}  .
\end{array} 
\end{equation}
The $T^{\sigma\bar{\sigma}}_{Y,Z\pm}$ are the Dirac traces found in Eq.(\ref{eq:leptrace}) while the $T_{Xnm\pm}^{\nu\mu\bar{\mu}\nu}$ are those found in (\ref{eq:trace8}). Again $\pm$ indicates whether the trace ends with a $\gamma^5$ ($-$) or not ($+$). We note that unlike in the signal case, when organized in this way (essentially by powers of $\gamma^5$) the gauge structure completely factors from the Lorentz structure.  This allows us to sum over all possible intermediate vector bosons at this stage to write,
\begin{equation}
\begin{array}{ccc}
\label{eq:sumgenbgXXexp}
\mathcal{M}^n_{X} \mathcal{M}^{m\ast}_{X} =\sum\limits_{ij\bar{i}\bar{j}}  \mathcal{M}^n_{Xij} \mathcal{M}^{m\ast}_{X\bar{i}\bar{j}}\\
=\sum\limits_{ij\bar{i}\bar{j}} \sum\limits_{abc} \mathcal{C}^{abc}_{XXij\bar{i}\bar{j}} L^{abc}_{XXnm}\\
=\sum\limits_{abc} (\sum\limits_{ij\bar{i}\bar{j}} \mathcal{C}^{abc}_{XXij\bar{i}\bar{j}}) L^{abc}_{XXnm} = \sum\limits_{abc} \mathcal{C}^{abc}_{XX} L^{abc}_{XXnm}~~.
\end{array}
\end{equation}
This simplifies things greatly and in particular the objects $\mathcal{C}^{abc}_{XX}$ now contain all of the information regarding the intermediate vector bosons including the interference effects between the different processes.   These will serve as overall coefficients for the various Lorentz structure pieces.   

We are now in a position to examine the interference terms.  Let us take the product of any two diagrams not in the same row.  One can show explicitly,
\begin{equation}
\begin{array}{ccc}
\label{eq:genbgampXY}
 \mathcal{M}^n_{Xij} \mathcal{M}^{m\ast}_{Y\bar{i}\bar{j}} = (D_{Yi} D_{Zj} D^\ast_{Z\bar{i}} D^\ast_{X\bar{j}})^{-1}\\
(g_{\mu\sigma} \mathcal{T}_{Xij\bar{j}n}^{\nu\mu\bar{\gamma}} g_{\bar{\nu}\bar{\gamma}})(g_{\nu\gamma} \mathcal{T}_{Yi\bar{i}\bar{j}m}^{\bar{\mu}\bar{\nu}\sigma} g_{\bar{\mu}\bar{\sigma}})\mathcal{T}_{Zj\bar{i}}^{\gamma\bar{\sigma}}
\end{array}
\end{equation}
where the $\mathcal{T}_{Zi\bar{i}}^{\gamma\bar{\sigma}}$ are as before and the new Dirac strings are given by,
\begin{equation}
\begin{array}{ccc}
\label{eq:trace6}
 \mathcal{T}_{Xij\bar{j}n}^{\nu\mu\bar{\gamma}} = (g^i_{XR}g^j_{XR}g^{\bar{j}}_{XR} + g^i_{XL}g^j_{XL} g^{\bar{j}}_{XL})/2\\
\cdot Tr(\not{p}_X\gamma^\nu \not k_{Xn} \gamma^\mu \not{p}_{\bar{X}} \gamma^{\bar{\gamma}}) ~+\\ 
(g^i_{XR}g^j_{XR}g^{\bar{j}}_{XR} - g^i_{XL}g^j_{XL} g^{\bar{j}}_{XL})\\
\cdot Tr(\not{p}_X\gamma^\nu \not k_{Xn} \gamma^\mu \not{p}_{\bar{X}} \gamma^{\bar{\gamma}}\gamma^5)/2. 
\end{array}
\end{equation}
The distinct Lorentz structure found here as compared to that found in Eq.(\ref{eq:genbgampXX}) is due to the different path taken when tracing over the fermonic strings.  

Again we expand out the terms in Eq.(\ref{eq:genbgampXY}) to obtain,
\begin{equation}
\begin{array}{ccc}
\label{eq:genbgXYexp}
 \mathcal{M}^n_{Xij} \mathcal{M}^{m\ast}_{Y\bar{i}\bar{j}} = \sum\limits_{abc} \mathcal{C}^{abc}_{XYij\bar{i}\bar{j}} L^{abc}_{XYnm} 
\end{array}
\end{equation}
where,
\begin{equation}
\begin{array}{ccc}
\label{eq:XYgaugestruc}
\mathcal{C}^{\pm\pm\pm}_{XYij\bar{i}\bar{j}} = (8D_{Yi} D_{Zj} D^\ast_{Z\bar{i}} D^\ast_{X\bar{j}})^{-1}(g^j_{ZR} g^{\bar{i}}_{ZR} \pm g^j_{ZL} g^{\bar{i}}_{ZL})\\
\cdot(g^i_{XR}g^j_{XR}g^{\bar{j}}_{XR} \pm g^i_{XL}g^j_{XL} g^{\bar{j}}_{XL})(g^i_{YR}g^{\bar{i}}_{YR}g^{\bar{j}}_{YR} \pm g^i_{YL}g^{\bar{i}}_{YL} g^{\bar{j}}_{YL})\\
\\
L^{\pm\pm\pm}_{XYnm} = (g_{\mu\sigma}T_{Xn}^{\nu\mu\bar{\gamma}} g_{\bar{\nu}\bar{\gamma}})(g_{\nu\gamma}T_{Ym}^{\bar{\mu}\bar{\nu}\sigma} g_{\bar{\mu}\bar{\sigma}})T_Z^{\gamma\bar{\sigma}}  .
\end{array} 
\end{equation}
\\
and the $T_{Xn}^{\nu\mu\bar{\gamma}}$ are the traces found in Eq.(\ref{eq:trace6}).~As mentioned above, since the gauge and Lorentz structure factor completely we are free to perform the sum over the intermediate vector bosons at this stage once again to obtain the various Lorentz structure coefficients,
\begin{equation}
\begin{array}{ccc}
\label{eq:sumgenbgXYexp}
\mathcal{M}^n_{X} \mathcal{M}^{m\ast}_{Y} = \sum\limits_{ij\bar{i}\bar{j}}  \mathcal{M}^n_{Xij} \mathcal{M}^{m\ast}_{Y\bar{i}\bar{j}} = \\
\\
\sum\limits_{abc} (\sum\limits_{ij\bar{i}\bar{j}} \mathcal{C}^{abc}_{XYij\bar{i}\bar{j}}) L^{abc}_{XYnm} = \sum\limits_{abc} \mathcal{C}^{abc}_{XY} L^{abc}_{XYnm}~~.
\end{array}
\end{equation}
Thus again all of the information concerning the intermediate vector bosons is contained in $\mathcal{C}^{abc}_{XY}$. We now have all of the pieces\footnote{Expressions for the various coefficients and Lorentz structure can be obtained by emailing the corresponding author.} necessary to build the total amplitude squared of the diagrams in Fig.~\ref{fig:XYZdiags} including all contributions from the intermediate vector bosons.  Explicitly we have,
\small{
\begin{equation}
\begin{array}{ccc}
\label{eq:Mq12amp2}
|\mathcal{M}_q + \mathcal{M}_1 + \mathcal{M}_2|^2 \\
= \sum\limits_{abc} \sum\limits_{nm} \Big( \Big( \mathcal{C}^{abc}_{qq} L^{abc}_{qqnm}+ \mathcal{C}^{abc}_{11} L^{abc}_{11nm}+ \mathcal{C}^{abc}_{22} L^{abc}_{22Œnm} \Big)\\
+ 2Re \Big( \mathcal{C}^{abc}_{q1} L^{abc}_{q1nm}+ \mathcal{C}^{abc}_{12} L^{abc}_{12nm}+ \mathcal{C}^{abc}_{2q} L^{abc}_{2qnm} \Big) \Big)
\end{array}
\end{equation}
}
\\
where the sum over intermediate vector bosons has been already implicitly performed and the sum over $n,m$ which includes the t and u channel contributions is shown explicitly (note that this also factors from the vector boson sum). The $\mathcal{C}^{abc}_{XY}$ coefficients are in general complex due to the factor of $i$ multiplying the decay width in the massive vector boson propogators.  The Lorentz structure is either purely real or purely imaginary depending on whether the term contains an even or odd number of traces ending in $\gamma^5$.  These traces give an overall factor of $i$ (and an epsilon tensor).  Thus if $L^{abc}_{XYnm}$ contains an even number of these traces, then it is purely real and if it contains an odd number it is purely imaginary.  The squared Lorentz structure $L^{abc}_{XXnm}$ however is strictly real as are the squared coefficients $\mathcal{C}^{abc}_{XX}$. Taking this into account, we can write for Eq.(\ref{eq:Mq12amp2}) the final amplitude squared as,
\begin{eqnarray}
\label{eq:Mq12amp2final}
&&|\mathcal{M}_{4\ell}|^2 = |\mathcal{M}_q + \mathcal{M}_1 + \mathcal{M}_2|^2\nonumber \\ 
&&=\frac{1}{2}\sum\limits_{abc}^{even} \Big( \mathcal{C}^{abc}_{qqR} L^{abc}_{qqR}+ \mathcal{C}^{abc}_{11R} L^{abc}_{11R}+ \mathcal{C}^{abc}_{22R} L^{abc}_{22R}\Big)\nonumber\\
&&+\sum\limits_{abc}^{even} \Big( \mathcal{C}^{abc}_{q1R}L^{abc}_{q1R}+\mathcal{C}^{abc}_{12R}L^{abc}_{12R}+ \mathcal{C}^{abc}_{2qR}L^{abc}_{2qR} \Big)\nonumber\\
&&-\sum\limits_{abc}^{odd} \Big( \mathcal{C}^{abc}_{q1I} L^{abc}_{q1I} + \mathcal{C}^{abc}_{12I}L^{abc}_{12I} + \mathcal{C}^{abc}_{2qI}L^{abc}_{2qI} \Big)
\end{eqnarray}
where we have now performed the sum over $t$ and $u$ channel diagrams and $\mathcal{C}^{abc}_{XYR,I} = \mathcal{C}^{abc}_{XY} \pm\mathcal{C}^{\ast abc}_{XY}$ respectively.  We have also implicitly included a factor of $1/4$ from averaging over initial state quark spins and a color factor of $1/3$. The sums labeled $even\equiv(+++,+--,-+-,--+)$ indicate terms with even powers of $\gamma^5$ and those with $odd\equiv(-++,+-+,++-,---)$ indicate terms with odd powers of $\gamma^5$.  Note that since the photon has vector like couplings where $g_L = g_R$ all coefficients $\mathcal{C}^{abc}_{XY}$ with $a$, $b$, or $c \equiv -$ are zero for the $\gamma\gamma$ intermediate state.  Thus $\gamma\gamma$ only contributes to the $\mathcal{C}^{+++}_{XY}$ coefficients (including of course when $X\equiv Y$).  

Previous calculations of the golden channel background, which include only the di-boson production process, are contained within the first term $\mathcal{C}^{abc}_{qq} L^{abc}_{qq}$ of Eq.(\ref{eq:Mq12amp2final}).  All the other terms arise from the resonant four lepton production process and the interference between it and the di-boson production process. Note that Eq.(\ref{eq:Mq12amp2final}) is also more general than for just the golden channel. In principal this expression holds for any process with the same topology and `$Z$-like' couplings to fermions. Since we have built the expression out of a generalized Lorentz structure with coefficients, it can easily be adapted to consider new physics contributions which may enter with the same topology and alter some of the coefficients by an observable amount. Thus one can imagine performing stringent tests of the SM using this parametrization to extract the various Lorentz structure coefficients. We leave an investigation of this to future work. 

The final fully differential cross section\footnote{This expression has been validated with the Madgraph matrix element squared.} is again obtained by combining the amplitude squared with the invariant four body phase space (see Eq.(\ref{eq:phasespace})),
\begin{equation}
\begin{array}{ccc}
\label{eq:bgdiffcxn}
\frac{d\sigma_{4\ell}}{dM_1^2dM_2^2d\Omega} =  \Large{\Pi}_{4\ell} |\mathcal{M}_{4\ell}|^2~.   
\end{array}
\end{equation}
The differential mass spectrum\footnote{An analytic expression for the dominant component to the background is given in Eq.(\ref{eq:bgm1m2spec}) of the Appendix.} is obtained again via,
\begin{equation}
\begin{array}{ccc}
\label{eq:bgm1m2}
\frac{d\sigma_{4\ell}}{dM_1^2dM_2^2} =  \Large{\Pi}_{4\ell} \int d\Omega |\mathcal{M}_{4\ell}|^2~.   
\end{array}
\end{equation}
We now examine how the various components of the background contribute to the differential mass spectrum.

\subsection{The Differential Spectra}
\label{sec:bgmassdiff}
In this section we examine how the individual components of the background contribute to the invariant mass spectrum of the four lepton system.  In addition we also study how including parton distribution functions (\emph{pdfs}) and NLO corrections change the differential spectra by comparing normalized projections obtained from our analytic expression to Monte Carlo generated by POWHEG~\cite{Melia:2011tj,Nason:2004rx,Frixione:2007vw,Alioli:2010xd} and Madgraph~\cite{Alwall:2011uj}. 

\begin{figure}
\includegraphics[width=0.5\textwidth]{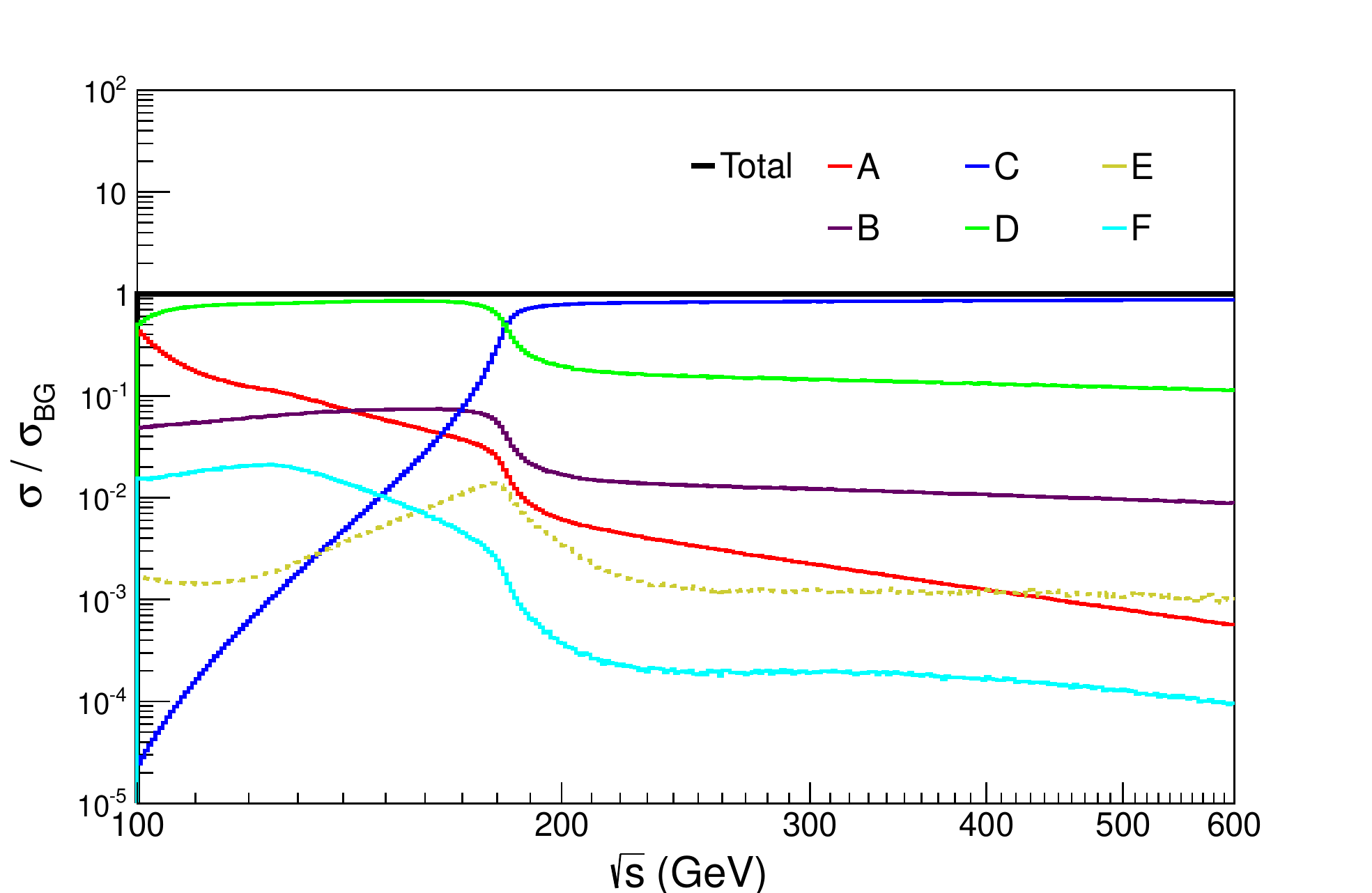}
\includegraphics[width=0.5\textwidth]{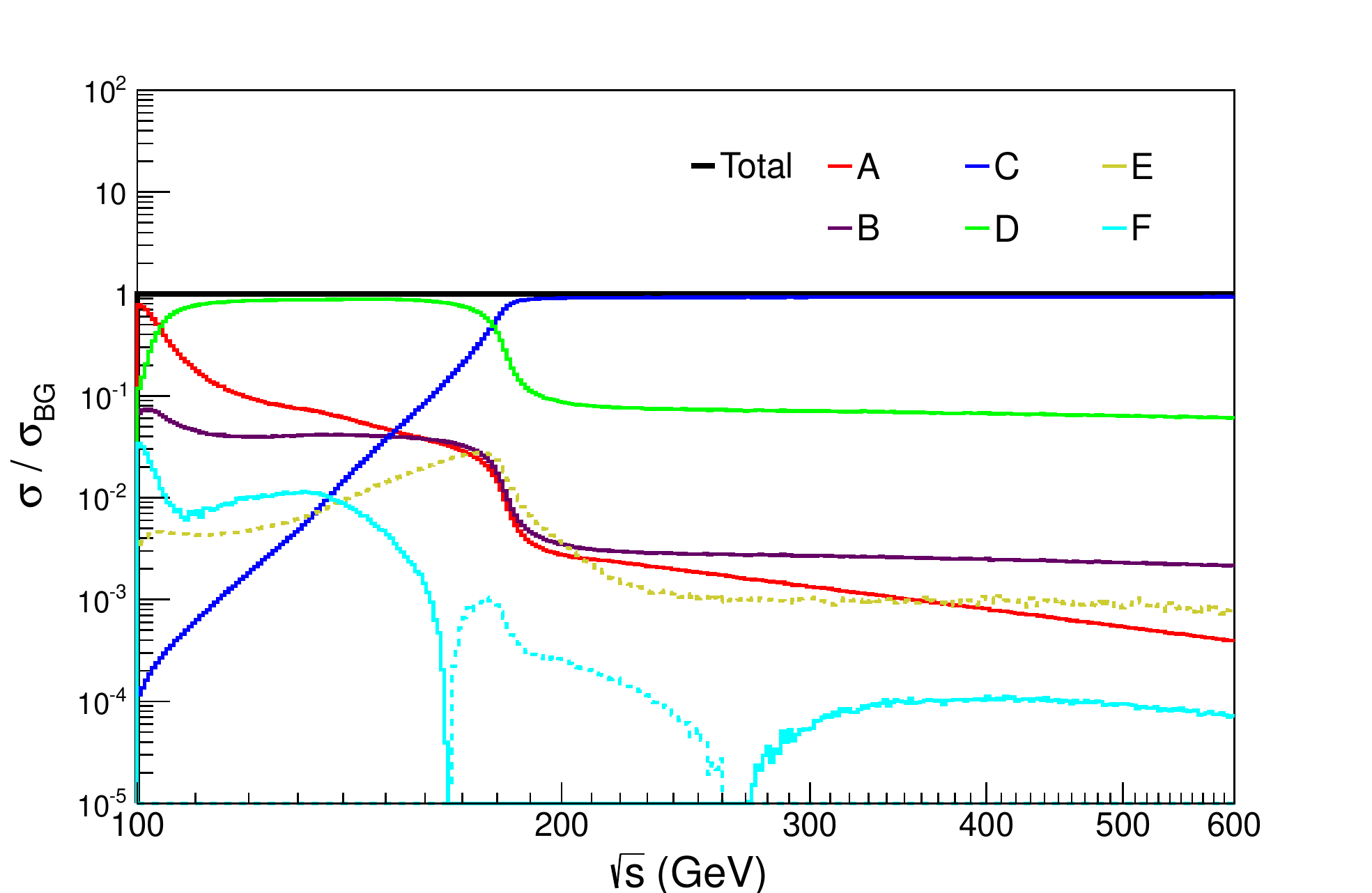}
\caption{The four lepton system invariant mass spectrum (without \emph{pdfs}) for the various components defined in the text.  The dotted lines indicate when the interference between components is destructive, thus giving a negative contribution. In the top plot we take the ranges 4~GeV$< M_{1,2} < 120$~GeV while in the bottom plot we take the range 40~GeV$ < M_1 < 120$~GeV and 10~GeV$< M_2 < 120$~GeV while taking $M_1>M_2$.}
\label{fig:moneyplot}
\end{figure}

We first separate the background into its various components which we define as the following,
\begin{itemize}
\item ${\bf A}$: s-channel $2e2\mu$ process
\item ${\bf B}$: $t+u$-channel $\gamma\gamma$
\item ${\bf C}$: $t+u$-channel $ZZ$
\item ${\bf D}$: $t+u$-channel $Z\gamma$ 
\item ${\bf E}$:  $t+u$-channel $ZZ/Z\gamma/\gamma\gamma$ interference only
\item ${\bf F}$: $ZZ+Z\gamma+\gamma\gamma$ $s/t$-channel interference only
\end{itemize}
where now $s$, $t$, and $u$ are used in the usual sense and the resonant s-channel $2e2\mu$ process can proceed through any combination of $Z$ and $\gamma$. We first consider the relative fractions of these components as a function of the invariant mass of the four lepton system for the range $100-600$ GeV in Fig.~\ref{fig:moneyplot}.  The dotted lines indicate when a contribution is negative, which of course only occurs for interference terms in certain energy ranges when the interference is destructive.  The solid black line at constant value of 1 is the total partonic level $q\bar{q}\rightarrow 2e2\mu$ ($q=u,d$) background including all interference and all intermediate vector bosons. From Fig.~\ref{fig:moneyplot} one can see how the relative contributions coming from the different components change as a function of energy.  

Component $C$ (the $ZZ~t+u$ channel) is the only piece of the background to have been previously calculated analytically~\cite{Gainer:2011xz,Hagiwara:1986vm}. This makes up the dominant contribution above the $ZZ$ threshold, but is negligible from 110~GeV $< \sqrt{s} <$ 140 GeV and in fact is even smaller than the interference terms. We also plot the spectrum if one requires a window around the $Z$ boson mass in the bottom plot of Fig.~\ref{fig:moneyplot}. The dominant component near the resonance mass of 125 GeV is $D$ regardless of the window on the $Z$ mass. Except for component $F$, one can see that the relative fractions are fairly insensitive to the $Z$ window requirement except in the range $\sim 100 - 110$ GeV.

\begin{figure}
\includegraphics[width=0.5\textwidth]{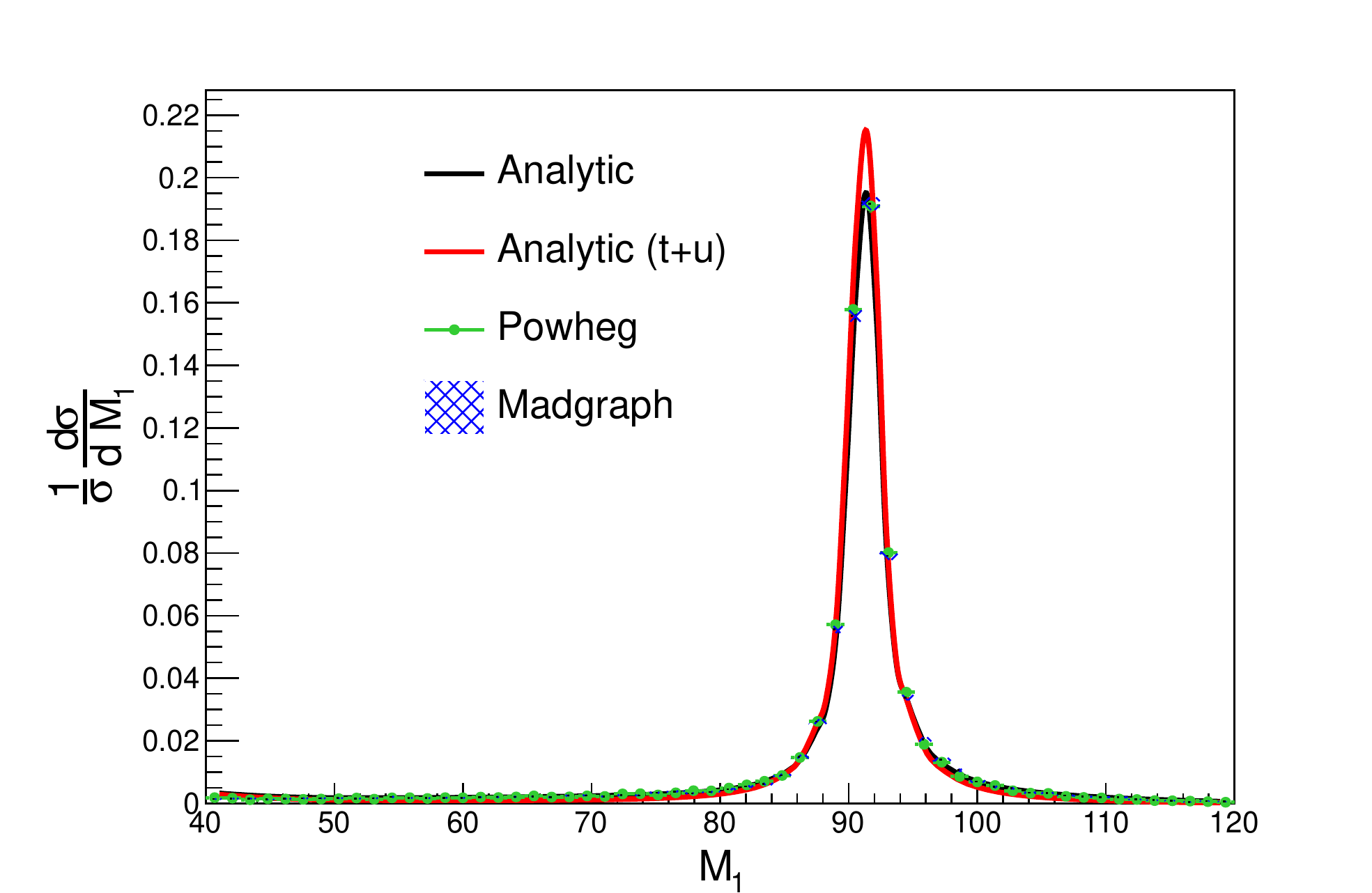}
\includegraphics[width=0.5\textwidth]{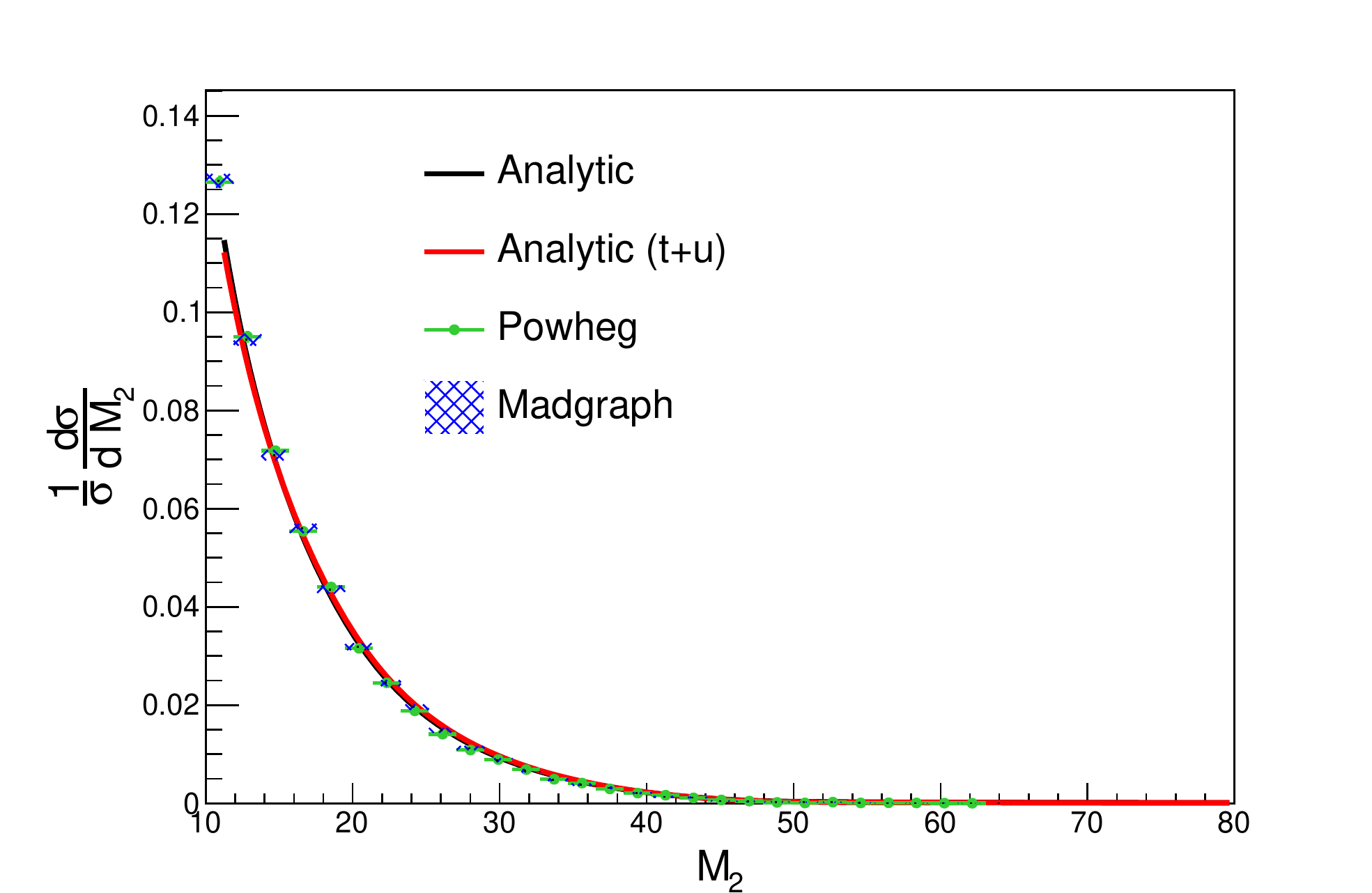}
\caption{Comparing the LO and NLO results for the $M_1$ and $M_2$ invariant mass spectra for the ranges 40~GeV $< M_1 < 120$~GeV and 10~GeV $< M_2 <$ 120~GeV. We take the range of the four lepton system invariant mass to be $110 < \sqrt{s} < 140$ GeV.}
\label{fig:NLOm1m2plots}
\end{figure}

The flexibility of the analytic expressions also allow us to easily isolate the contribution coming from interference terms. Component $E$ for example is due to the interference between the intermediate gauge bosons in the $t+u$ channel and is destructive over the entire range regardless of the $Z$ window.  The interference between the resonant s-channel and the t-channel pair production processes is shown in $F$ and switches between constructive and destructive if one requires a window around the $Z$, but otherwise is constructive. Though these components are small it is possible for them to have subtle effects on the angular distributions such as in the modulation of the azimuthal angle $\Phi$ (See Fig.~\ref{fig:dPhi}) and may be particularly interesting to study in the range 100~GeV $\lesssim \sqrt{s} \lesssim$ 110~GeV. The expressions for most of the components themselves are too cumbersome to write here, but in the Appendix we give the expression for the doubly differential ($M_1, M_2$) mass spectrum of the full $t+u$ (the sum of $B$-$E$) component which as we can see in Fig.~\ref{fig:moneyplot} and Fig.~\ref{fig:NLOm1m2plots} provides a very good approximation above $\sqrt{s} \sim$ 110~GeV.  

To examine the effects of NLO contributions and \emph{pdfs} we compare our parton level result for $q\bar{q}\rightarrow 2e2\mu$ ($q=u,d$) to Monte Carlo data generated by the NLO  POWEG and LO Madgraph codes which include \emph{pdfs}~\cite{Pumplin:2002vw}.  For this we define our phase space as 40~GeV $< M_1 < 120$ GeV and 10~GeV $< M_2 <$ 120~GeV for the energy range 110~GeV$< \sqrt{s} <$ 140~GeV. We also plot the $t+u$ component only (defined as the sum of $B$-$E$) to examine what affects neglecting the resonant $2e2\mu$ process has. 

\begin{figure}
\label{fig:bgpolarangles}
\includegraphics[width=0.45\textwidth]{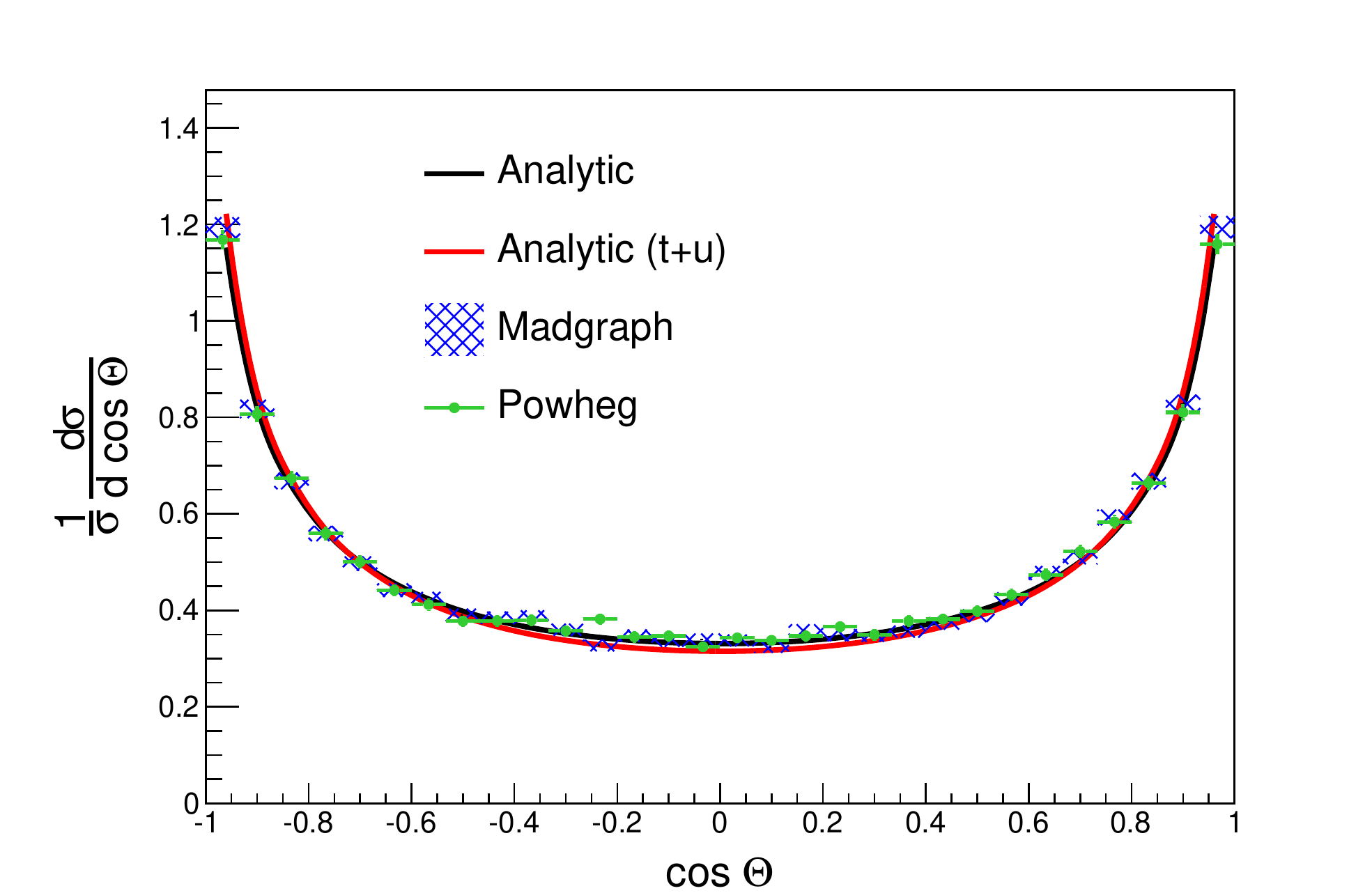}
\includegraphics[width=0.45\textwidth]{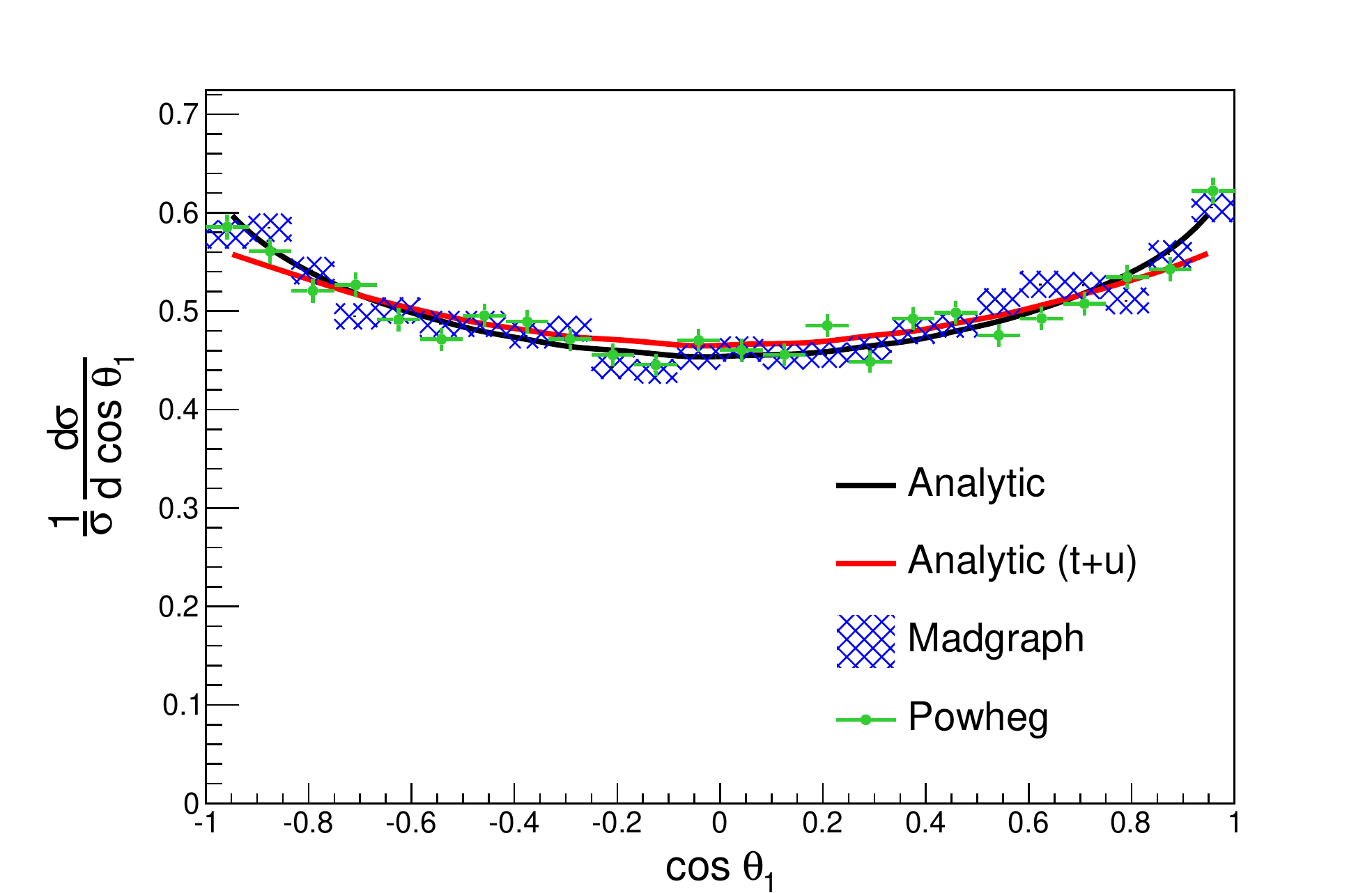}
\includegraphics[width=0.45\textwidth]{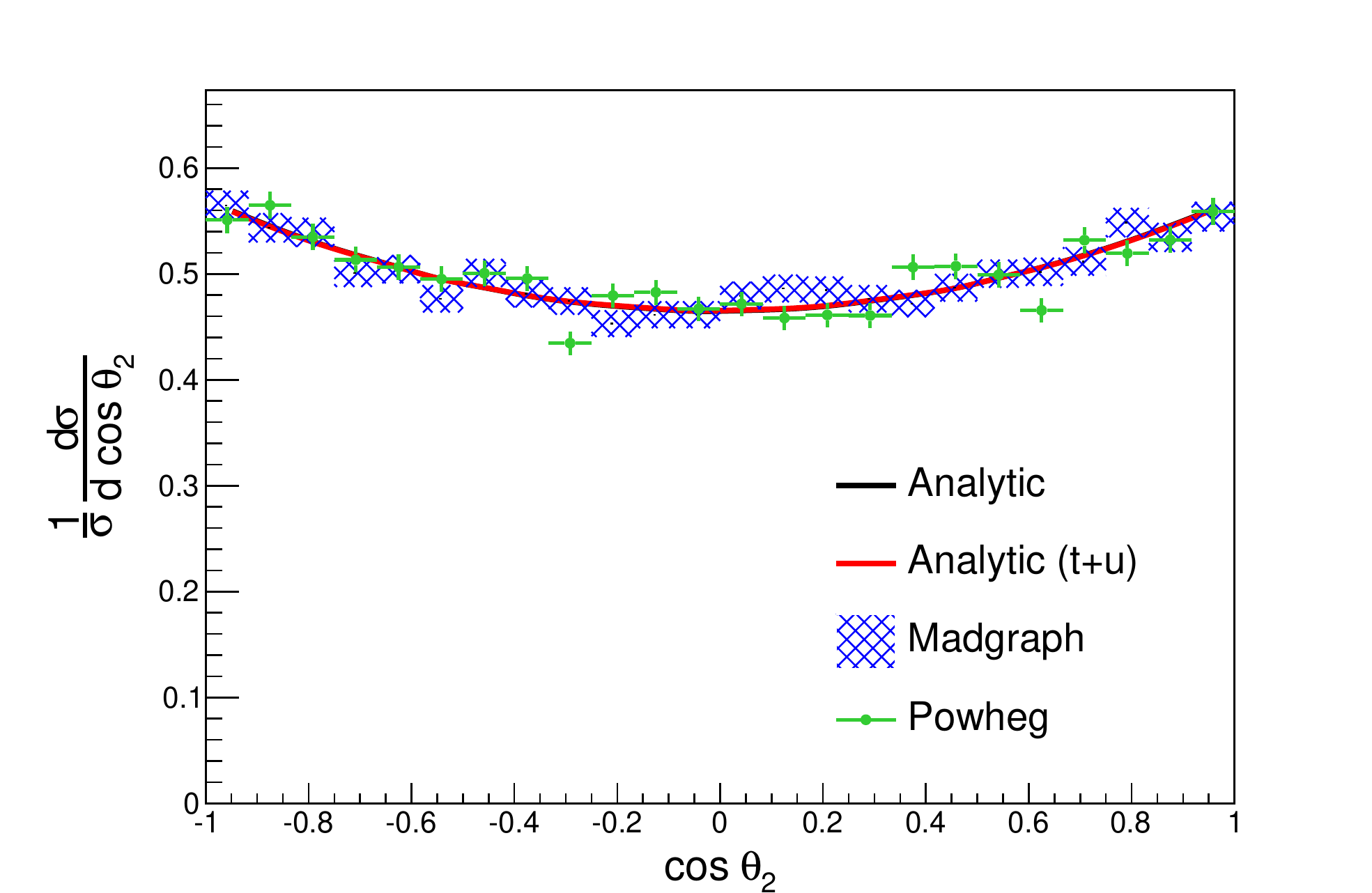}
\caption{Comparing the LO and NLO results for the polar angles $\cos{\Theta}$,~$\cos{\theta_1}$,~$\cos{\theta_2}$ for the ranges 40~GeV $< M_1 < 120$~GeV and 10~GeV $< M_2 <$ 120~GeV. We take the range of the four lepton system invariant mass to be $110 < \sqrt{s} < 140$ GeV.}
\end{figure}

\begin{figure}
\includegraphics[width=0.45\textwidth]{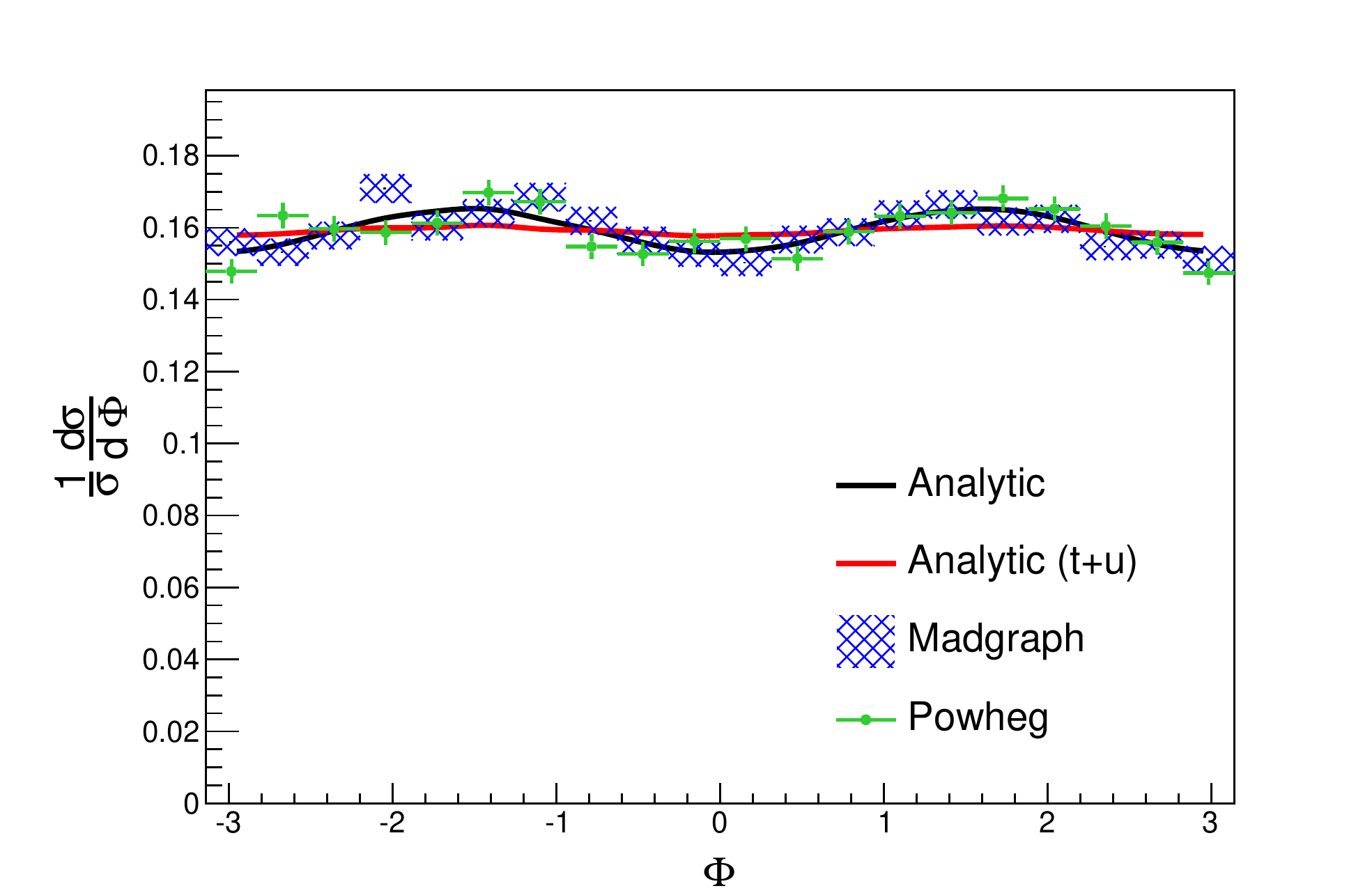}
\includegraphics[width=0.45\textwidth]{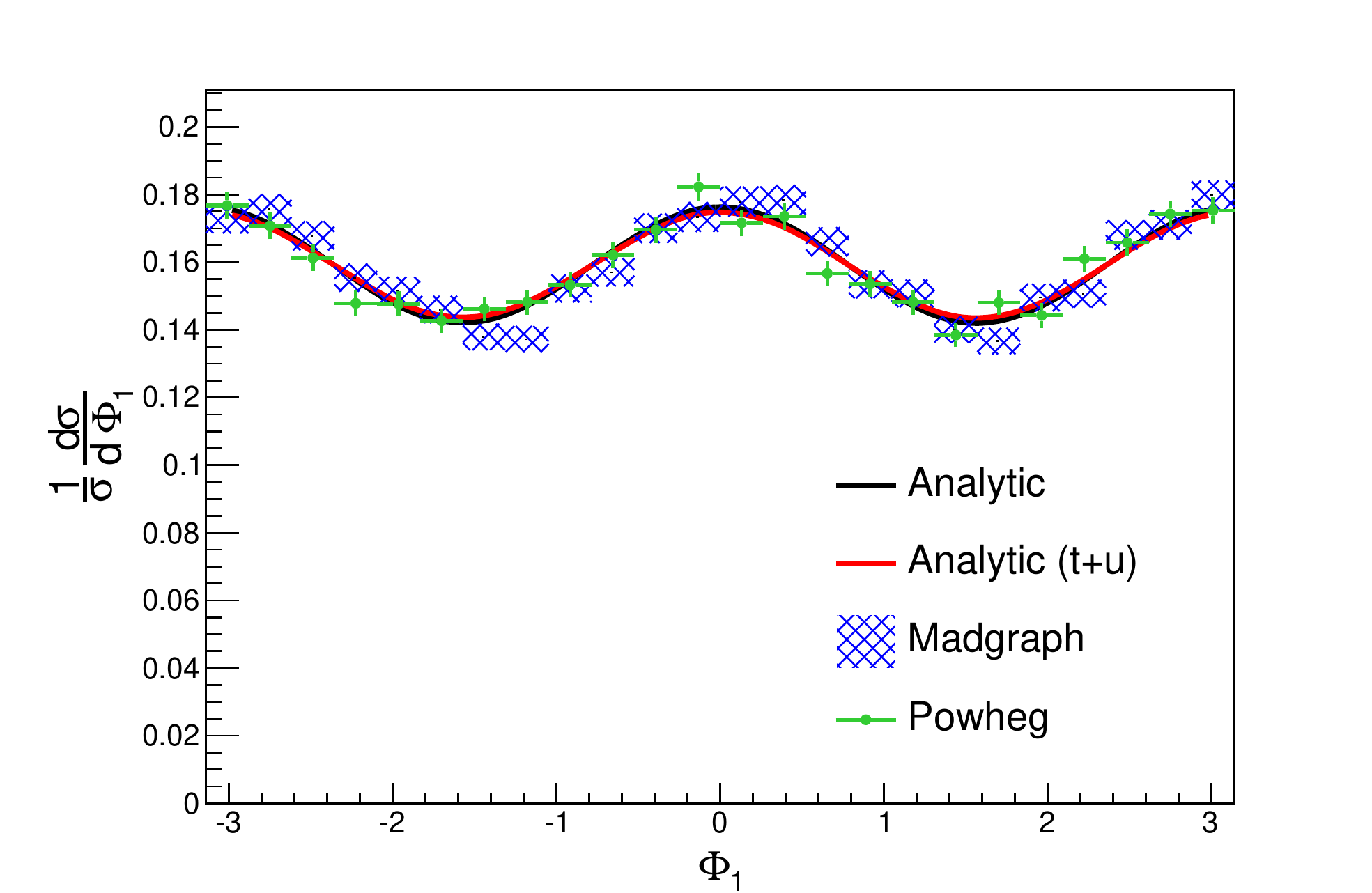}
\caption{Comparing the LO and NLO results for the azimuthal angles $\Phi$ and $\Phi_1$ for the ranges 40~GeV $< M_1 < 120$~GeV and 10~GeV $< M_2 <$ 120~GeV. We take the range of the four lepton system invariant mass to be $110 < \sqrt{s} < 140$ GeV.}
\label{fig:bgazangles}
\end{figure}

In Fig.~\ref{fig:NLOm1m2plots}-\ref{fig:bgazangles} we show the kinematic distributions where it can be seen that NLO and \emph{pdf} contributions affect the normalized spectra negligibly. In addition we can see that neglecting the resonant process also has little effect on all the kinematic variables except $\Phi$, where it affects the modulation and in the forward regions of $\cos{\theta_1}$.  As we will see in the Appendix, the modulation is due almost entirely to the resonant process. These distributions simply reflect the fact that the various kinematic distributions are not highly correlated with $\sqrt{s}$ allowing us to take $\sqrt{\hat{s}}$ essentially as an input from the \emph{pdfs}. To build a complete hadronic differential cross section one could convolve the $\sqrt{s}$ spectrum obtained from Madgraph or POWHEG with the partonic differential cross section obtained analytically.  This of course is what would be done for an LHC analysis, but we do not do that here and instead simply integrate our partonic differential cross section over $\sqrt{\hat{s}}$.   

From Figs.~\ref{fig:moneyplot} and~\ref{fig:NLOm1m2plots} we expect the doubly differential spectrum obtained from the $t+u$ component only to be a good approximation which could be useful for a simplified analysis.  We give an explicit expression for this component in Eq.~(\ref{eq:bgm1m2spec}) of the Appendix. Though it does not use all of the kinematic variables, it should still have strong discriminating power and can be used with the methods proposed in~\cite{Boughezal:2012tz} to form a powerful simplified study.  


\section{Conclusions and Outlook}
We have calculated and presented analytic fully general differential cross sections for the golden channel signal and background in the $2e2\mu$ final state including all intermediate vector bosons and interference effects. We have presented various singly and doubly differential spectra and examined how the different interference effects manifest themselves in these distributions and in correlations between the different kinematic variables.  We have also emphasized the need to push the `off-shell' invariant mass ($M_2$) reach as low as possible as well as relaxing the `$Z$-window' to maximize the discriminating power when testing different signal hypothesis. We have shown that the expressions can aid in distinguishing between different signal hypotheses and because signal and background are provided, both can be included into one likelihood, as should be done when performing simple hypothesis tests of different signals.  These expressions can be implemented into an MEM analysis to perform detailed studies of the spin and CP properties of any scalar resonance which has been or may be discovered at the LHC.
\label{sec:conc} 

\vskip 0.2 cm

\noindent
{\bf Acknowledgements:} We thank Joe Lykken, Ian Low, Andrei Gritsan, and Maria Spiropulu for helpful discussions.  R.V.M. is especially grateful to Kunal Kumar for help with validation and numerous discussions.  N.T. is grateful to Andrew Whitbeck and Ian Anderson for consultation.  R.V.M. is supported by the Fermilab Graduate Student Fellowship program.  Y.C. is supported by the Weston Havens Foundation and DOE grant DE-FG02-92-ER-40701.  This research is also partially supported by Fermi Research Alliance, LLC under Contract No. De-AC02-07CH11359 with the United States Department of Energy.

\section{Appendix}
\label{sec:Appendix}
The general scalar and background differential spectra are too cumbersome to write in one page\footnote{The distributions will be made public in the near future, but can be obtained from the corresponding author in the meantime.} for most of the different components.  We give a couple of the simplest ones here, but that are not found in literature.  We also examine how the different signal hypotheses and background components contribute to the various kinematic distributions.  We show a multitude of singly and doubly differential distributions for both signal and background.  Of course none of these plots can show the discriminating power of the fully differential cross section, but one can visually get a sense for the discriminating power of these kinematic variables.  Detector effects will also shape these distributions and deserve careful study, but it is clear that the golden channel is a powerful probe of the underlying physics.

\subsection{Analytic Expressions}
\label{sec:anexp}
We give here a pair of analytic expressions for the differential mass spectra for one of the signal and one of the background components which are simple enough to fit on one page.  Although not as powerful as using the fully differential cross section, with just these two relatively simple expressions one can perform robust analyses of the newly discovered scalar and its coupling to neutral gauge bosons as suggested in~\cite{Boughezal:2012tz}. For the signal we give the $\varphi \rightarrow ZZ + Z\gamma \rightarrow 2e2\mu$ differential mass spectrum including interference.  For the $ZZ$ coupling we take only the `SM-like' coupling $A_{1ZZ}$ to be non-zero.  For the $Z\gamma$ coupling we allow for both $A_{2Z\gamma}$ and $A_{3Z\gamma}$ to be non-zero, thus allowing for CP violation. Using Eq.(\ref{eq:diffmassspec}) we obtain,
\begin{widetext}
\begin{eqnarray}
\label{eq:sigm1m2ZA}
&&\frac{d\Gamma_{SM+Z\gamma}}{dM_1^2dM_2^2}=\nonumber \\
&&\Big(\sqrt{M_1^4 + (M_2^2 - s)^2 - 2 M_1^2(M_2^2+s)}\Big(6 A_{1ZZ} A_{2ZA} e_l(g_L+g_R)(g_L^2 + g_R^2) M_1 M_2 m_z^2 (2 M_1^2 M_2^2-(M_1^2 + M_2^2) m_z^2)(-1+\beta_1^2)\nonumber \\
&&(1+\beta_1\beta_2)(-1+\beta_2^2)+ A_{1ZZ}^2 (g_L^2 + g_R^2)^2 M_1^2 M_2^2 m_z^4 \sqrt{1 - \beta_1^2} \sqrt{1 - \beta_2^2} (3 + 2 \beta_1 \beta_2 - 2 \beta_2^2 + \beta_1^2 (-2 + 3 \beta_2^2)) + 2 e_l^2 \sqrt{1-\beta_1^2} \sqrt{1 - \beta_2^2} \nonumber \\
&&(2 A_{3ZA}^2 (\beta_1 + \beta_2)^2 + A_{2ZA}^2 (3 - \beta_1^2 + 4 \beta_1 \beta_2 + (-1 + 3 \beta_1^2) \beta_2^2)) \Big(2 g_L g_R M_1^2 M_2^2 ((M_1 - m_z) (M_2 - m_z) (M_1 + m_z) (M_2 + m_z) + m_z^2 \Gamma_z^2) \nonumber \\
&&+ g_L^2 (M_2^4 m_z^2 (m_z^2 + \Gamma_z^2)+M_1^2 M_2^2 m_z^2 (-3 M_2^2 + m_z^2 + \Gamma_z^2)+M_1^4 (3 M_2^4 - 3 M_2^2 m_z^2 + m_z^4 + m_z^2 \Gamma_z^2)) + g_R^2 (M_2^4 m_z^2 (m_z^2 + \Gamma_z^2)\nonumber \\
&&+ M_1^2 M_2^2 m_z^2 (-3 M_2^2 + m_z^2 + \Gamma_z^2) + M_1^4 (3 M_2^4 - 3 M_2^2 m_z^2 + m_z^4 +  m_z^2 \Gamma_z^2))\Big)\Big)\Big)\nonumber \\
&&\Big/\Big(4608 \pi^4 s^2 v_h^2 (1 - \beta_1^2)^{3/2} (1 - \beta_2^2)^{3/2} ((M_1^2 - m_z^2)^2 + m_z^2 \Gamma_z^2) ((M_2^2 - m_z^2)^2 +m_z^2 \Gamma_z^2)\Big)
\end{eqnarray}
\normalsize
\end{widetext}
where we define,
\begin{eqnarray}
\label{eq:betagamma1}
\beta_{1,2}=\sqrt{1-\frac{4M_1^2}{\Big(1\pm(M_1^2-M_2^2)/s\Big)^2s}}~.
\end{eqnarray}
\\
This expression is frame invariant and can accommodate a Higgs-like particle with SM couplings to $ZZ$, but with perhaps new physics contributions through its couplings to $Z\gamma$.  The $e_l$ are the photon couplings to charged leptons while $g_{L,R}$ are the leptonic $Z$ couplings.  $M_1$ and $M_2$ are the final state lepton pair invariant masses while $m_Z$ is the mass of the $Z$ boson and $\sqrt{s}$ is the four lepton system invariant mass. The doubly differential mass spectrum for the full $t+u$ component of the background (sum of components $B$-$E$) can be obtained analytically via Eq.~(\ref{eq:bgm1m2}) to give,
\begin{widetext}
\begin{eqnarray}
\label{eq:bgm1m2spec}
&&\frac{d\sigma^{BG}_{t+u}}{dM_1^2dM_2^2}=\nonumber \\
&&-\Big(((g_{qL}^4 + g_{qR}^4) (g_{L}^2 + g_{R}^2)^2 M_1^4 M_2^4 + 2 e_l e_q (g_{qL}^3 + g_{qR}^3)(g_{L} + g_{R}) (g_{L}^2 + g_{R}^2) M_1^2 M_2^2 (2 M_1^2 M_2^2 - (M_1^2 + M_2^2) m_z^2)\nonumber \\  
&&~ + 8 e_l^4 e_q^4 ((M_1^2 - m_z^2)^2 + m_z^2 \Gamma_z^2) ((M_2^2 - m_z^2)^2 + m_z^2 \Gamma_z^2) + 4 e_l^3 e_q^3 (g_{qL} + g_{qR}) (g_{L} + g_{R}) ((M_2 - m_z)(M_2 + m_z) (-M_1^2 + m_z^2)\nonumber \\ 
&&(-2 M_1^2 M_2^2 + (M_1^2 + M_2^2) m_z^2) + m_z^2 (M_1^4 +  M_2^4 - (M_1^2 + M_2^2) m_z^2) \Gamma_z^2) + 2 e_l^2 e_q^2(g_{qL}^2 + g_{qR}^2) (4 g_{L} g_{R} M_1^2 M_2^2 (M_1 - m_z) (M_2 - m_z)\nonumber\\  
&& (M_1 + m_z) (M_2 +m_z)+ g_{L}^2 ((-2 M_1^2 M_2^2 + (M_1^2 + M_2^2) m_z^2)^2 + (M_1^4 + M_2^4) m_z^2 \Gamma_z^2) + g_{R}^2 ((-2 M_1^2 M_2^2 + (M_1^2 + M_2^2)\nonumber \\ 
&& m_z^2)^2 + (M_1^4 + M_2^4) m_z^2 \Gamma_z^2))) \Big(4 (M_1^2 + M_2^2 - s) \sqrt{M_1^4 + (M_2^2 - s)^2 - 2 M_1^2 (M_2^2 + s)} - ((M_1^2 + M_2^2)^2 + s^2)\nonumber \\ 
&&\Big(\log{\Big[\Big(M_1^2 + M_2^2 - s + \sqrt{M_1^4 + (M_2^2 - s)^2 - 2 M_1^2 (M_2^2 + s)}\Big)^2\Big]}- 2 \log{\Big[-M_1^2 - M_2^2 + s + \sqrt{ M_1^4 + (M_2^2 - s)^2 - 2 M_1^2 (M_2^2 + s)}}\Big]\Big)\Big)\Big) \nonumber \\ 
&&\Big/  \Big(27648~M_1^2 M_2^2 \pi^5 (M_1^2 + M_2^2 - s) s^2 ((M_1^2 - m_z^2)^2 + m_z^2 \Gamma_z^2) ((M_2^2 - m_z^2)^2 + m_z^2 \Gamma_z^2)\Big)~.
\end{eqnarray}
\end{widetext}
This expression includes the $ZZ$, $Z\gamma$ and $\gamma\gamma$ contributions including all interference and can be combined with \emph{pdfs} or be used for a leptonic initial state.  The $e_q$ are the photon couplings to the initial state fermions while the $g_{qR/L}$ are the initial state fermion couplings to $Z$ bosons. Note that these expressions have not been normalized and should be thought of as at fixed $s$.

\subsection{Singly Differential Angular Distributions}
\label{sec:sigdists}
\begin{figure*}[ht!]
\includegraphics[width=0.45\textwidth]{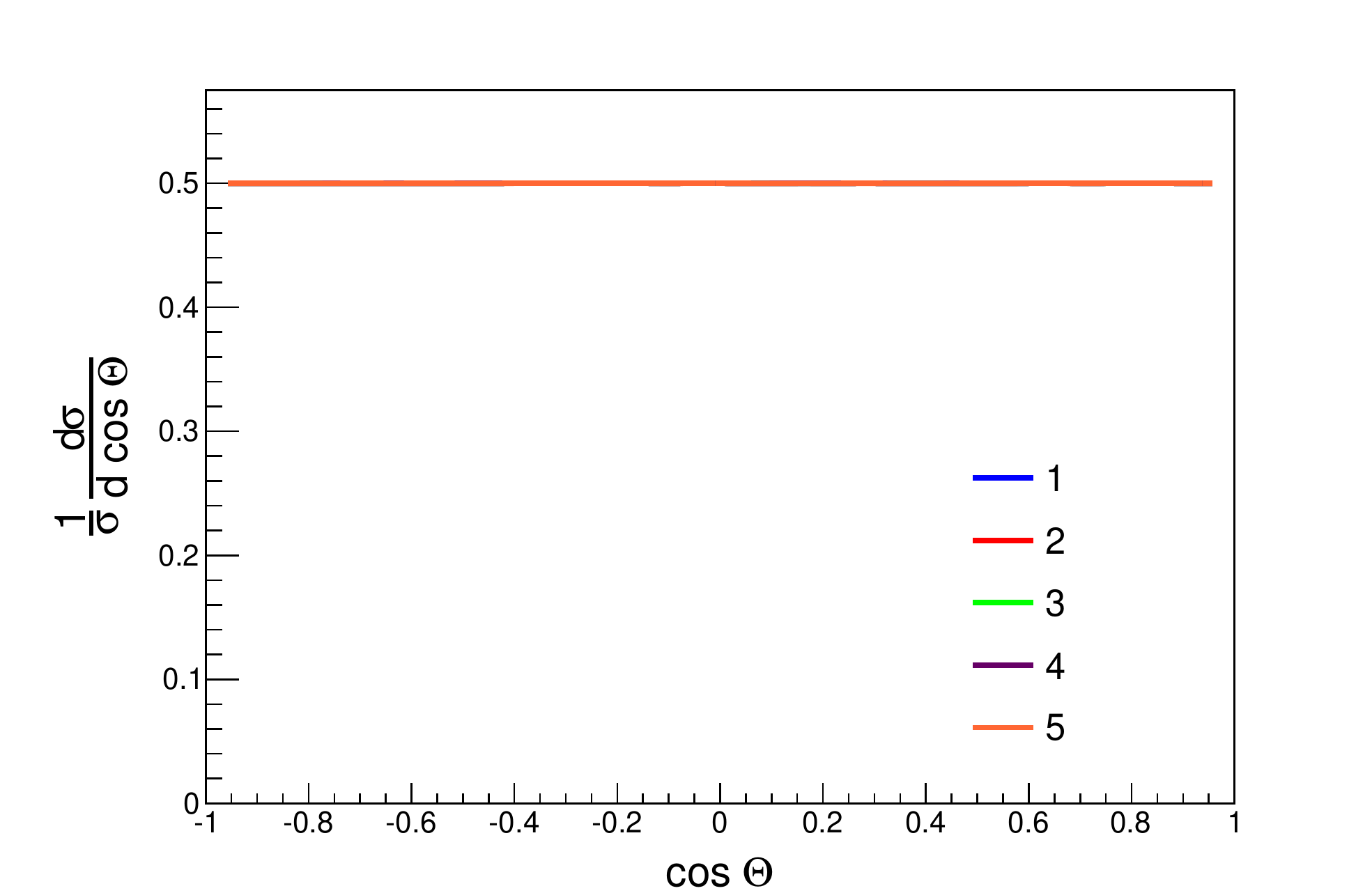}
\includegraphics[width=0.45\textwidth]{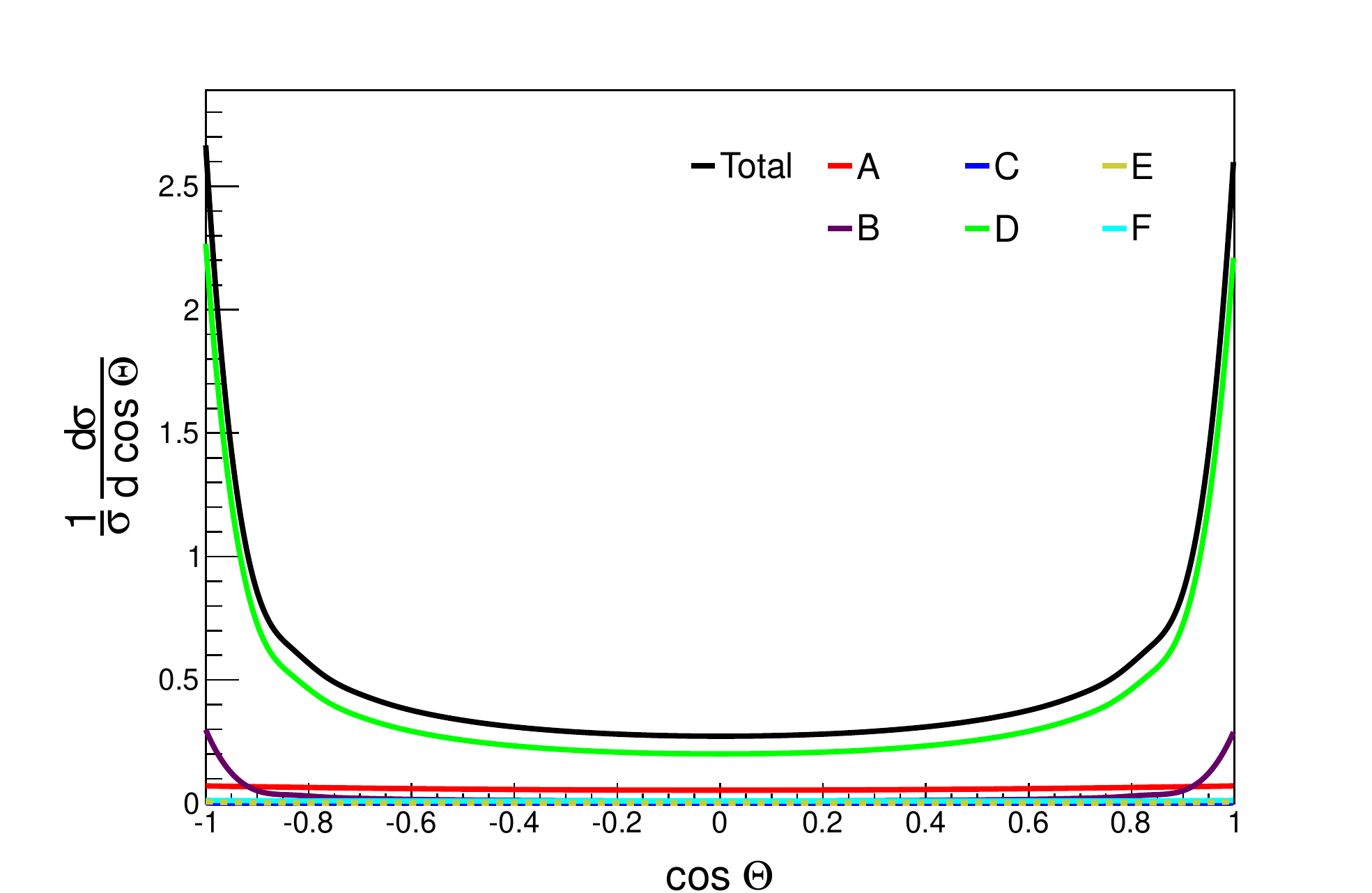}
\caption{On the left hand side we have plotted the $\cos{\Theta}$ angular distributions for hypotheses 1-5 (hypothesis 1~$\equiv$ SM) defined in Sec.\ref{sec:SigSinglyDoubly}. On the right hand side we plot the components A-F of the background defined in Sec.\ref{sec:bgmassdiff}.}
\label{fig:dTheta}
\end{figure*}

\begin{figure*}[ht!]
\includegraphics[width=0.45\textwidth]{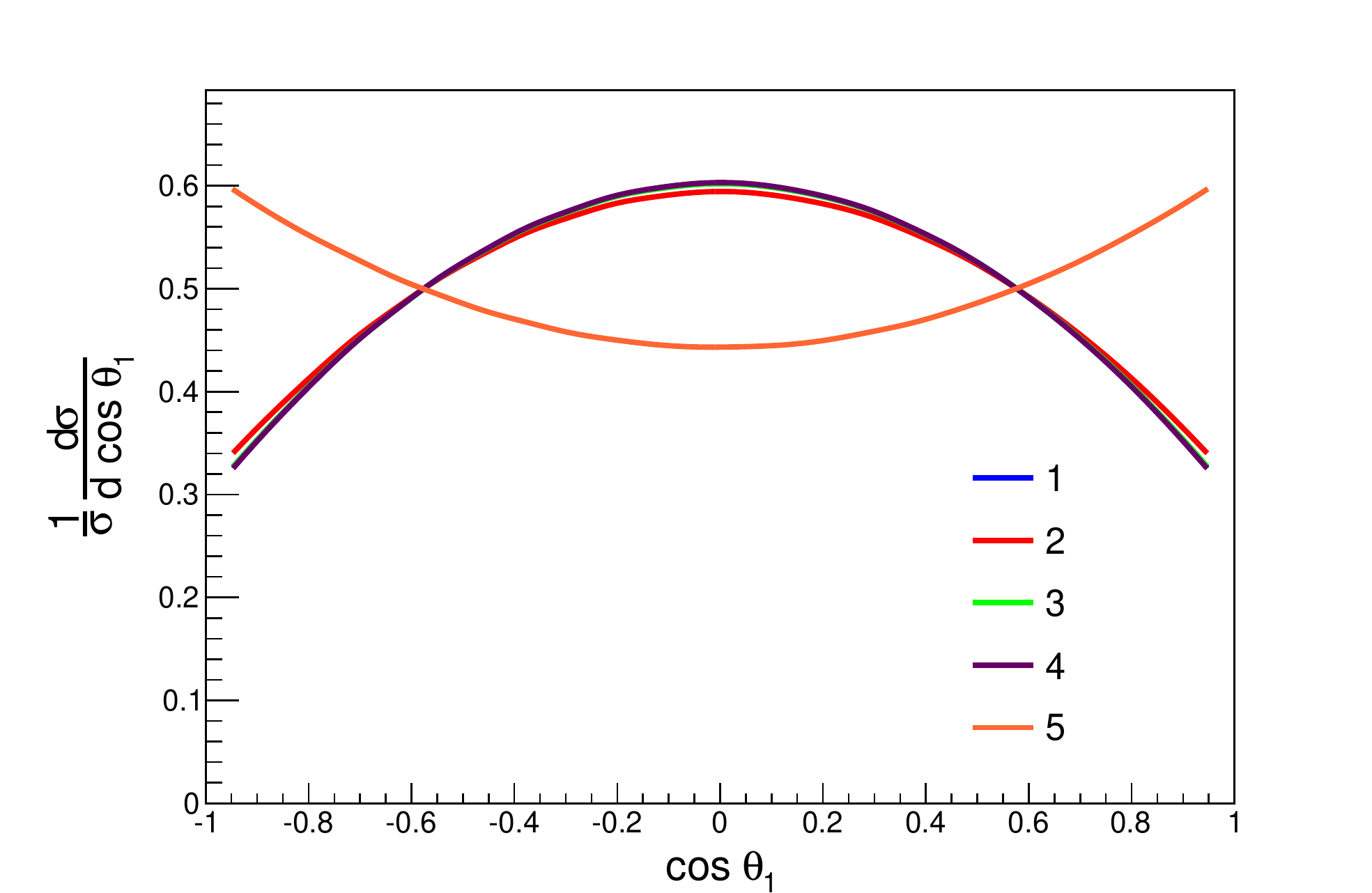}
\includegraphics[width=0.45\textwidth]{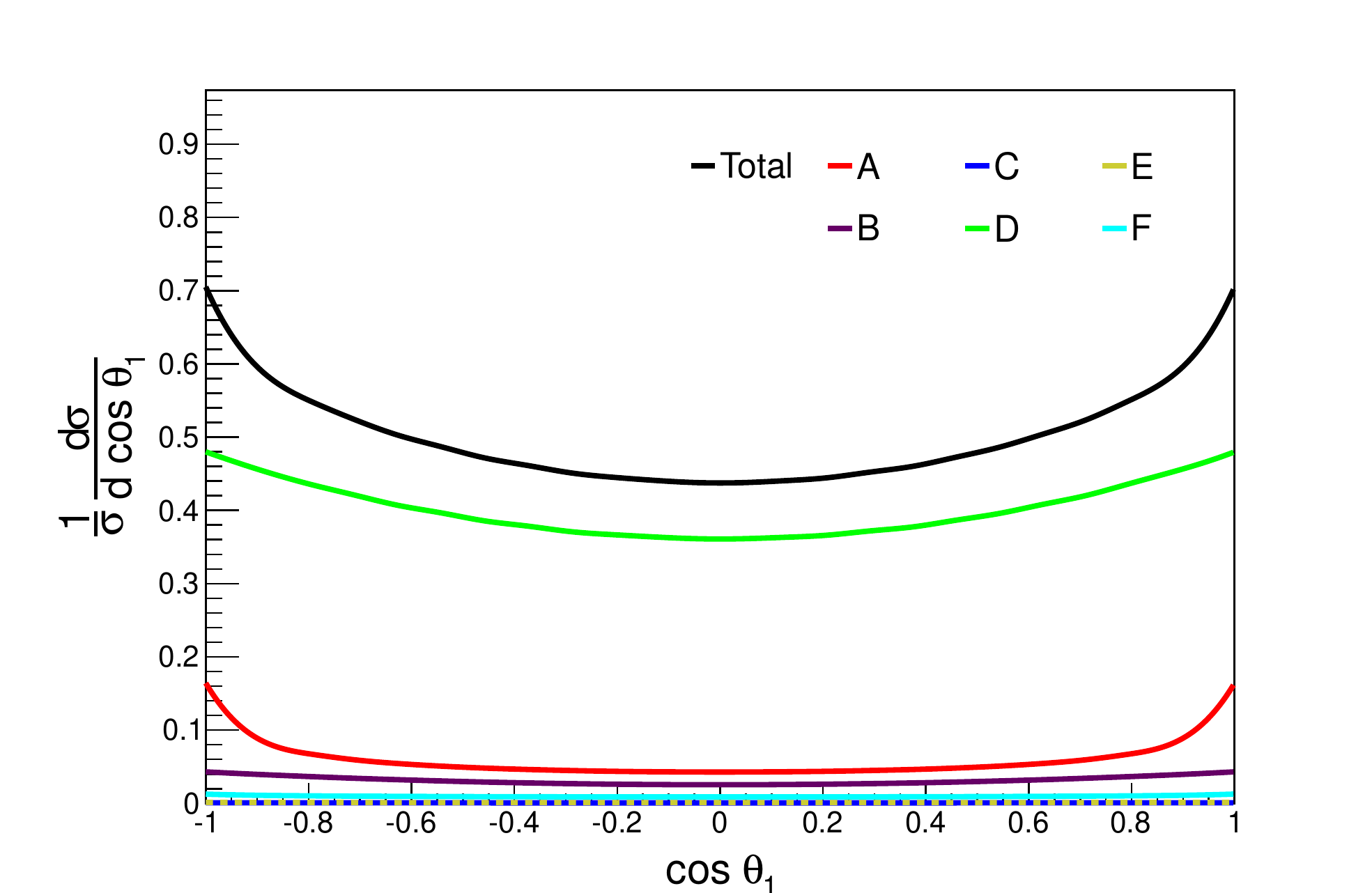}
\caption{On the left hand side we have plotted the $\cos{\theta_1}$ angular distributions for hypotheses 1-5 (hypothesis 1~$\equiv$ SM) defined in Sec.\ref{sec:SigSinglyDoubly}. On the right hand side we plot the components A-F of the background defined in Sec.\ref{sec:bgmassdiff}.}
\label{fig:dth1}
\end{figure*}

\begin{figure*}[ht!]
\includegraphics[width=0.45\textwidth]{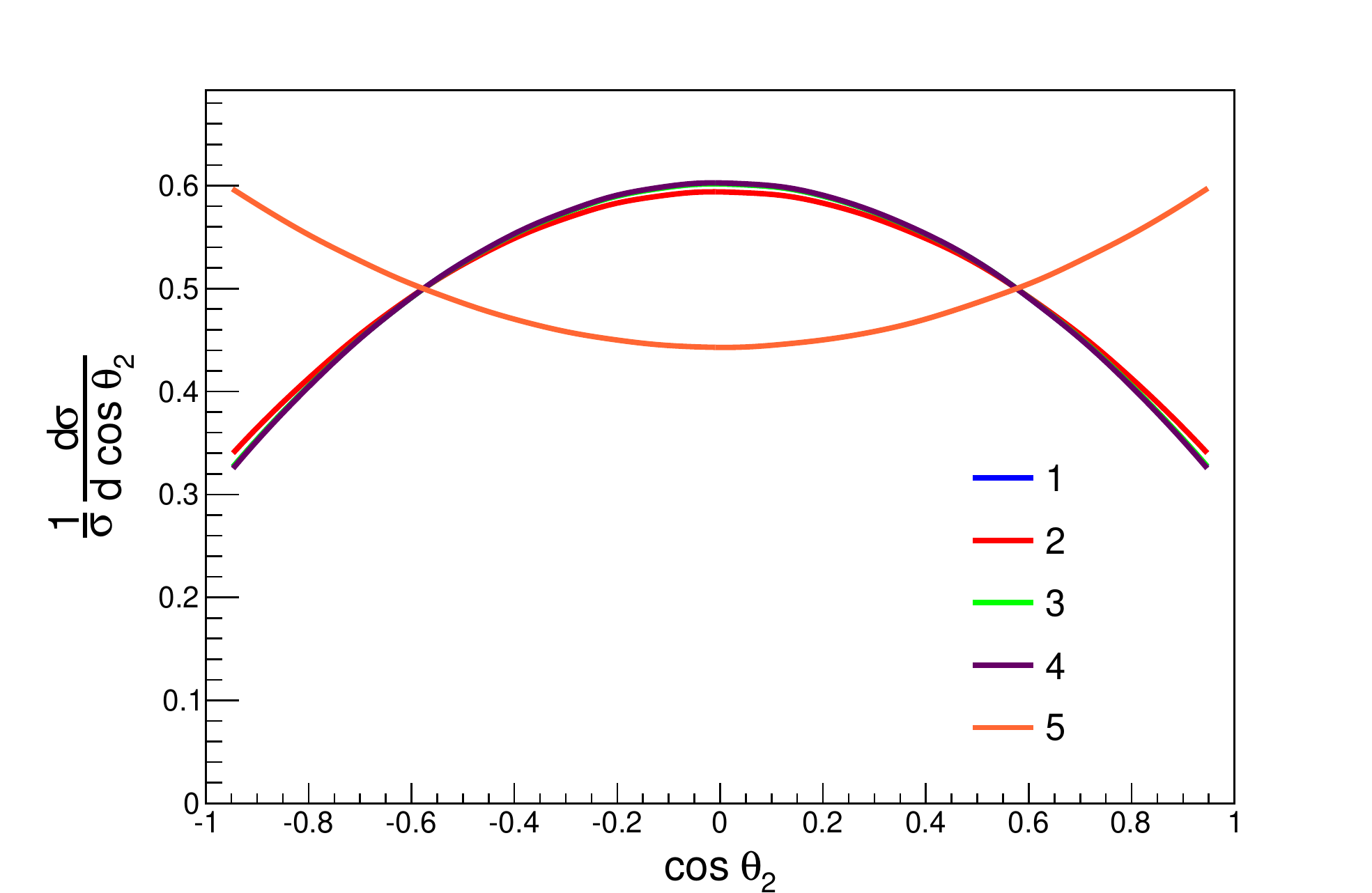}
\includegraphics[width=0.45\textwidth]{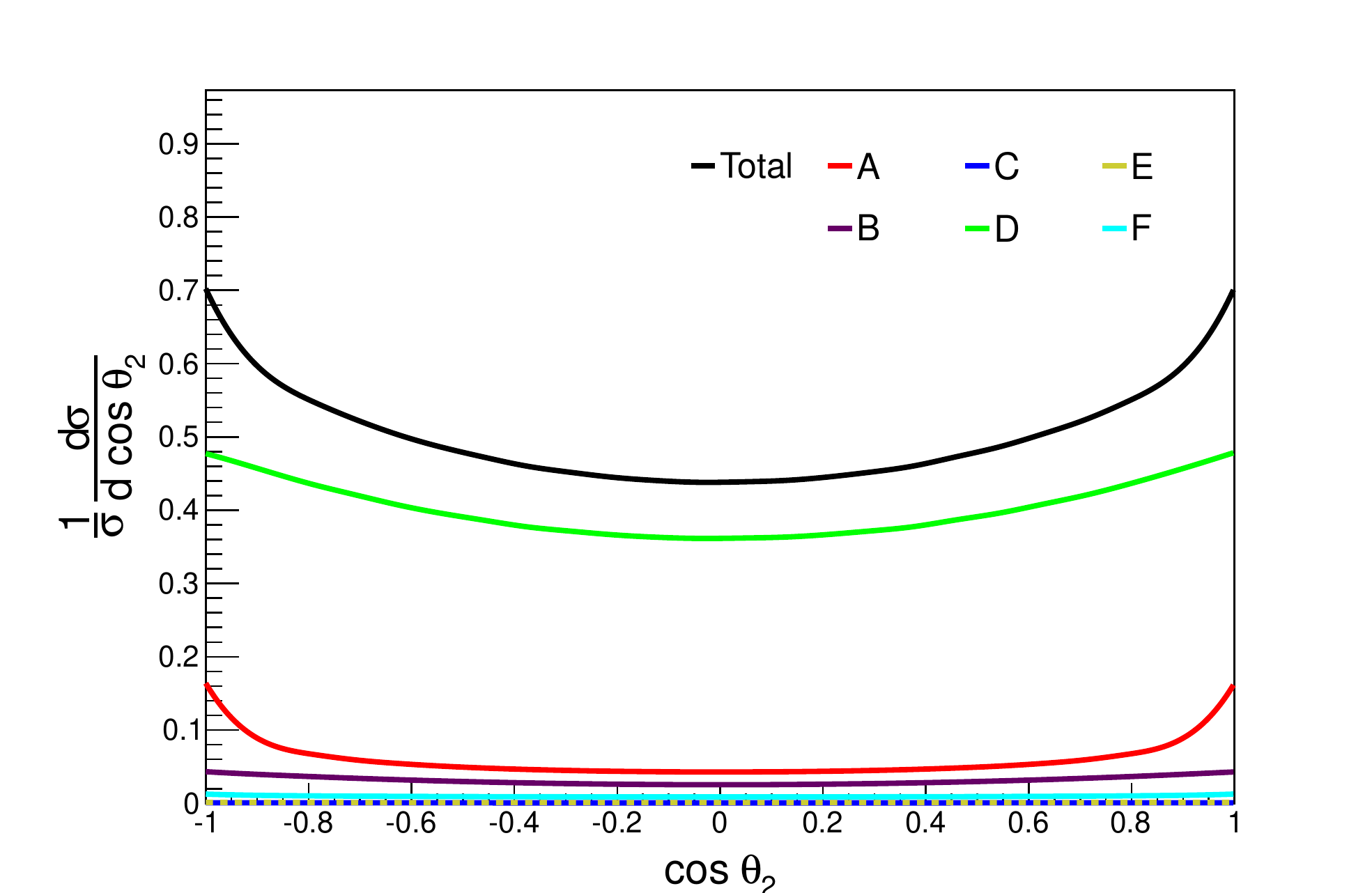}
\caption{On the left hand side we have plotted the $\cos{\theta_2}$ angular distributions for hypotheses 1-5 (hypothesis 1~$\equiv$ SM) defined in Sec.\ref{sec:SigSinglyDoubly}. On the right hand side we plot the components A-F of the background defined in Sec.\ref{sec:bgmassdiff}.}
\label{fig:dth2}
\end{figure*}

\begin{figure*}[ht!]
\includegraphics[width=0.45\textwidth]{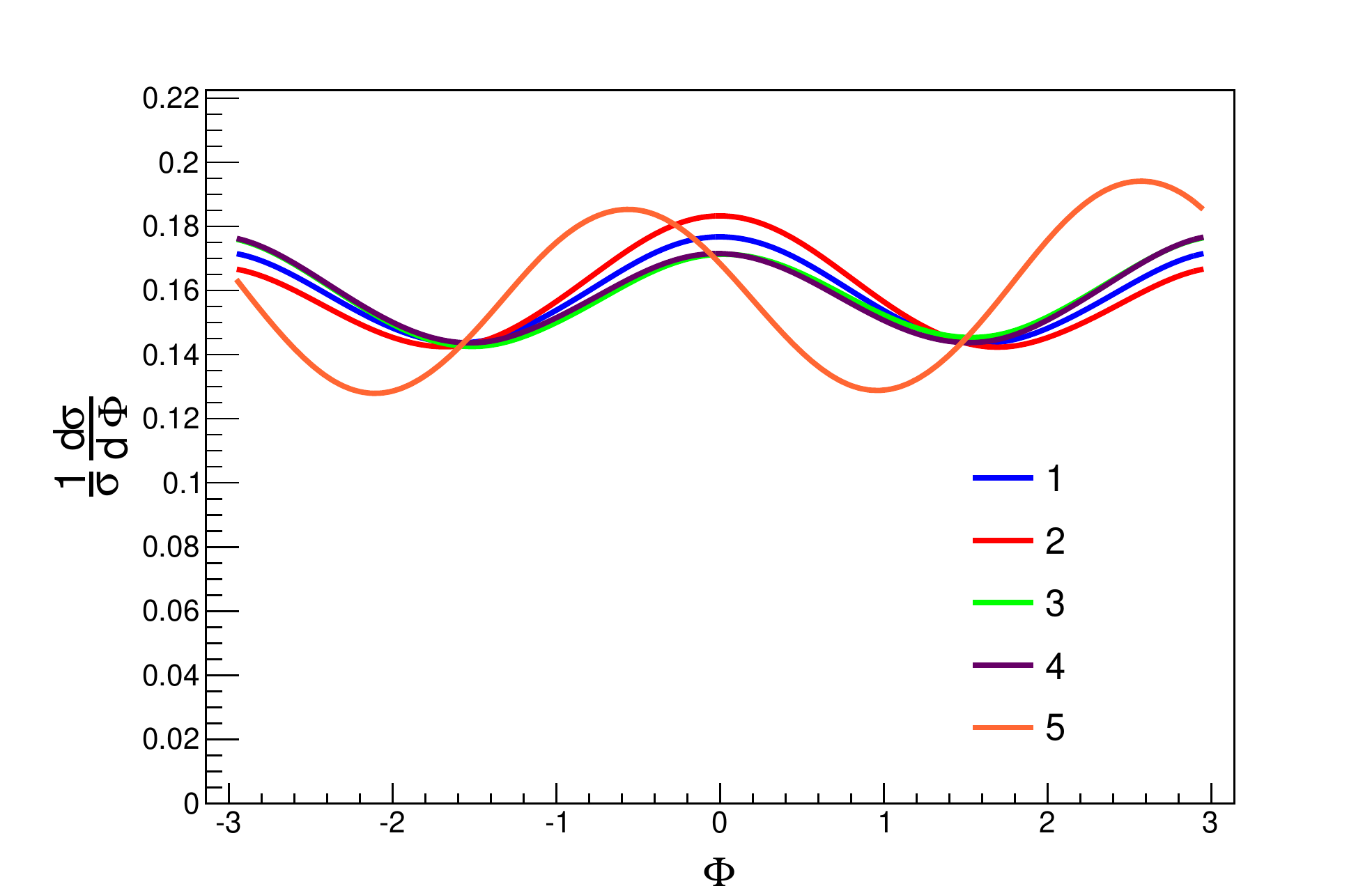}
\includegraphics[width=0.45\textwidth]{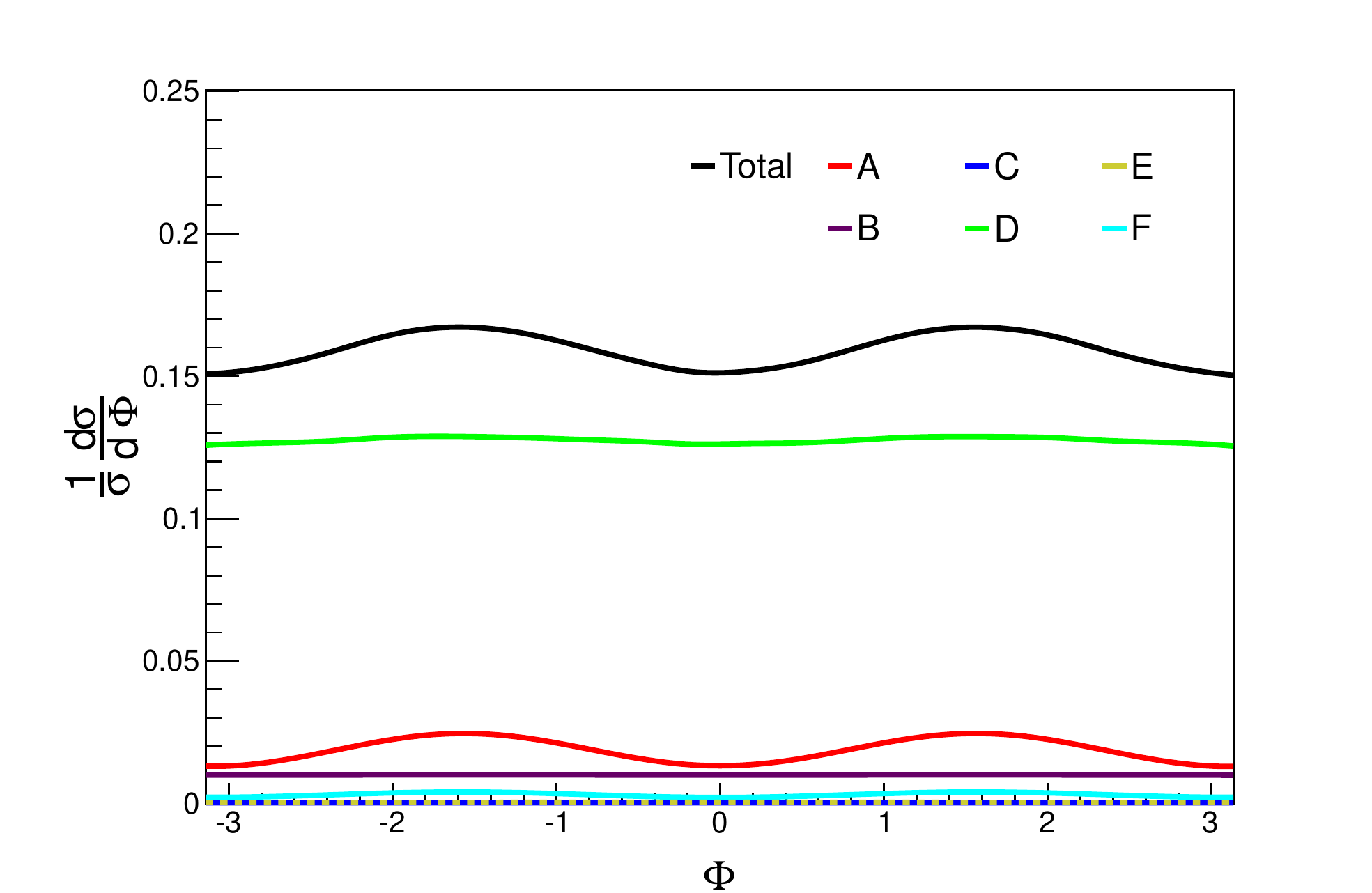}
\caption{On the left hand side we have plotted the $\Phi$ angular distributions for hypotheses 1-5 (hypothesis 1~$\equiv$ SM) defined in Sec.\ref{sec:SigSinglyDoubly}. On the right hand side we plot the components A-F of the background defined in Sec.\ref{sec:bgmassdiff}.}
\label{fig:dPhi}
\end{figure*}

\begin{figure*}[ht!]
\includegraphics[width=0.45\textwidth]{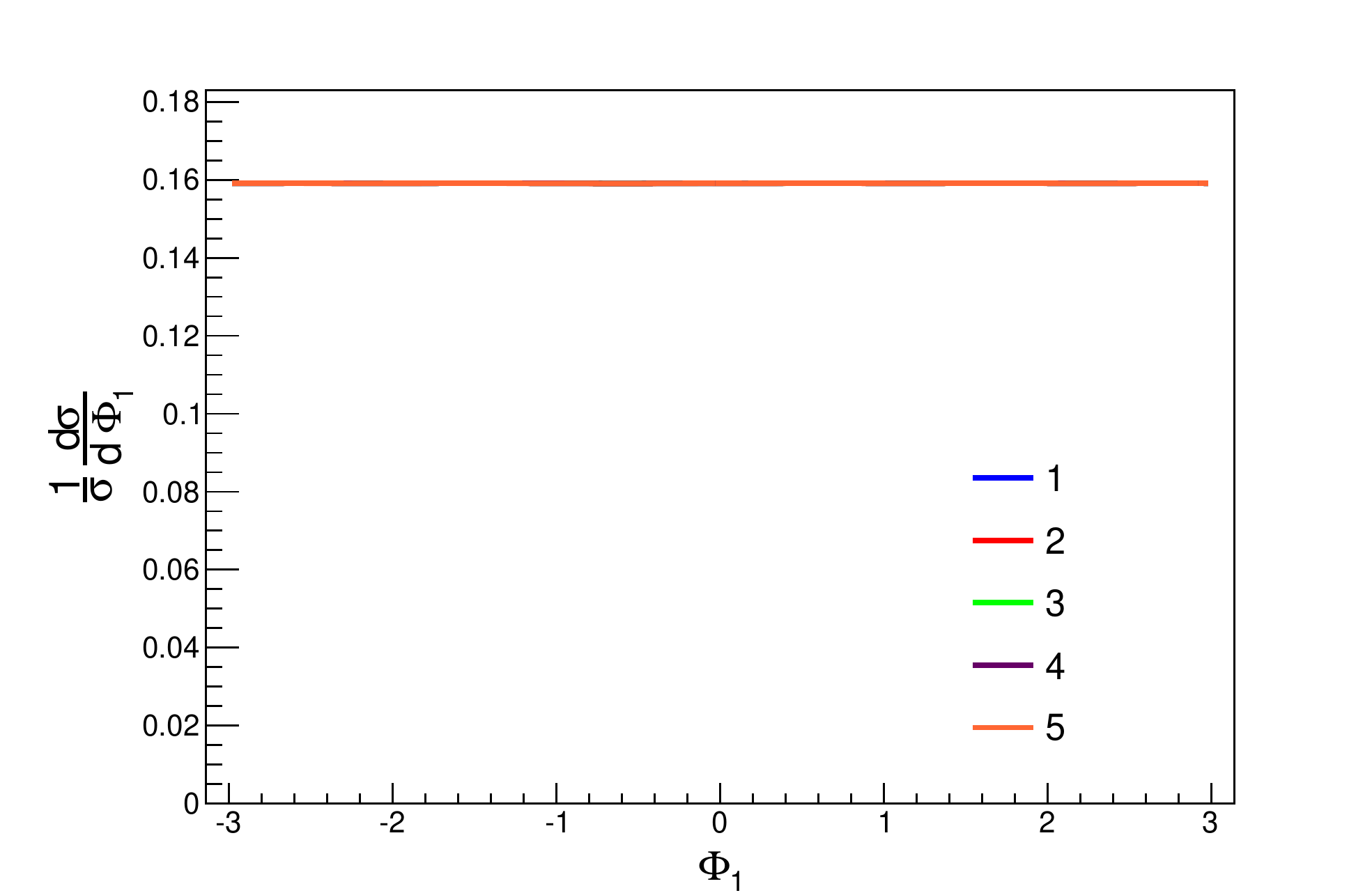}
\includegraphics[width=0.45\textwidth]{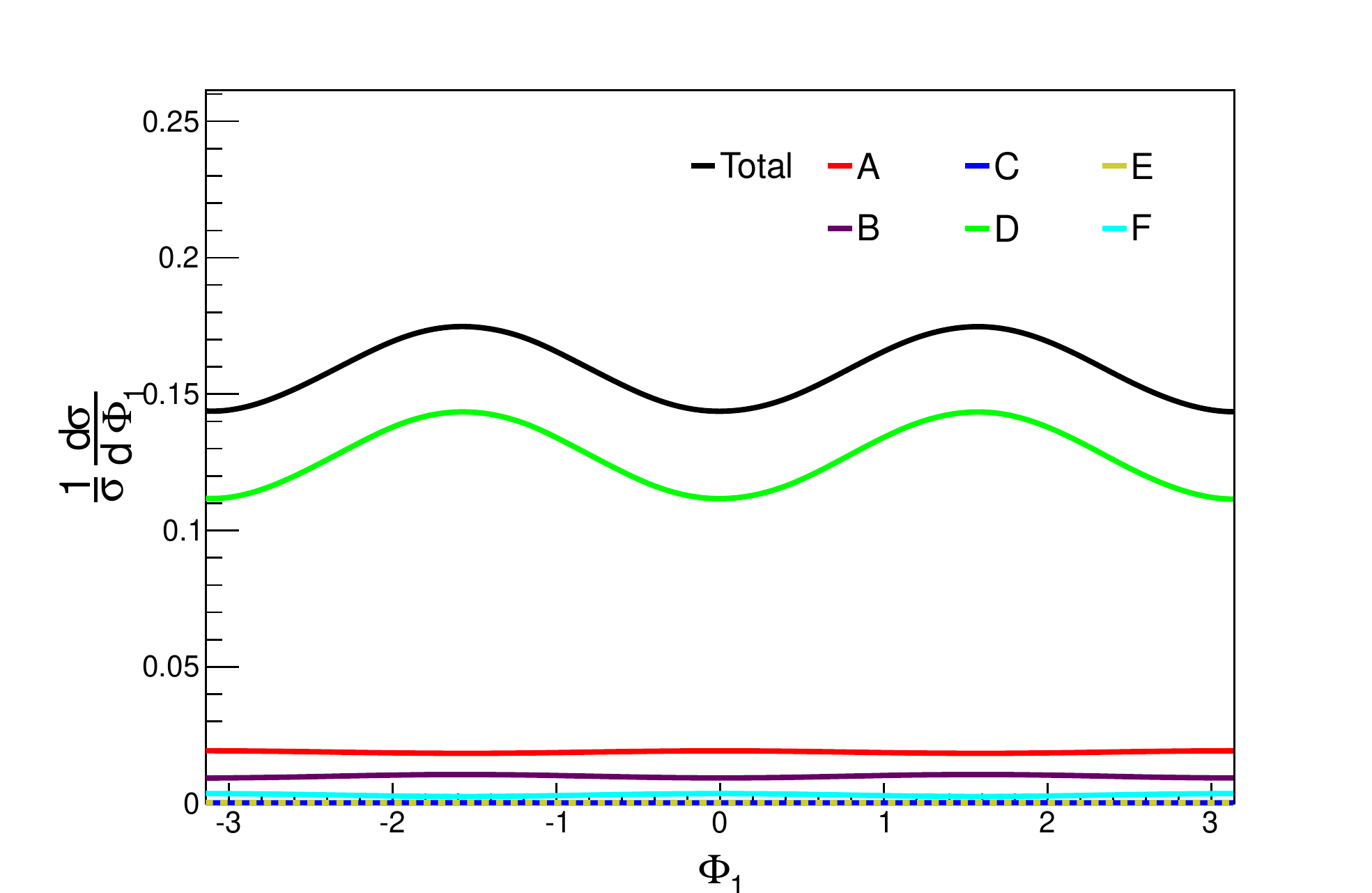}
\caption{On the left hand side we have plotted the $\Phi_1$ angular distributions for hypotheses 1-5 (hypothesis 1~$\equiv$ SM) defined in Sec.\ref{sec:SigSinglyDoubly}. On the right hand side we plot the components A-F of the background defined in Sec.\ref{sec:bgmassdiff}.}
\label{fig:dPhi1}
\end{figure*}

In Fig.~\ref{fig:dTheta}-\ref{fig:dPhi1} we show the angular distributions for the 5 angles ($\cos{\Theta},\cos{\theta_1},\cos{\theta_2},\Phi_1,\Phi$) found in the four lepton system and defined in Sec.~\ref{sec:events}.  We plot the angular distributions for signal hypotheses 1-5 defined in Sec.~\ref{sec:SigSinglyDoubly} and also show the various background components defined in Sec.~\ref{sec:bgmassdiff}.  For all distributions the phase space is defined as 4~GeV $< M_1 < 120$~GeV and 4~GeV $< M_2 <$ 120~GeV with $\sqrt{\hat{s}}=125$~GeV for signal and 110~GeV $< \sqrt{\hat{s}} <$ 140~GeV for background.

Since we are considering a spin-0 scalar as our signal, the $\cos{\Theta}$ and $\Phi_1$ are of course flat, but are still useful for discriminating between signal and background.  A particularly interesting variable is the azimuthal angle between the lepton decay planes, $\Phi$. This is especially sensitive to the various interference effects as well as the CP properties of the decaying scalar, as was pointed out in~\cite{Cao:2009ah}.  One can see that the different signal hypothesis affect the modulation of $\Phi$ while an extreme case like the CP violating hypothesis 5 can lead to a striking signal in the form of an asymmetric modulation and phase shift relative to the SM prediction. 

For the background we can see how the various components contribute to the different kinematic variables. It is clear that the $Z\gamma~t+u$ component~($D$) is the dominant contribution for our defined phase space. Note however, that the s-channel component ($A$) also contributes and in particular is the dominant contribution to the modulation of $\Phi$. We can also see that the resonant process affects $\cos{\theta_1}$ and $\cos{\theta_2}$, especially in the forward regions.  It is also interesting to comment that the $\gamma\gamma$ contribution~($B$) is featureless in all of the distributions except for a small upward slope in the extreme forward directions of $\cos{\Theta}$. Note that for the $\Phi_1$ azimuthal angle, the modulation is due entirely to the $Z\gamma~t+u$ component~($D$). Whether these different effects can still be seen once detector effects are included requires careful study which we leave for future work.



\subsection{Doubly Differential spectra}
\label{sec:bgdists}
\begin{figure*}
\includegraphics[width=0.32\textwidth]{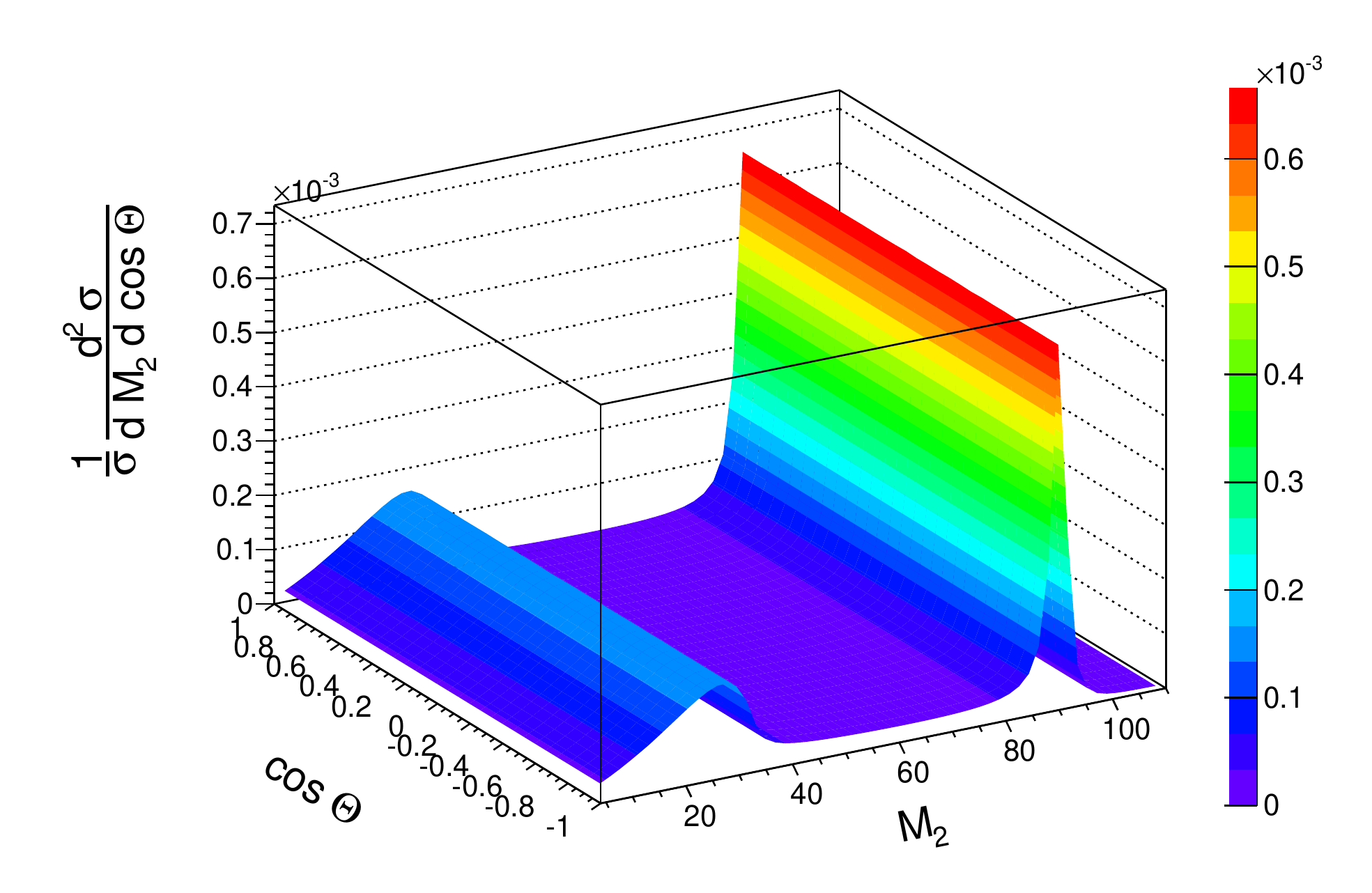}
\includegraphics[width=0.32\textwidth]{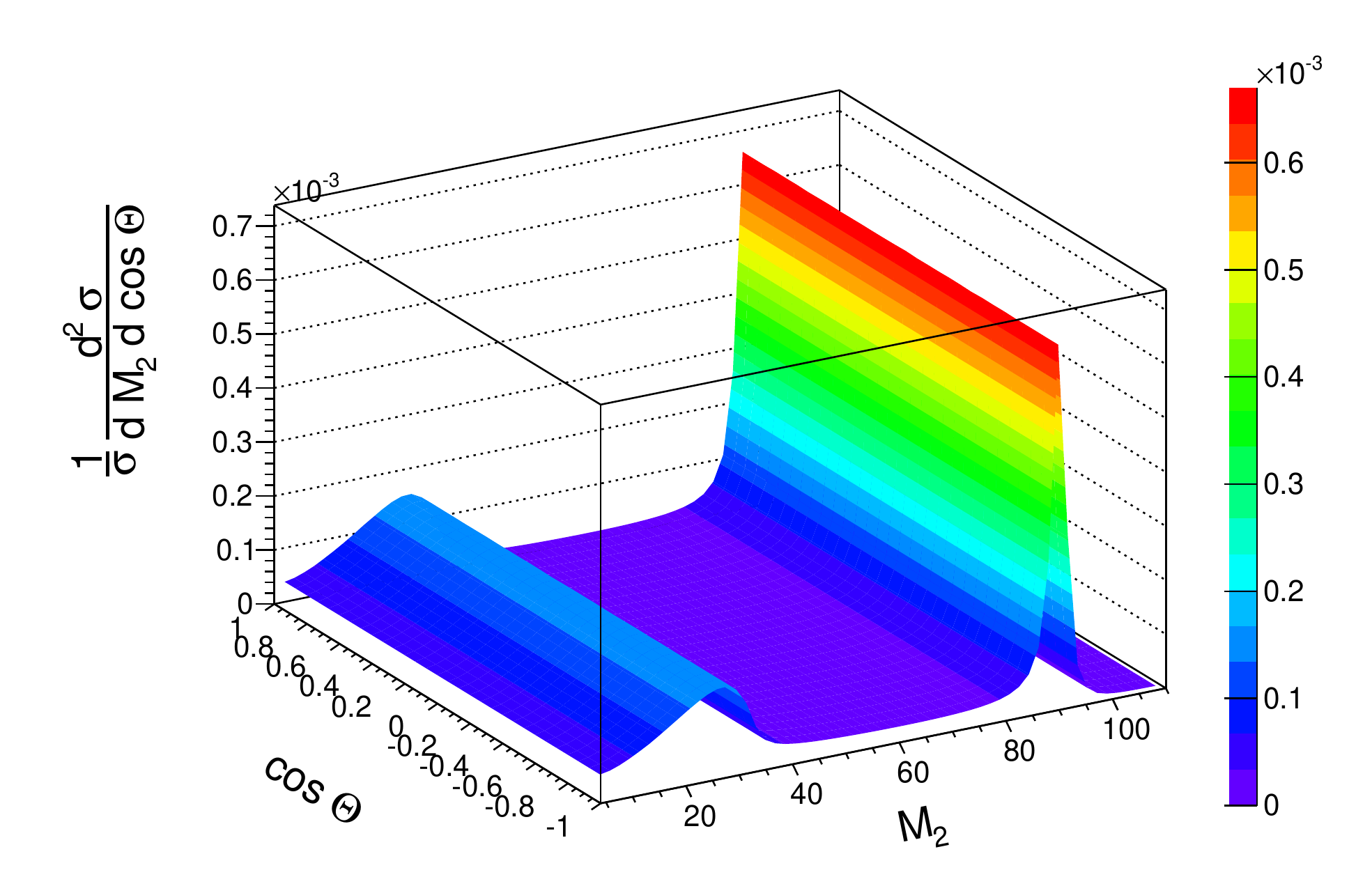}
\includegraphics[width=0.32\textwidth]{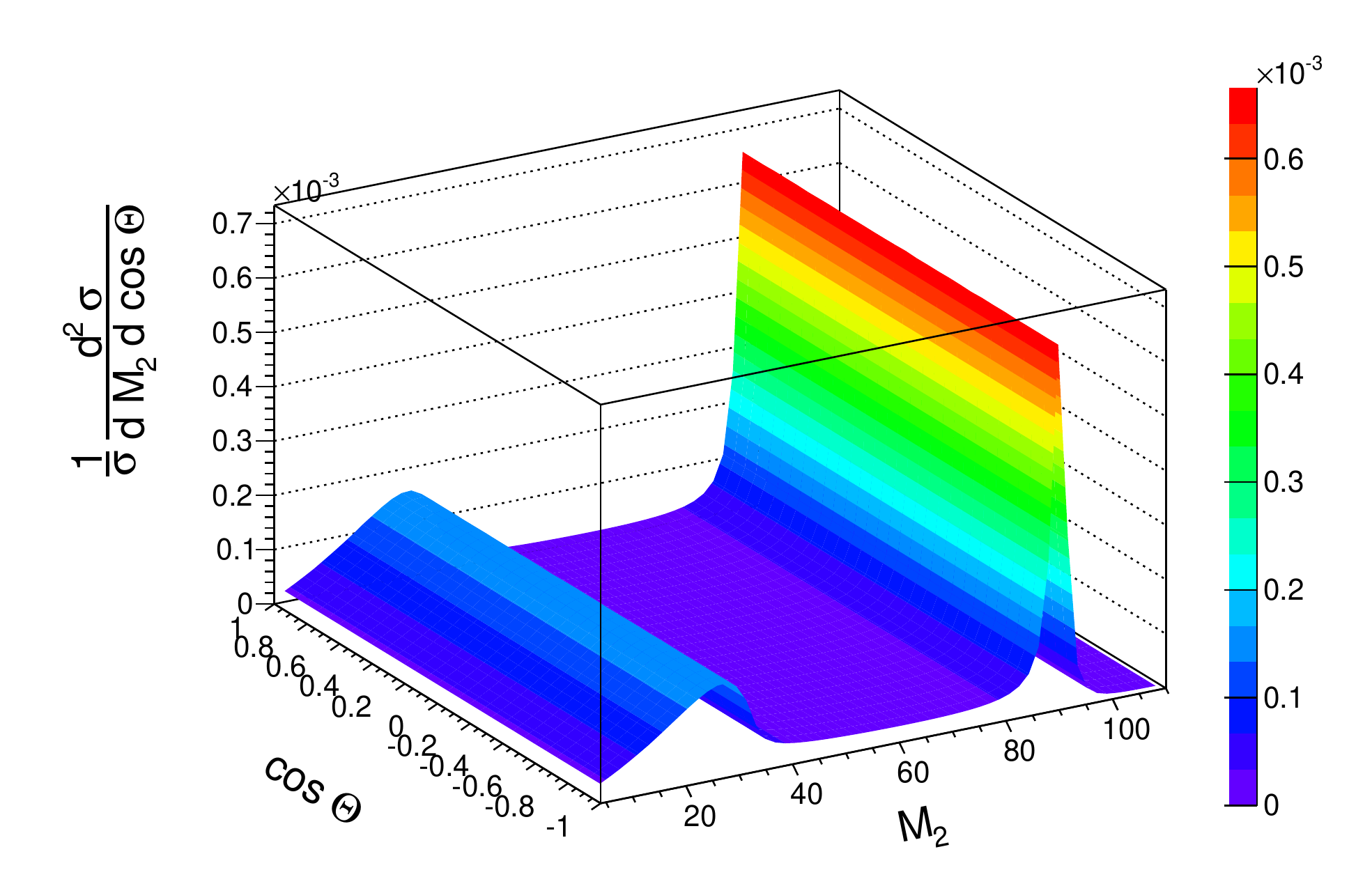}
\includegraphics[width=0.32\textwidth]{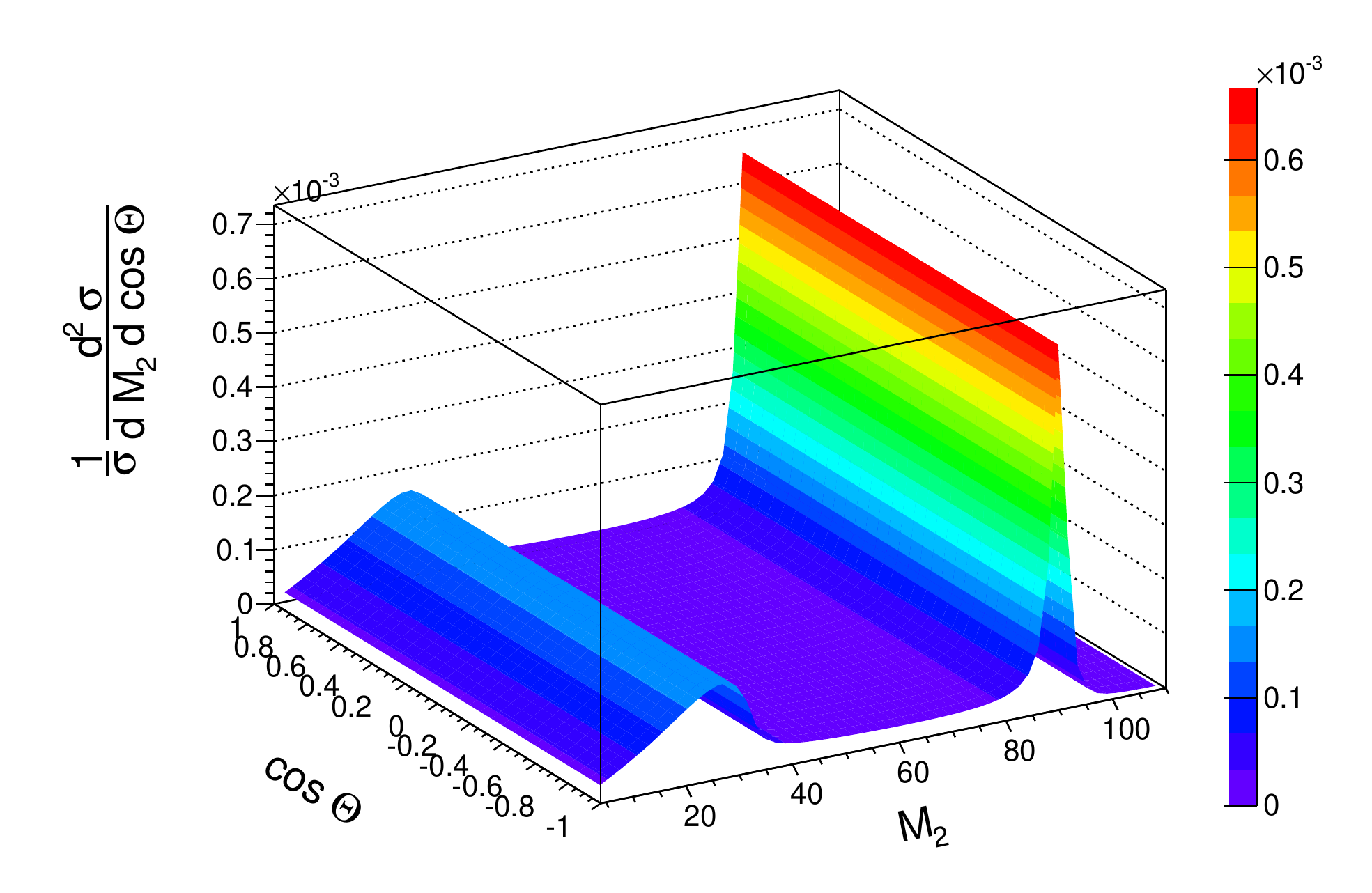}
\includegraphics[width=0.32\textwidth]{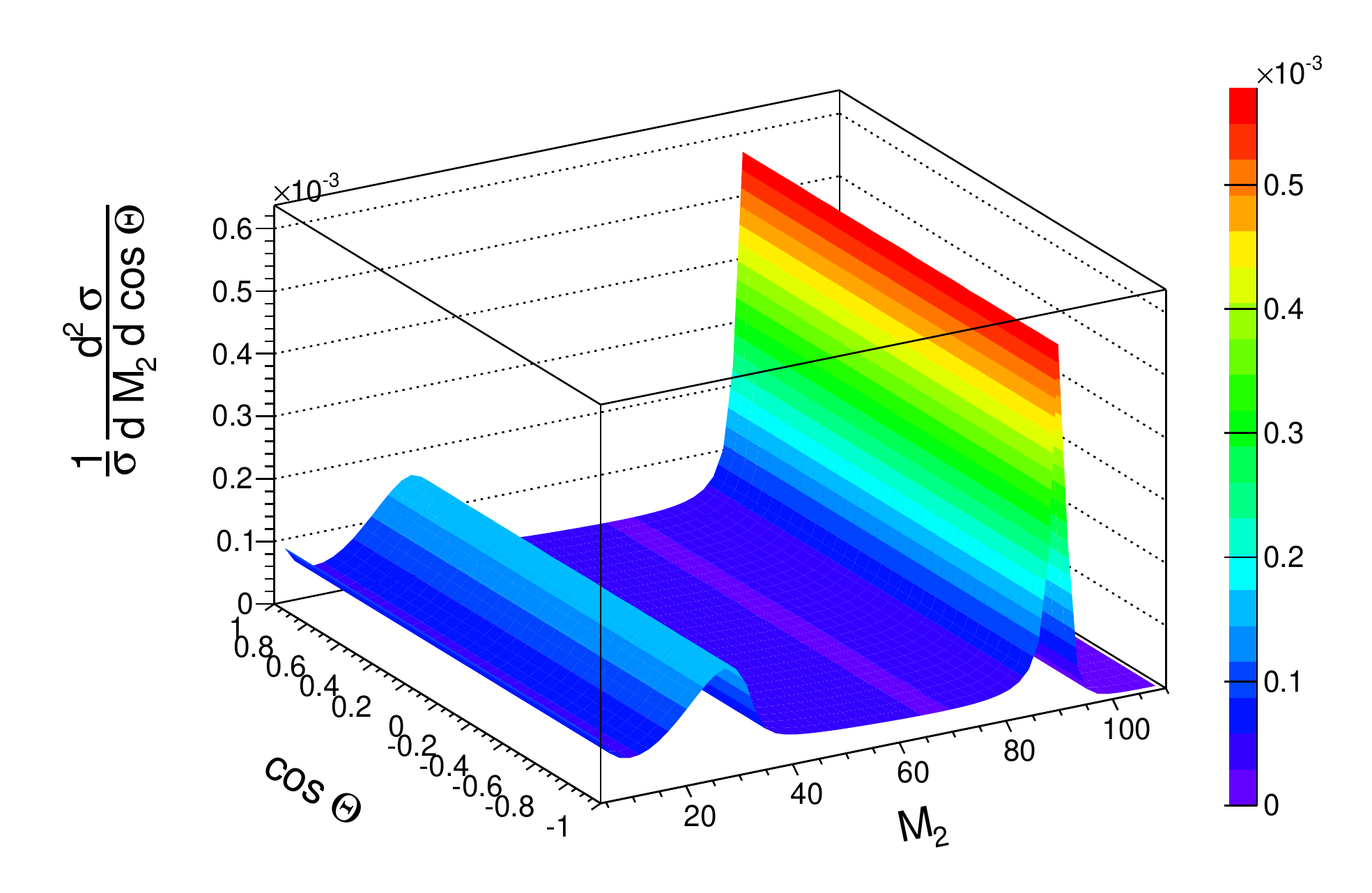}
\includegraphics[width=0.32\textwidth]{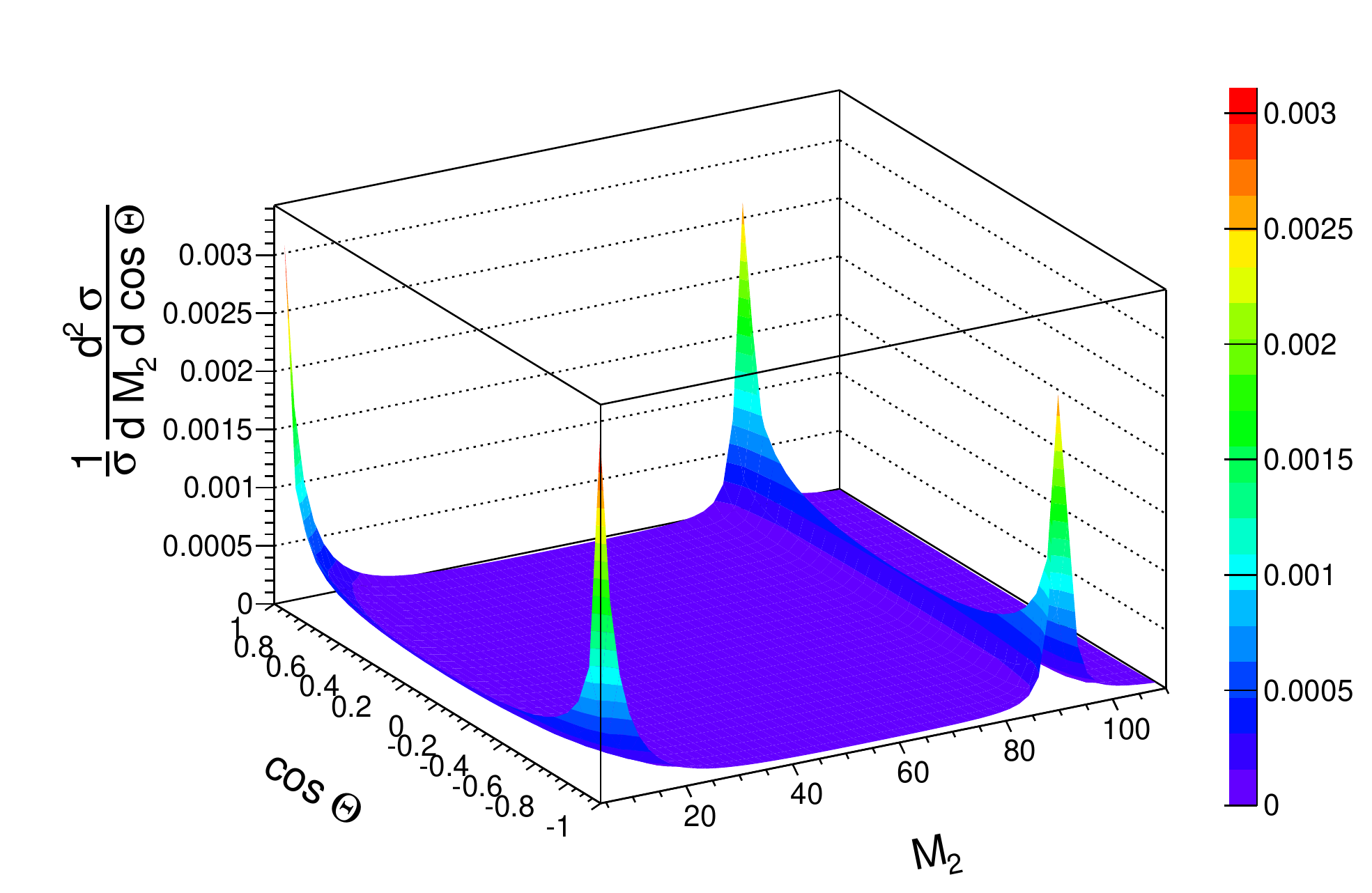}
\caption{The $(M_2, \cos{\Theta})$ doubly differential spectrum. The first five distributions are for signal hypotheses 1-5 (hypothesis 1~$\equiv$ SM in top left) defined in Sec.\ref{sec:SigSinglyDoubly} while the bottom right plot is for the full background.}
\label{fig:m2Thdiff}
\end{figure*}

\begin{figure*}
\includegraphics[width=0.32\textwidth]{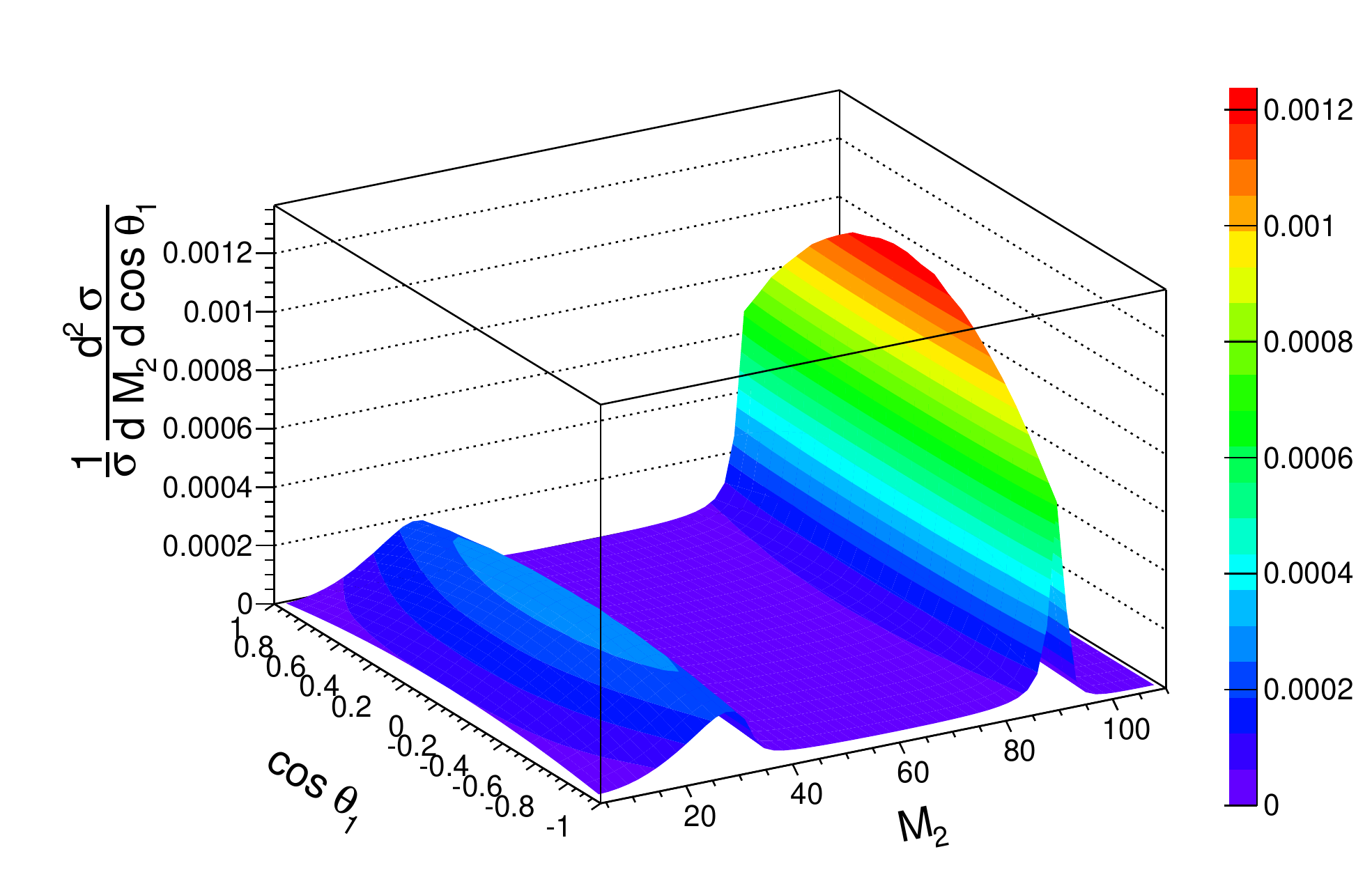}
\includegraphics[width=0.32\textwidth]{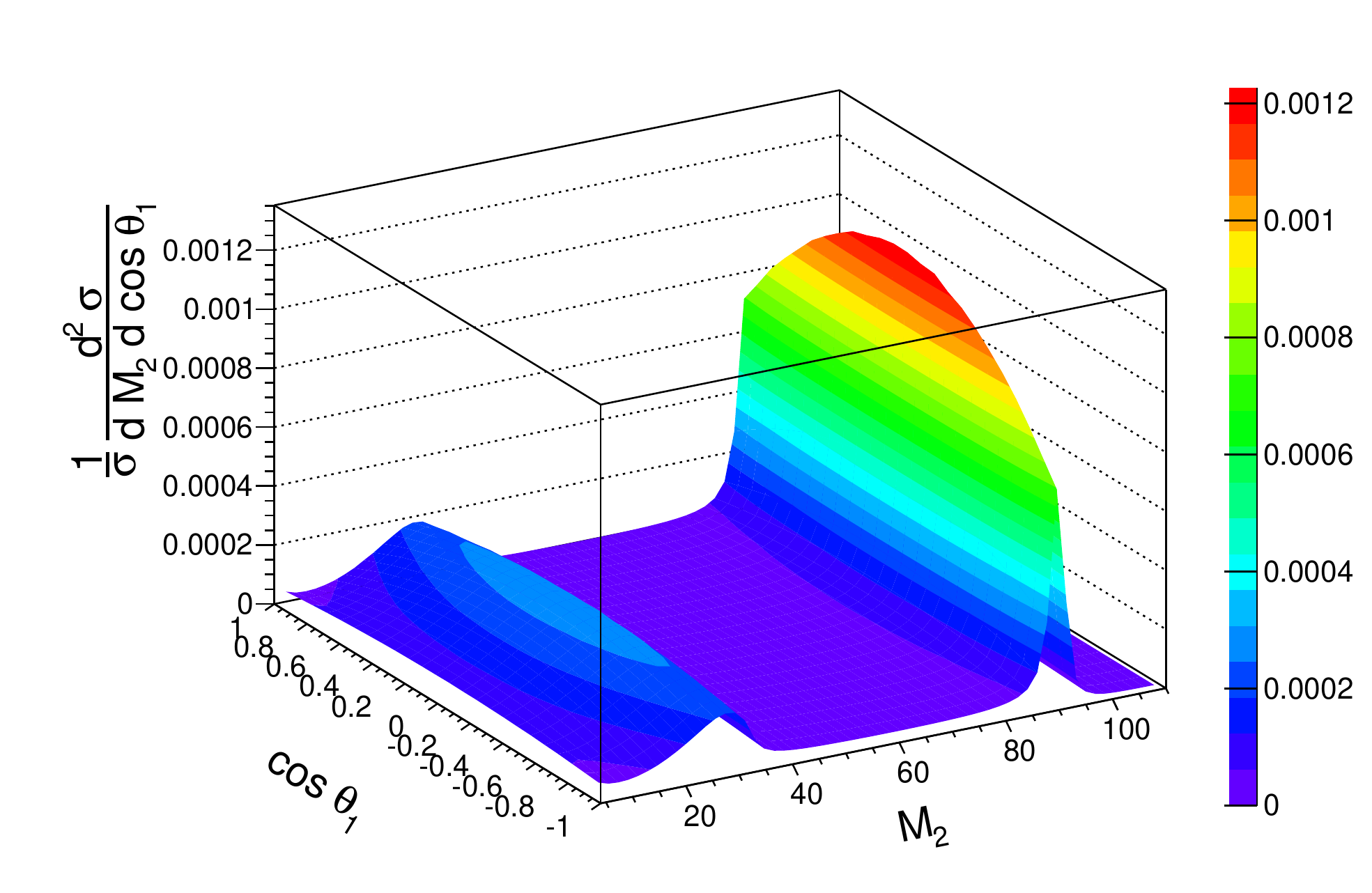}
\includegraphics[width=0.32\textwidth]{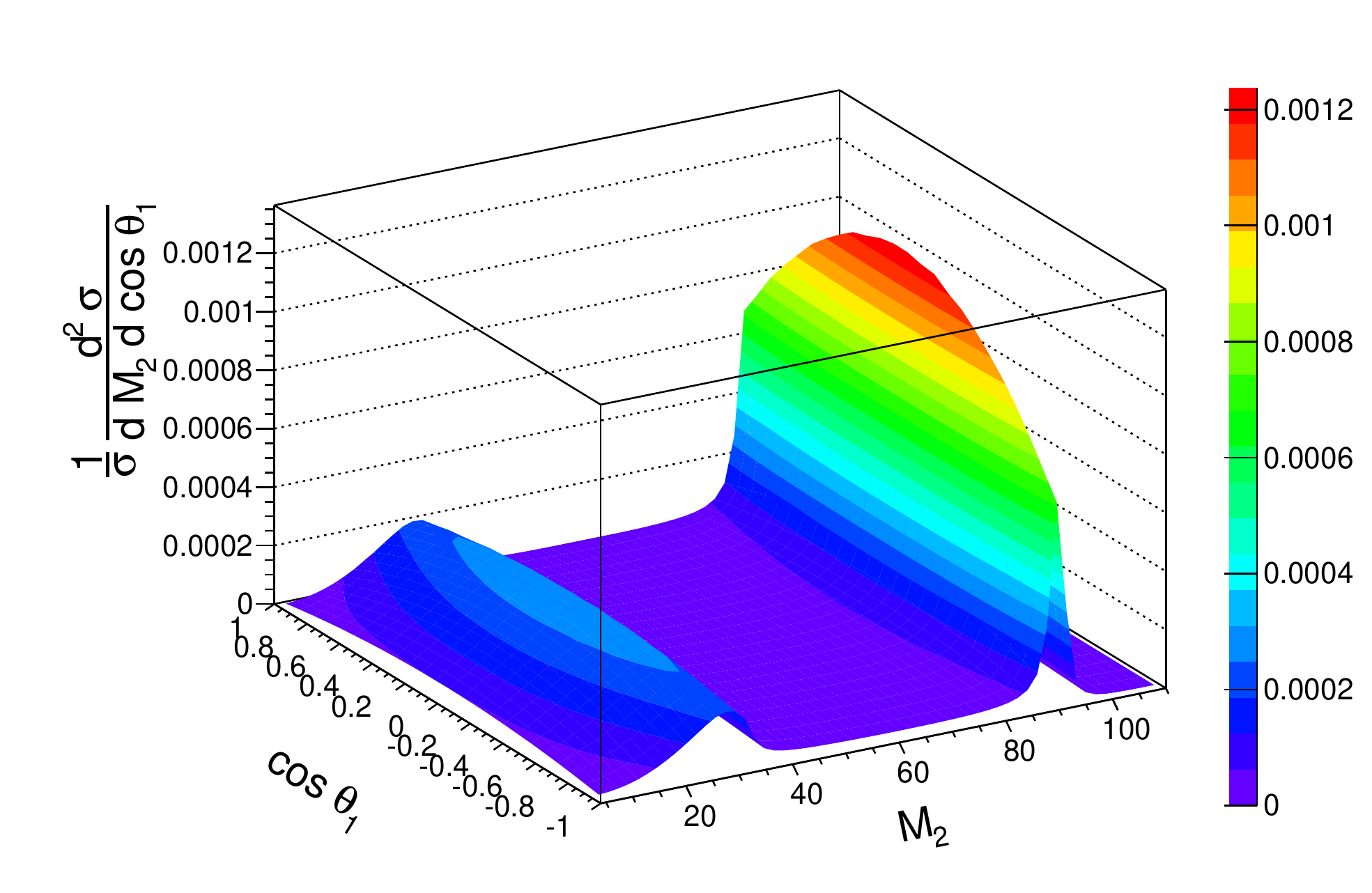}
\includegraphics[width=0.32\textwidth]{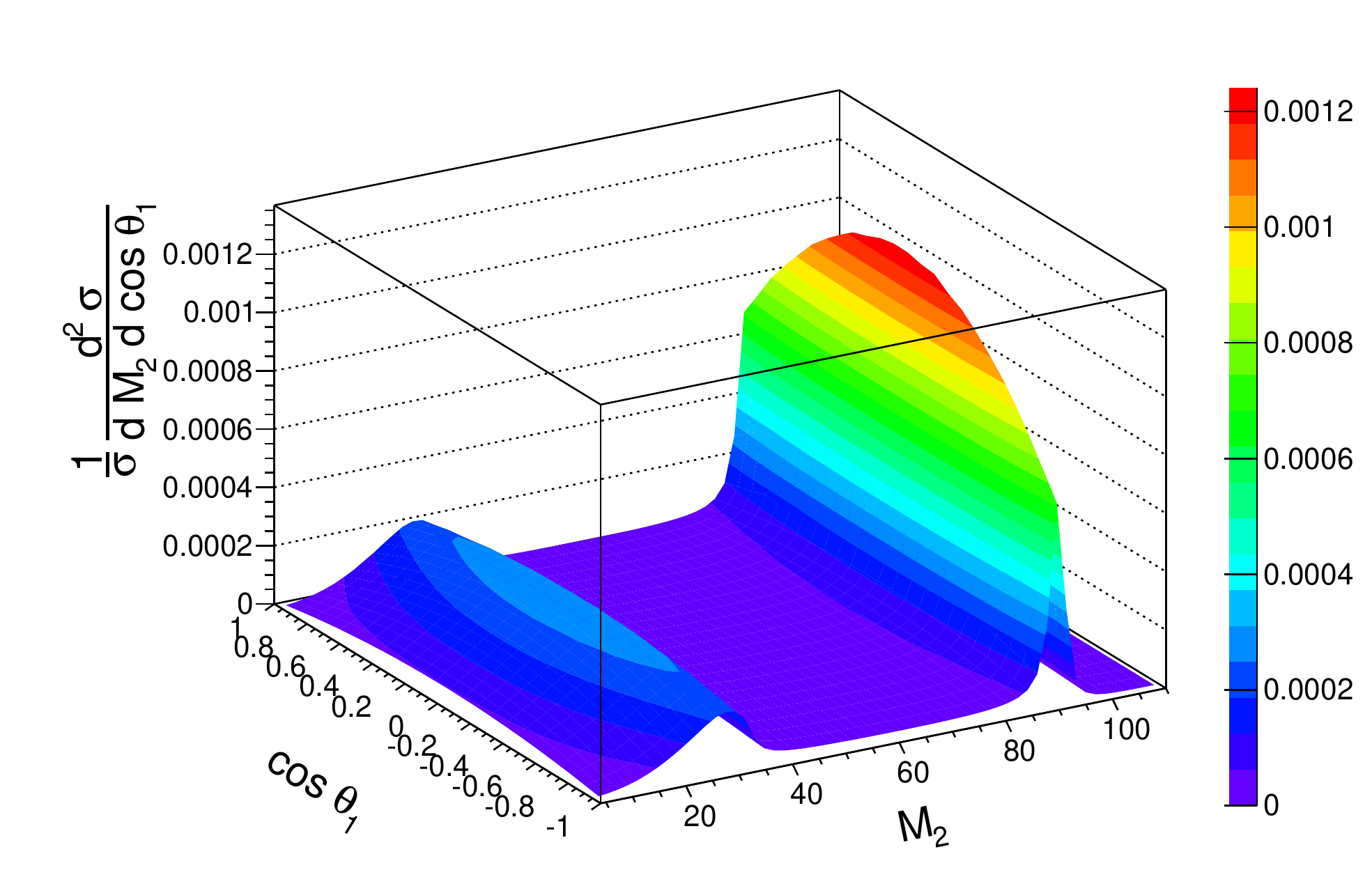}
\includegraphics[width=0.32\textwidth]{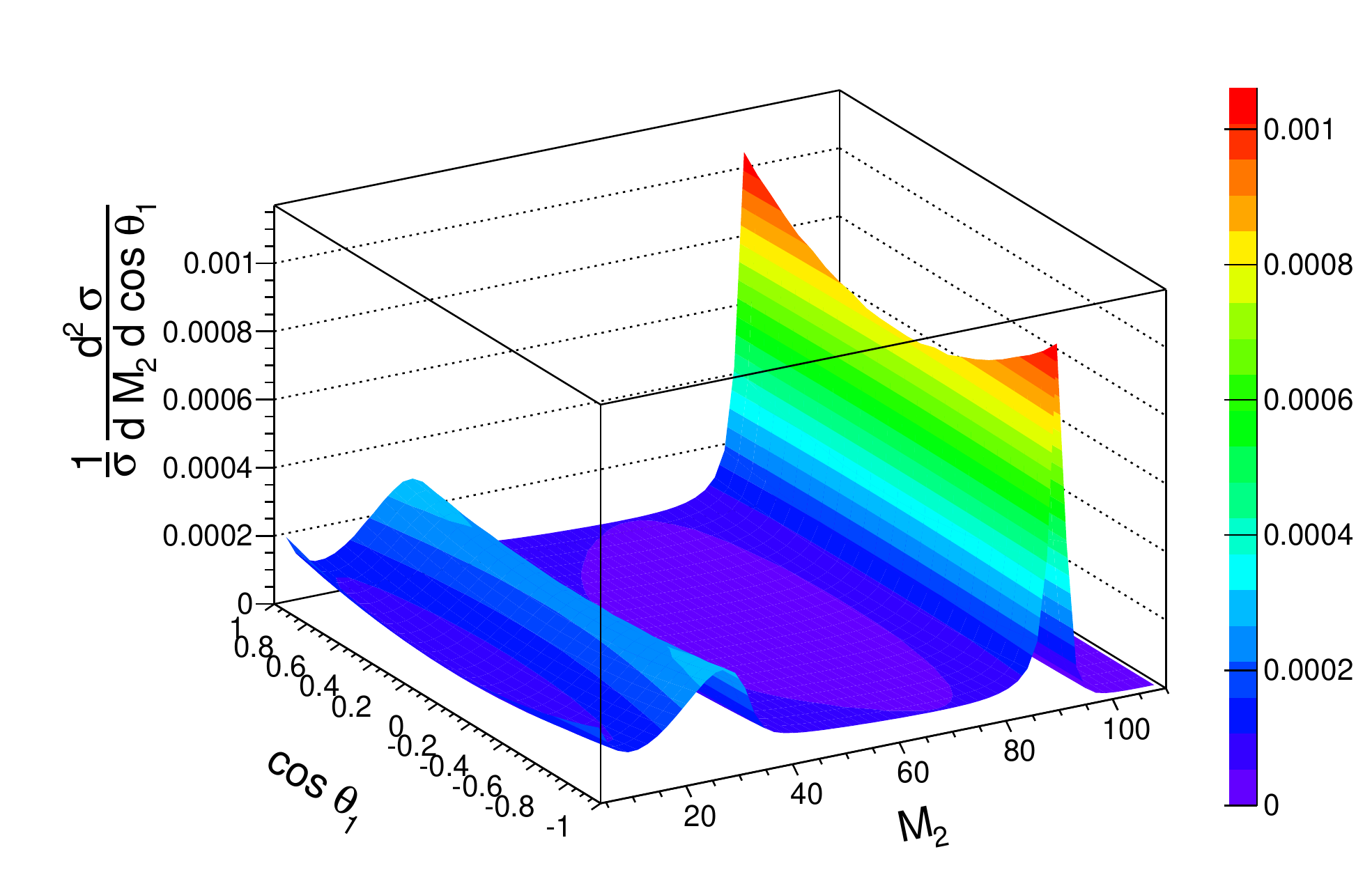}
\includegraphics[width=0.32\textwidth]{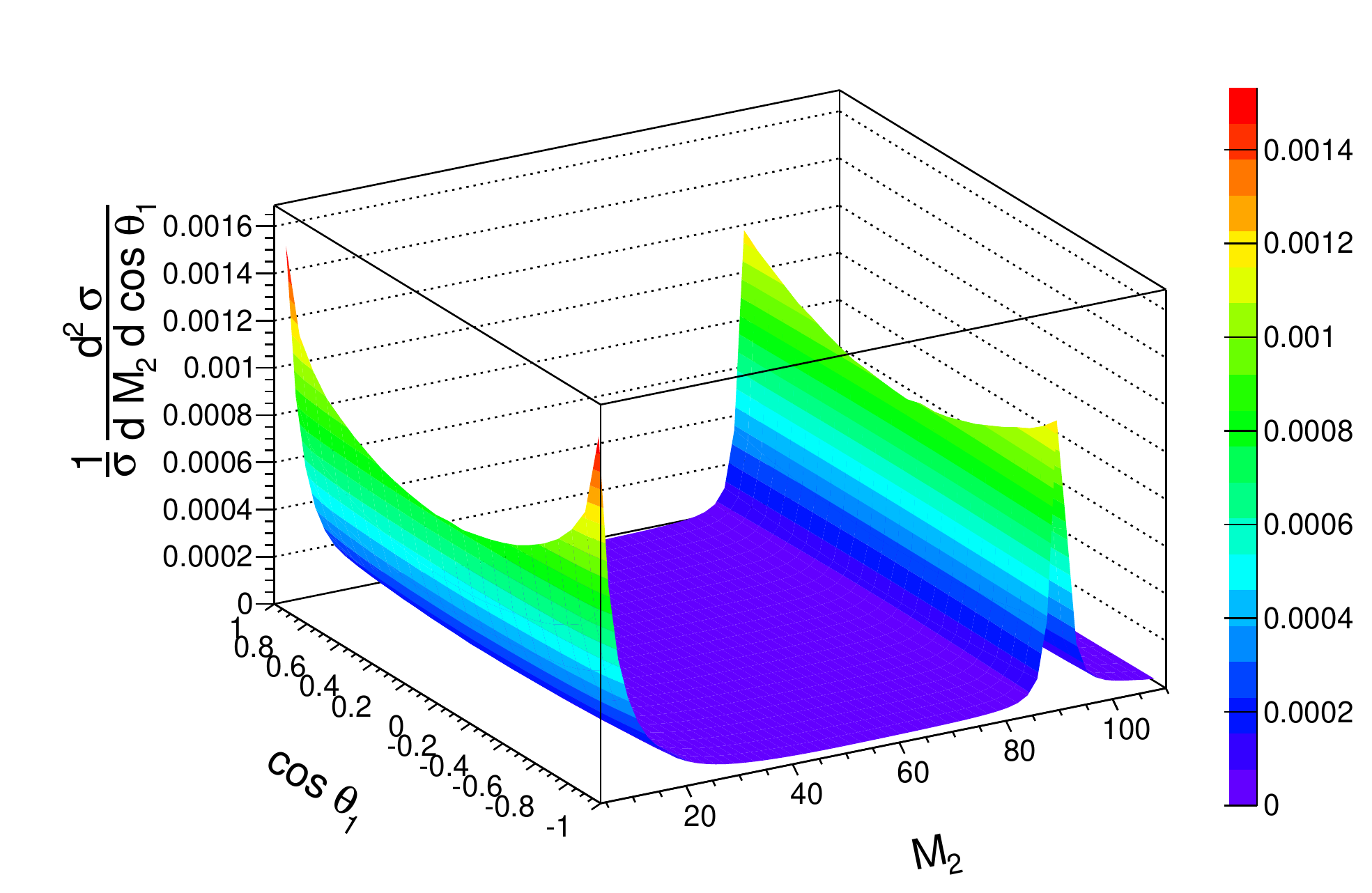}
\caption{The $(M_2, \cos{\theta_1})$ doubly differential spectrum. The first five distributions are for signal hypotheses 1-5 (hypothesis 1~$\equiv$ SM in top left) defined in Sec.\ref{sec:SigSinglyDoubly} while the bottom right plot is for the full background.}
\label{fig:m2th1diff}
\end{figure*}

\begin{figure*}
\includegraphics[width=0.32\textwidth]{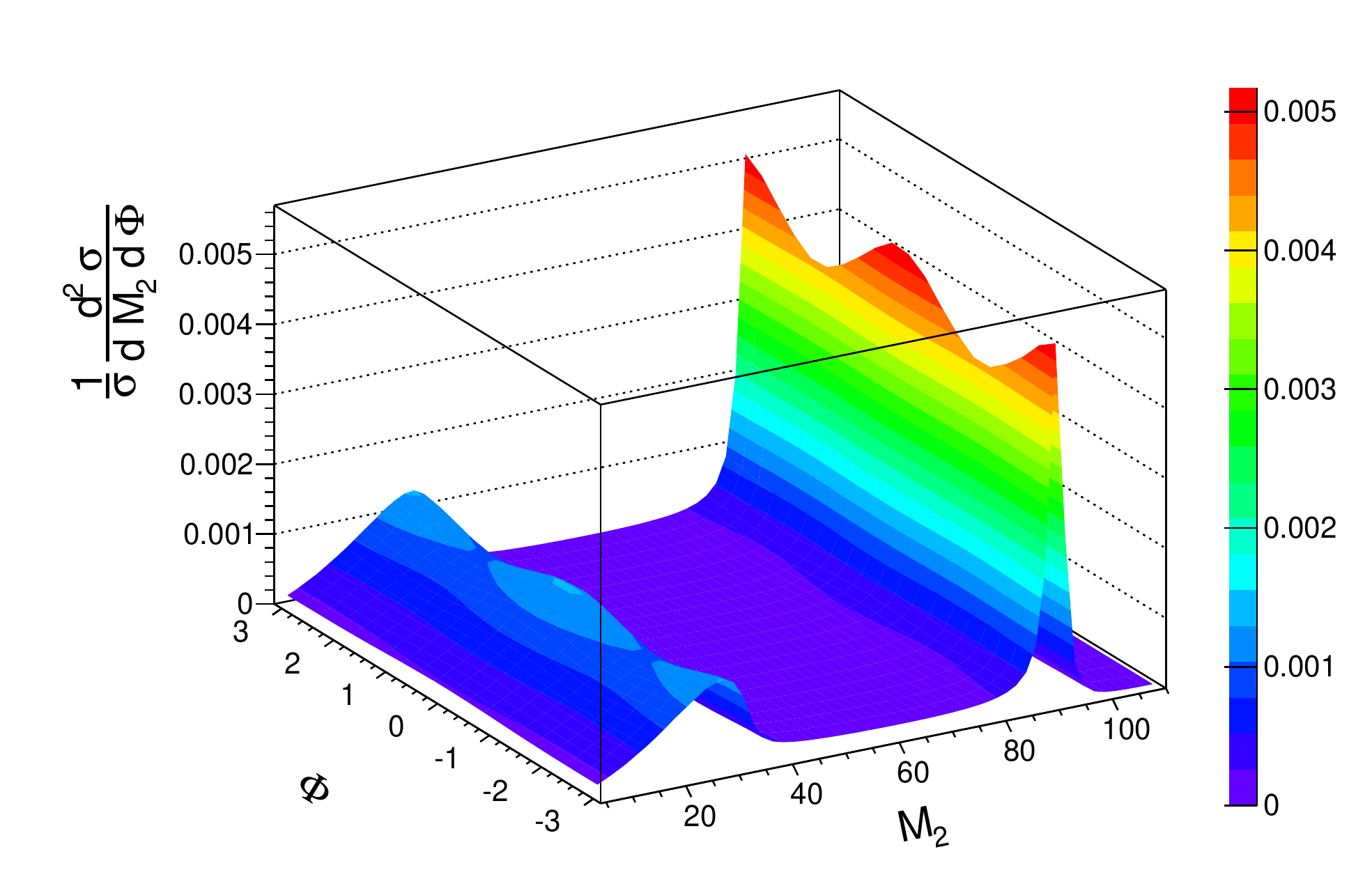}
\includegraphics[width=0.32\textwidth]{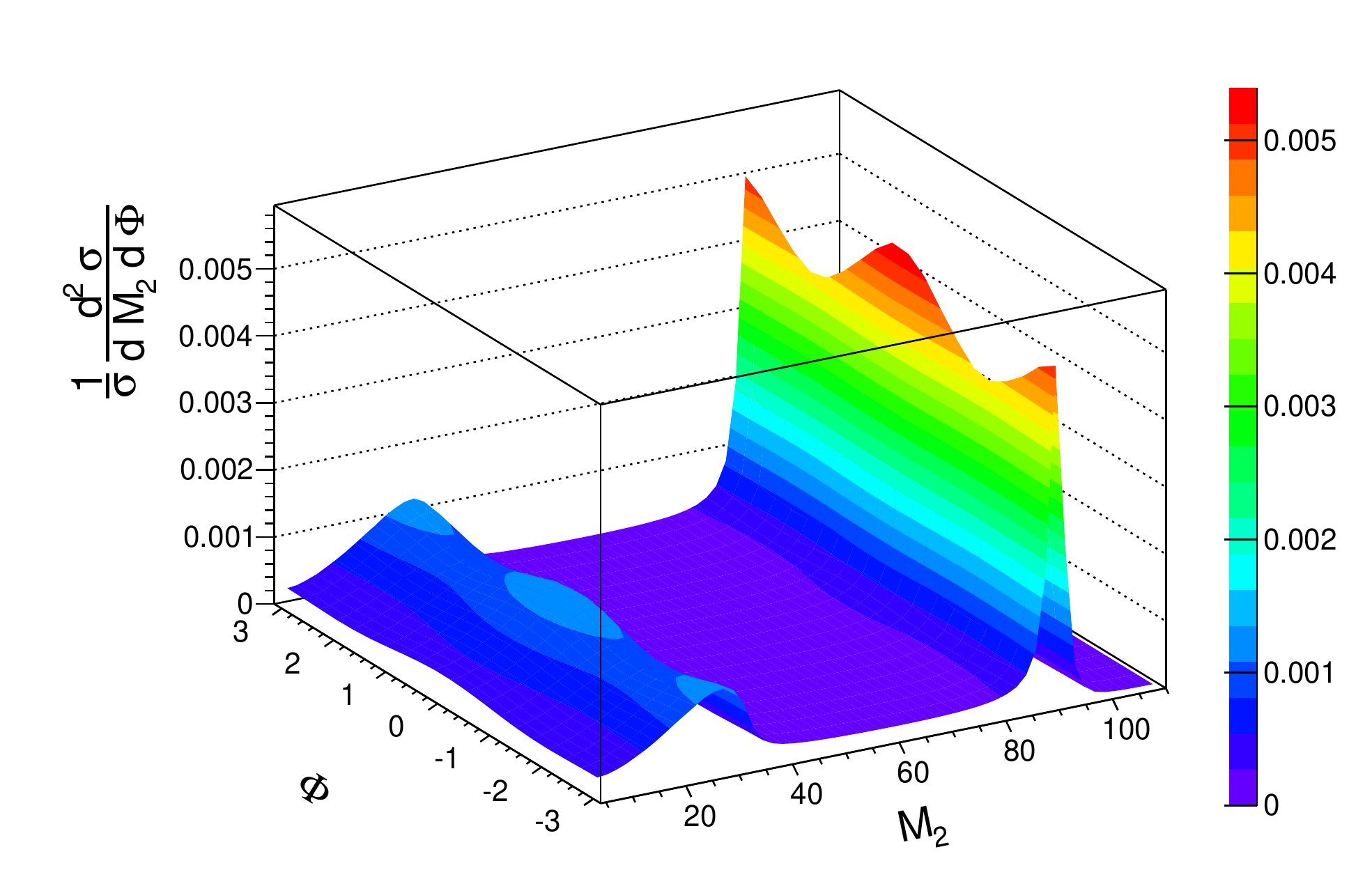}
\includegraphics[width=0.32\textwidth]{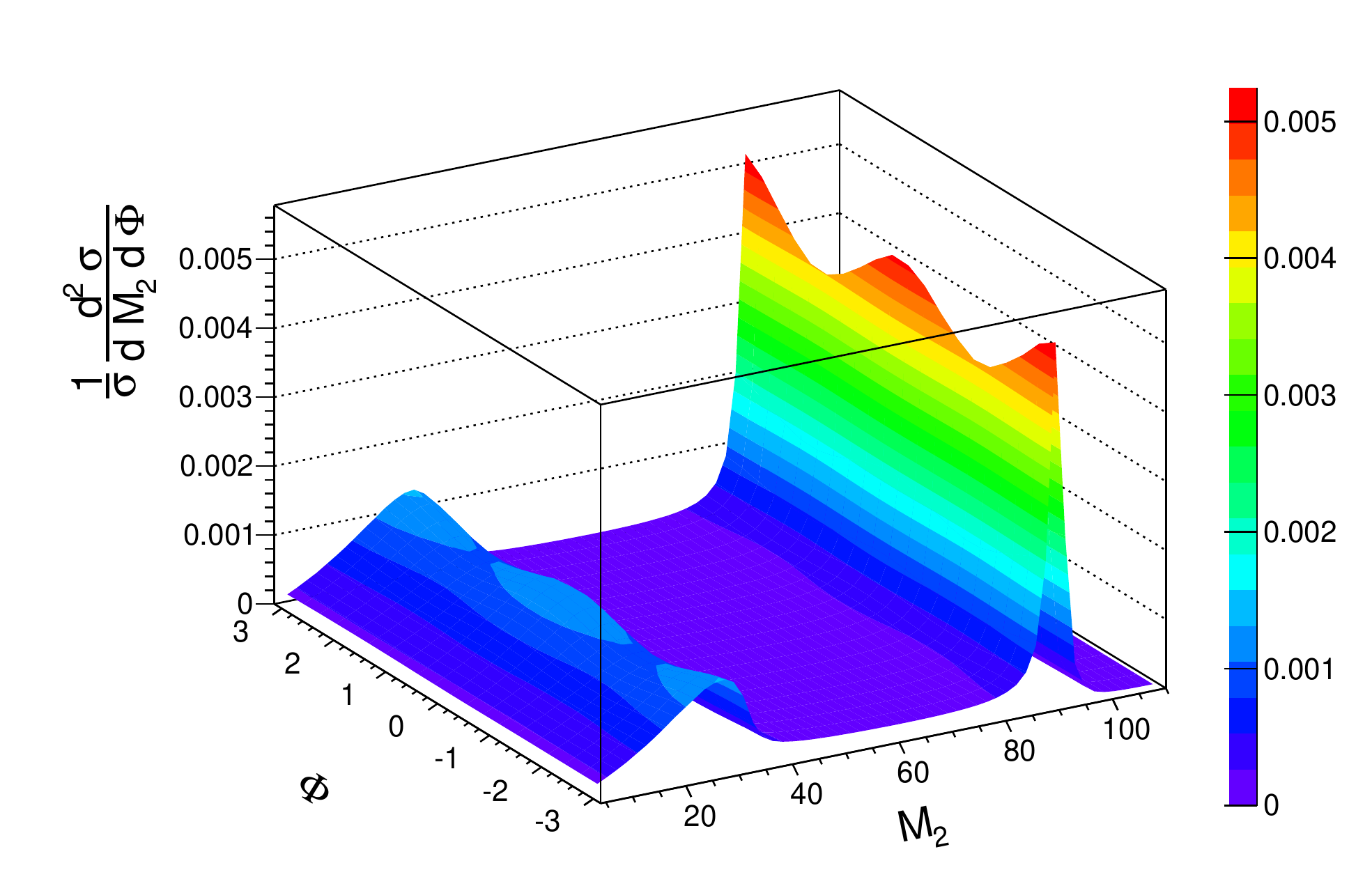}
\includegraphics[width=0.32\textwidth]{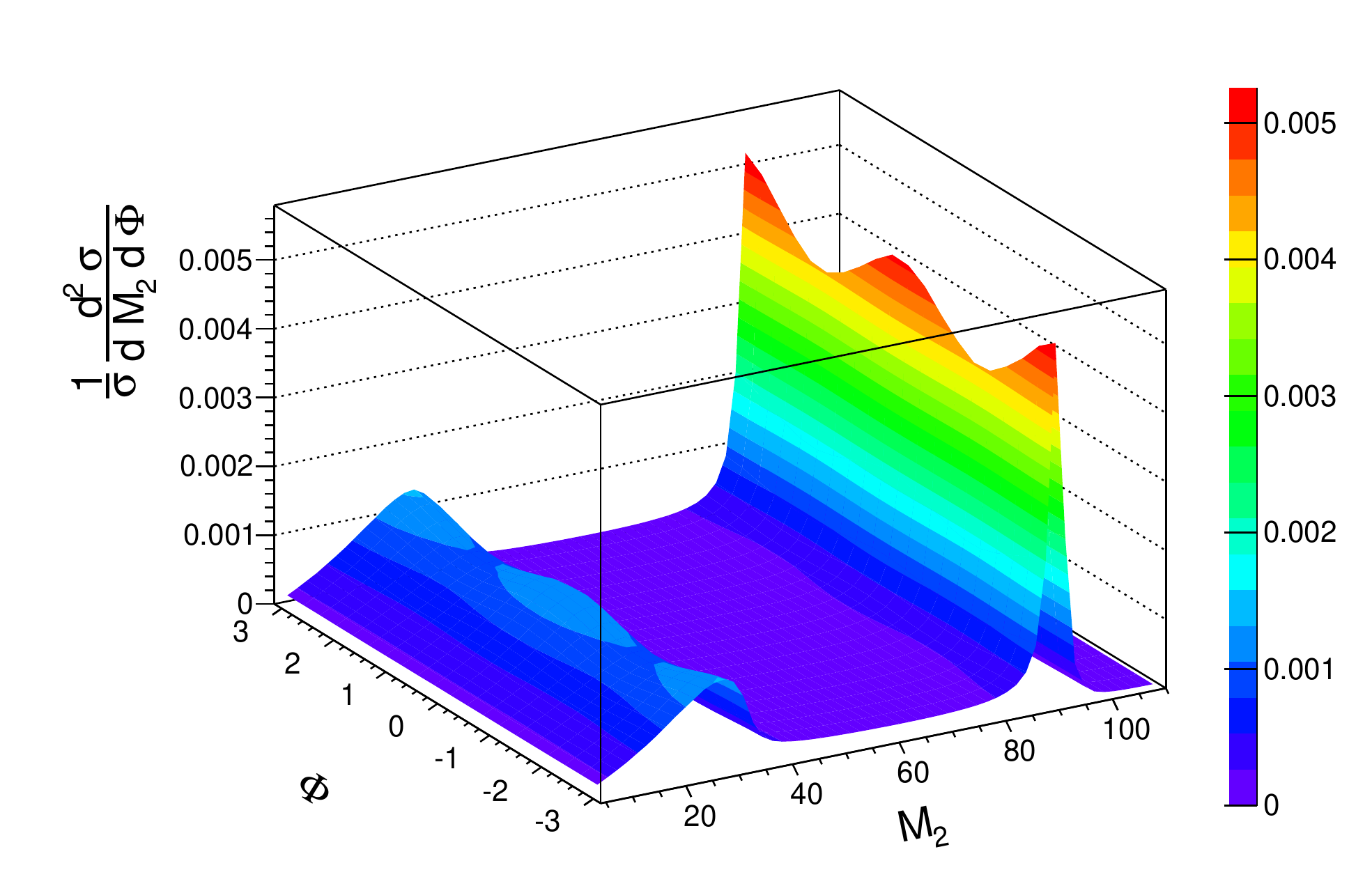}
\includegraphics[width=0.32\textwidth]{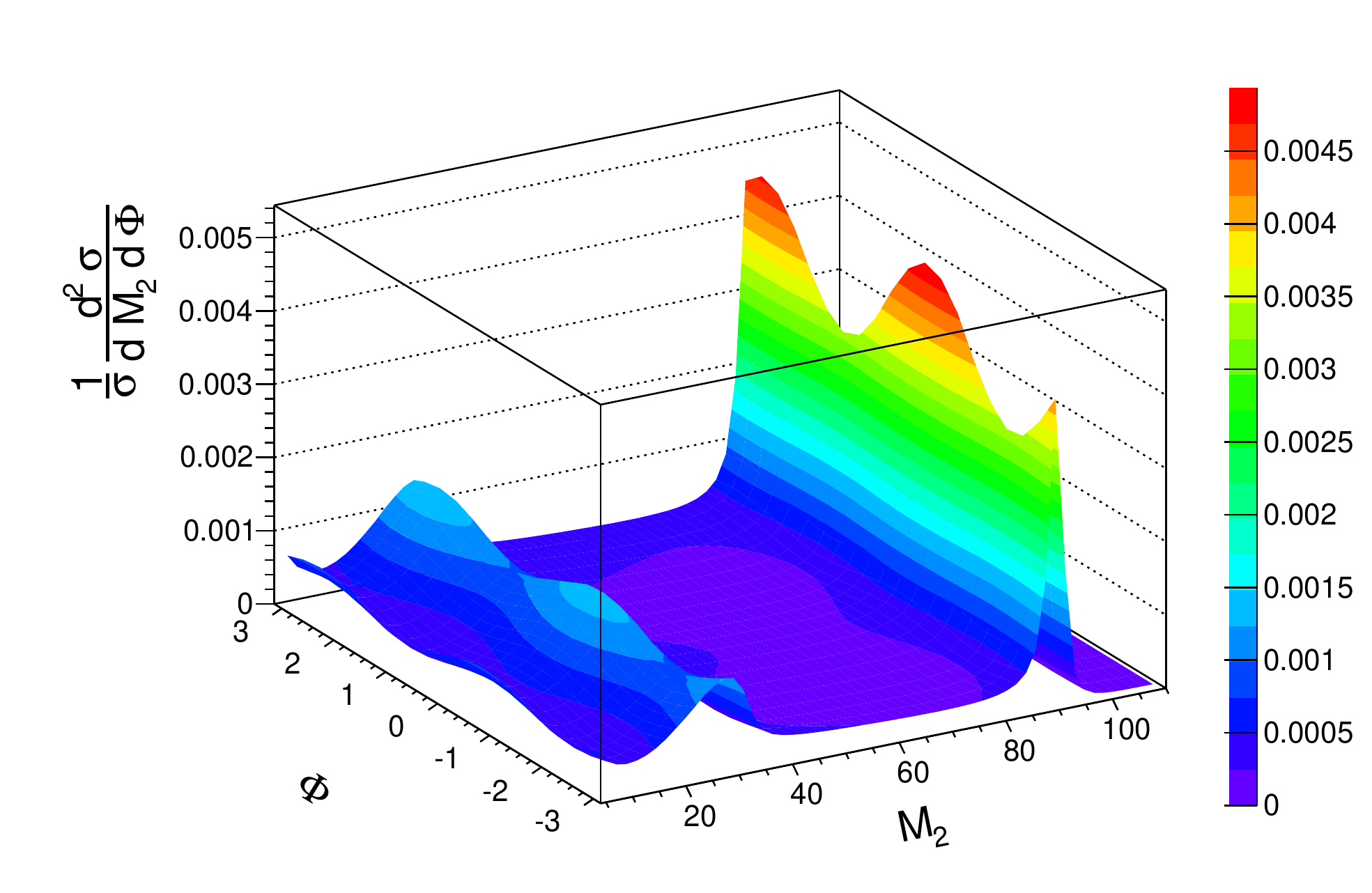}
\includegraphics[width=0.32\textwidth]{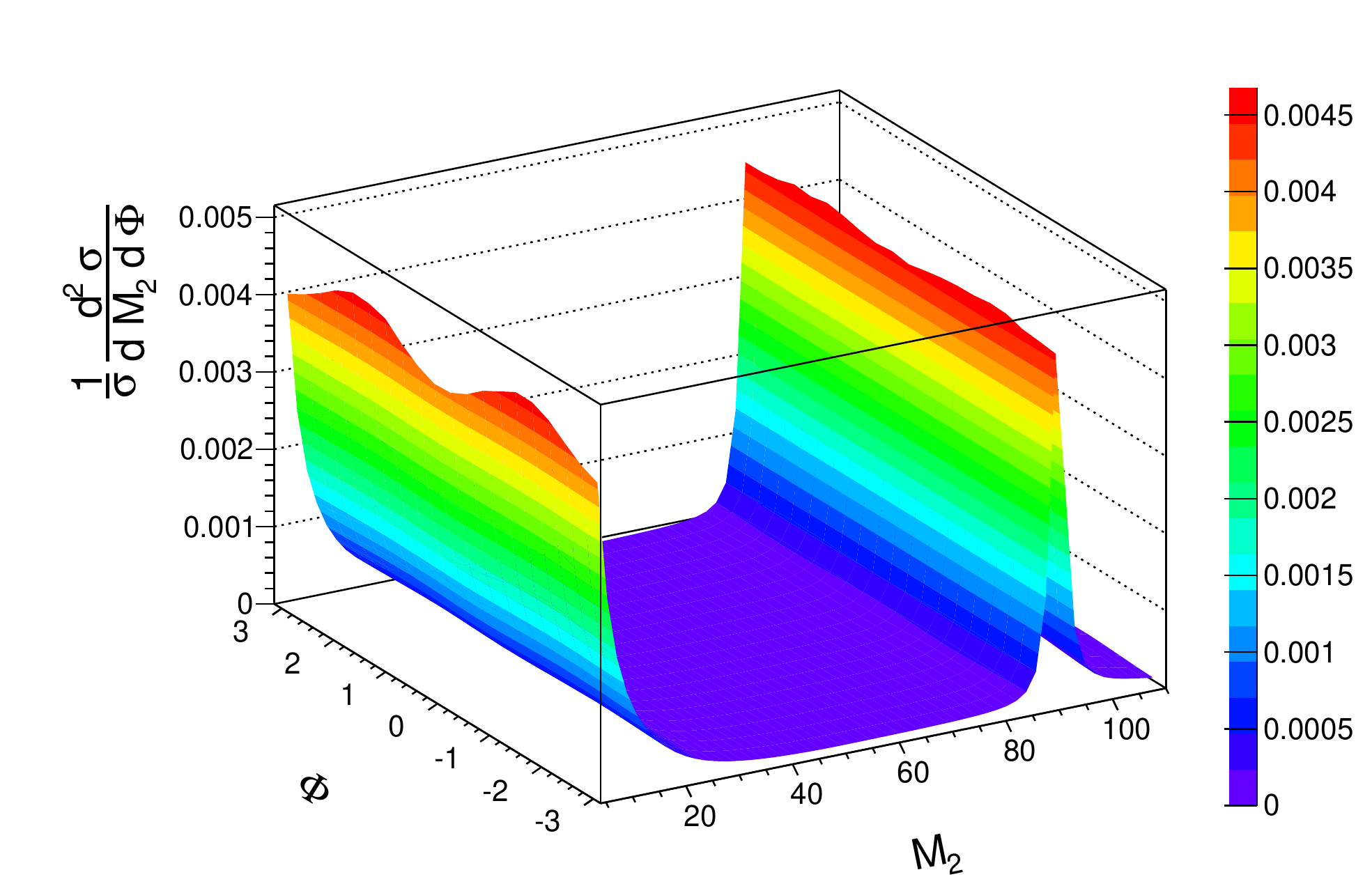}
\caption{The $(M_2, \Phi)$ doubly differential spectrum. The first five distributions are for signal hypotheses 1-5 (hypothesis 1~$\equiv$ SM in top left) defined in Sec.\ref{sec:SigSinglyDoubly} while the bottom right plot is for the full background.}
\label{fig:m2Phdiff}
\end{figure*}

\begin{figure*}
\includegraphics[width=0.32\textwidth]{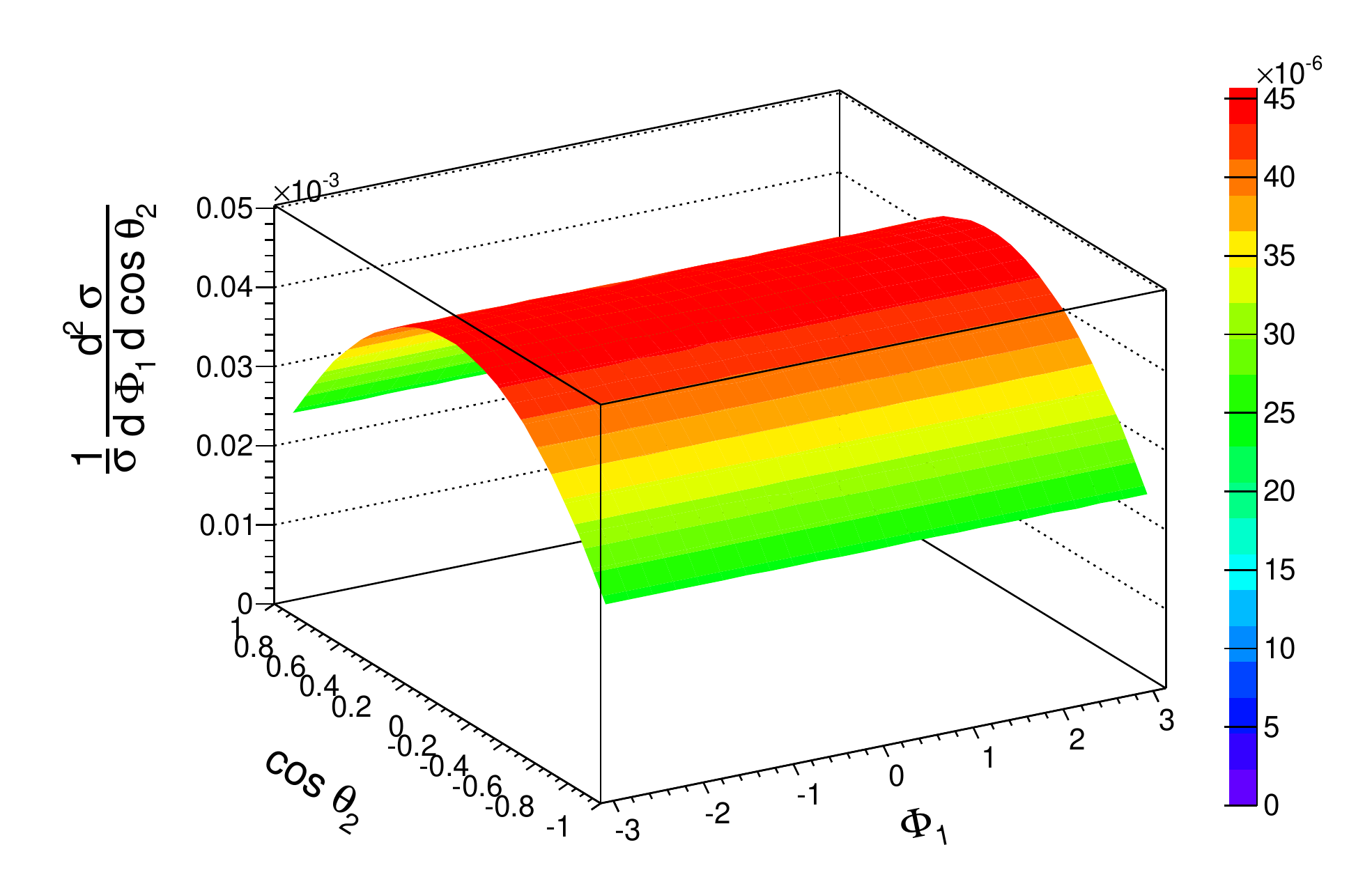}
\includegraphics[width=0.32\textwidth]{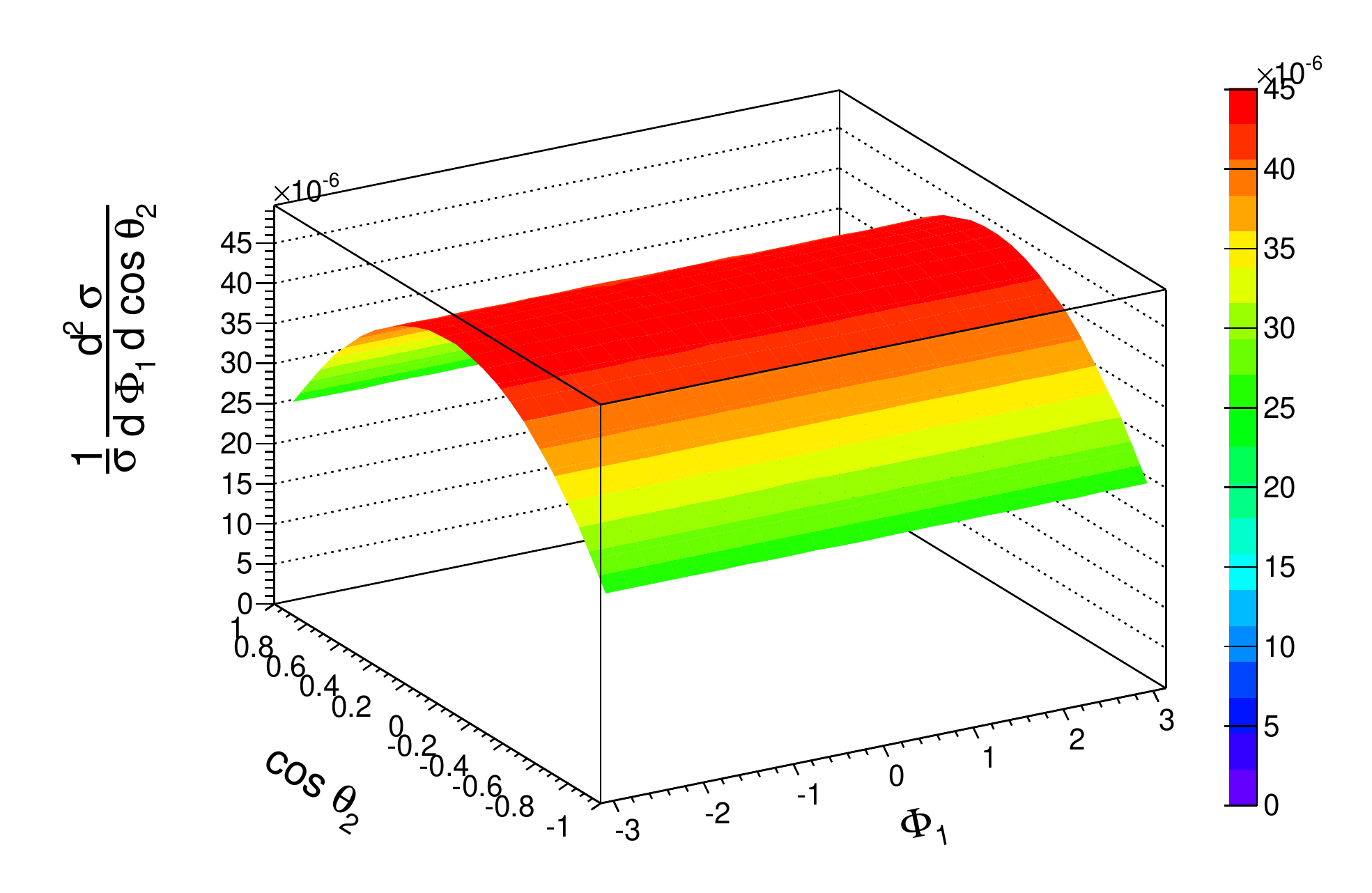}
\includegraphics[width=0.32\textwidth]{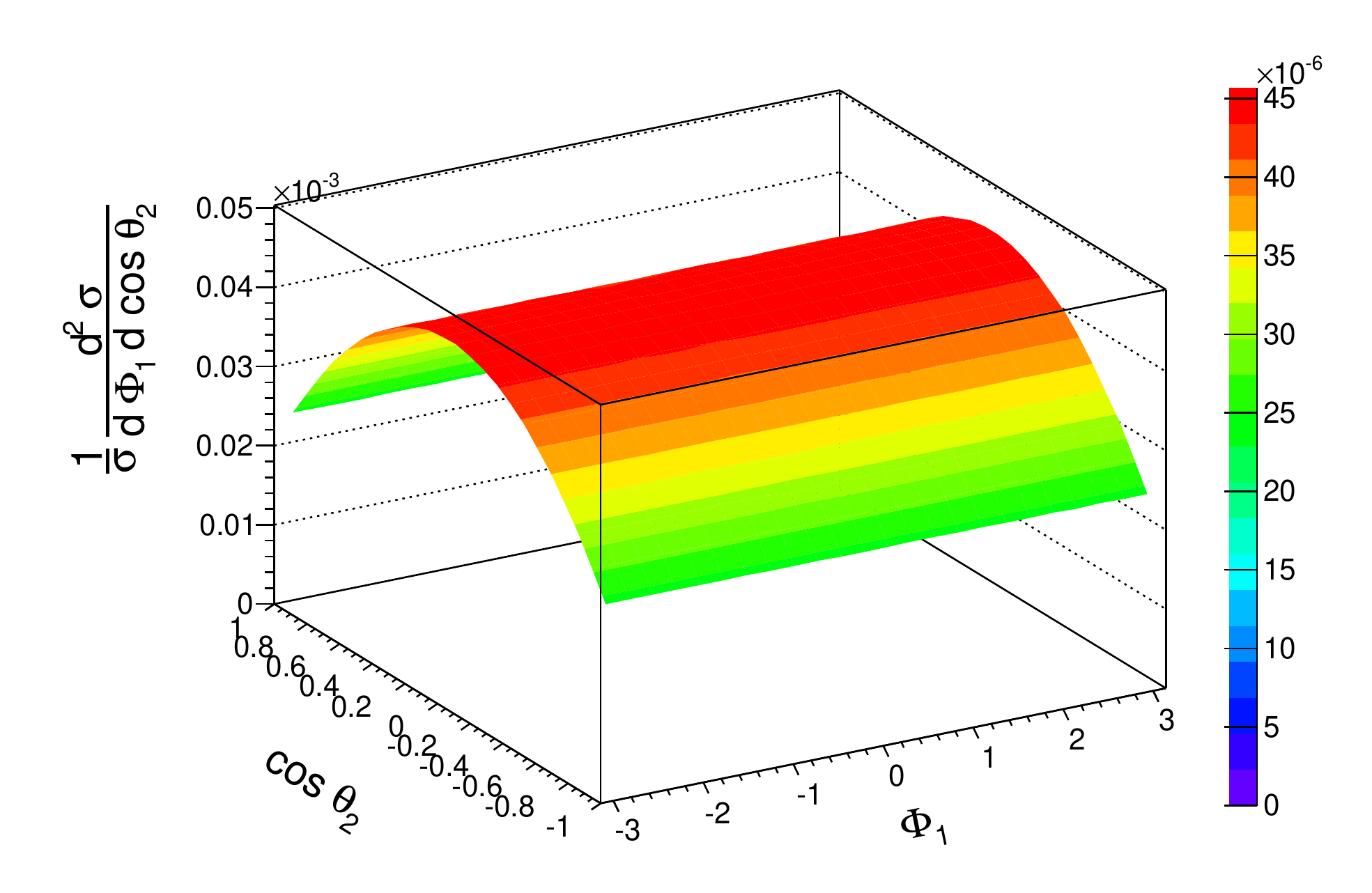}
\includegraphics[width=0.32\textwidth]{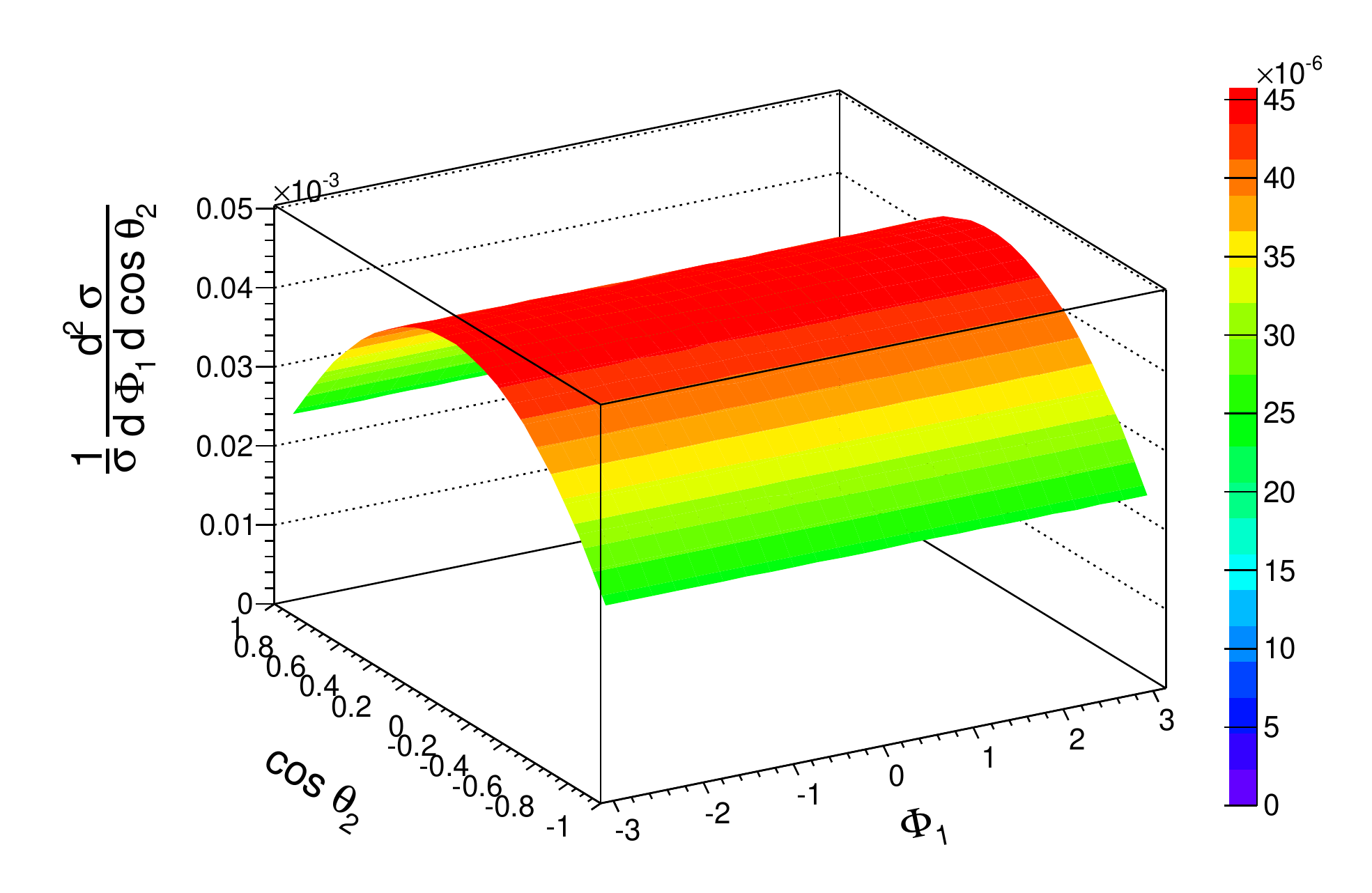}
\includegraphics[width=0.32\textwidth]{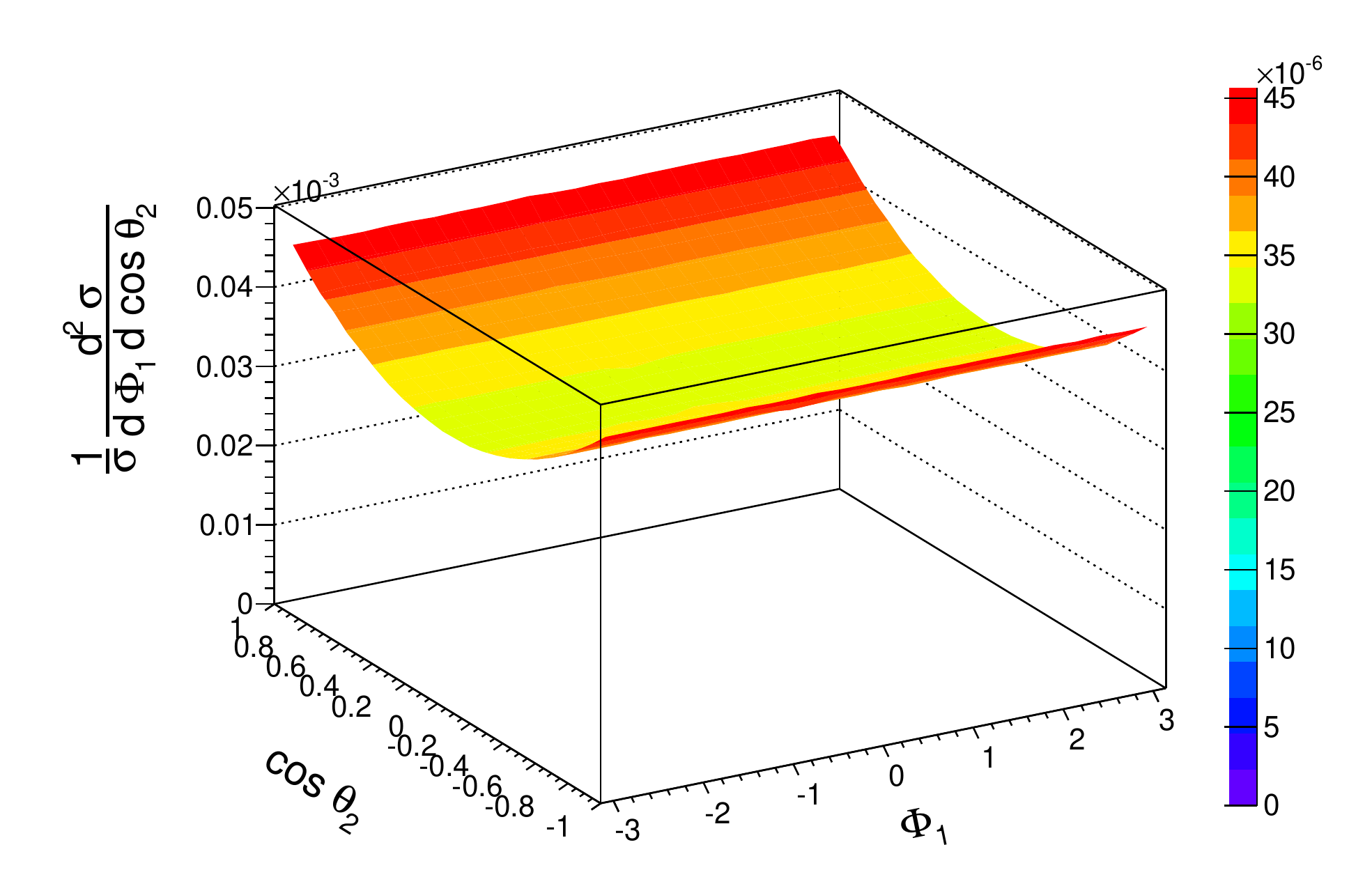}
\includegraphics[width=0.32\textwidth]{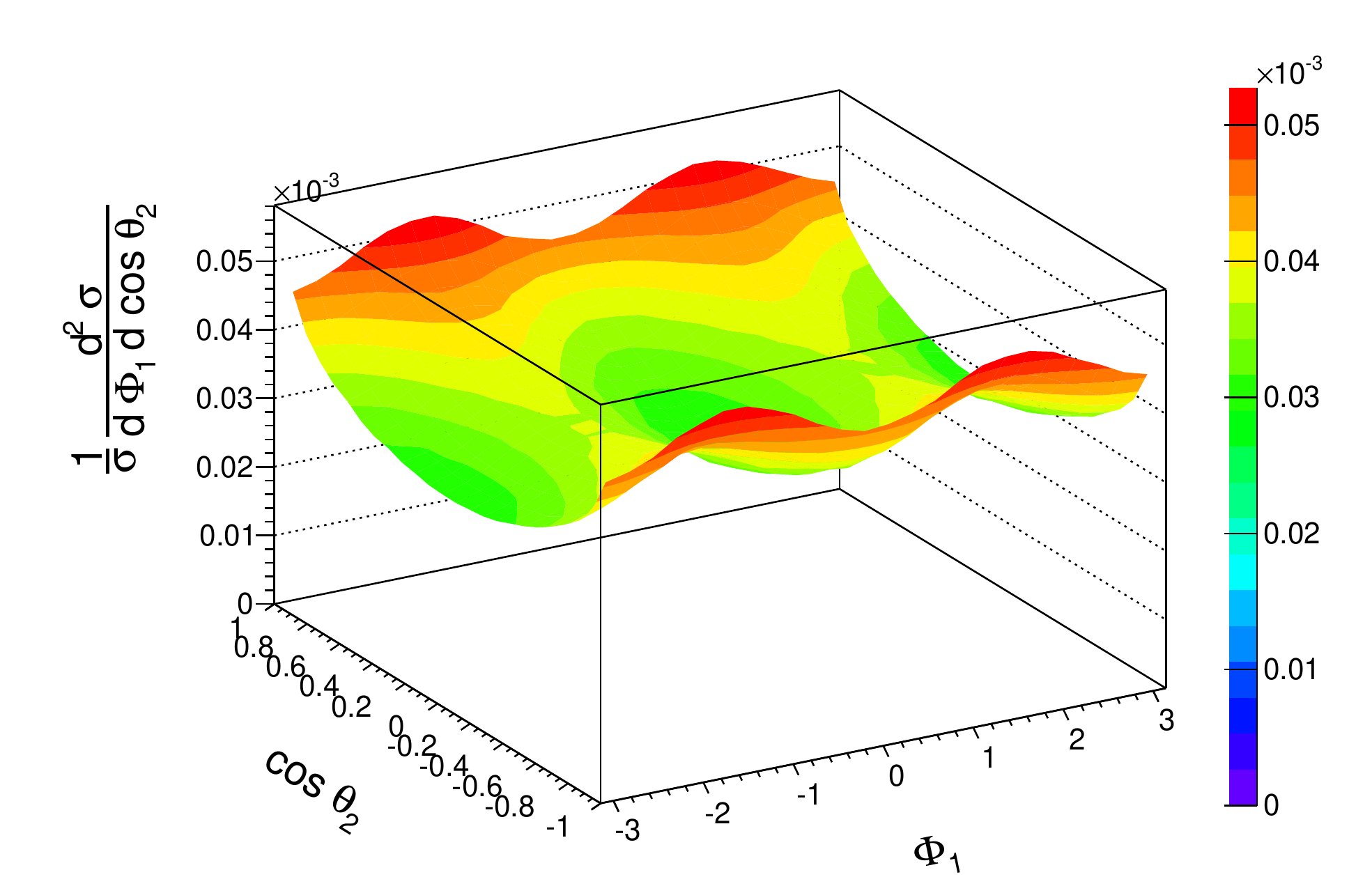}
\caption{The $(\cos{\theta_2}, \Phi_1)$ doubly differential spectrum. The first five distributions are for signal hypotheses 1-5 (hypothesis 1~$\equiv$ SM in top left) defined in Sec.\ref{sec:SigSinglyDoubly} while the bottom right plot is for the full background.}
\label{fig:m2Ph1diff}
\end{figure*}

\begin{figure*}
\includegraphics[width=0.32\textwidth]{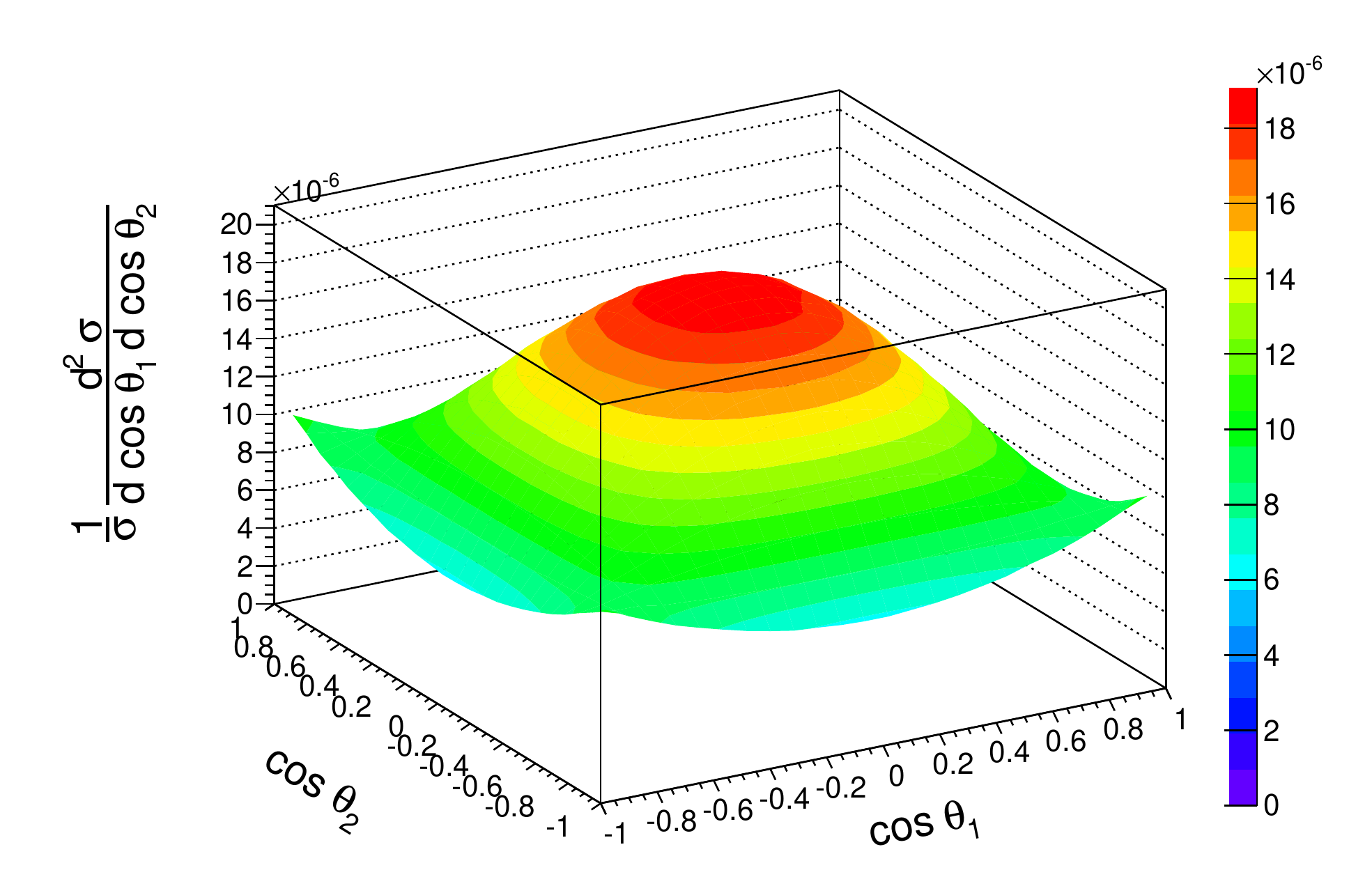}
\includegraphics[width=0.32\textwidth]{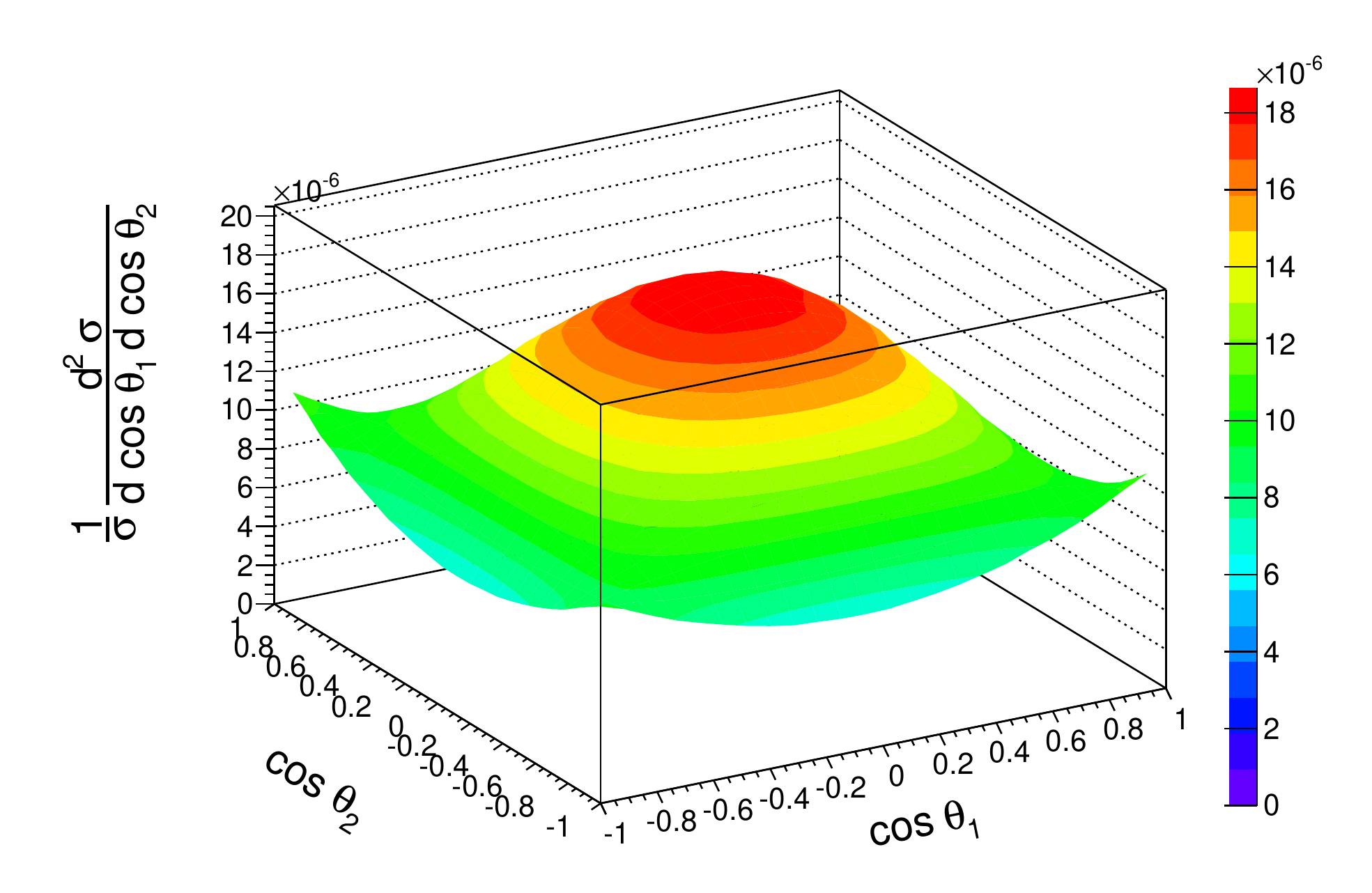}
\includegraphics[width=0.32\textwidth]{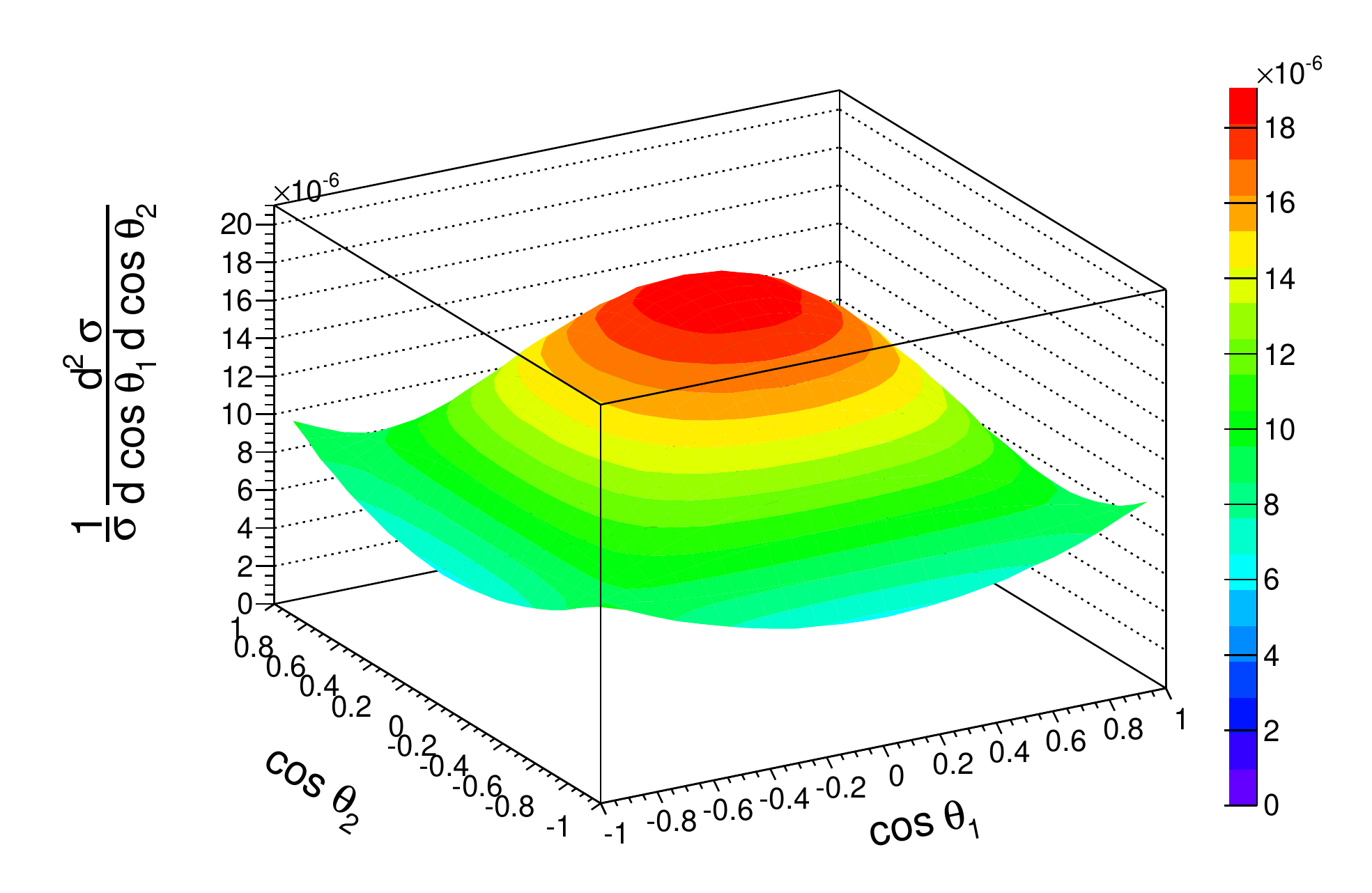}
\includegraphics[width=0.32\textwidth]{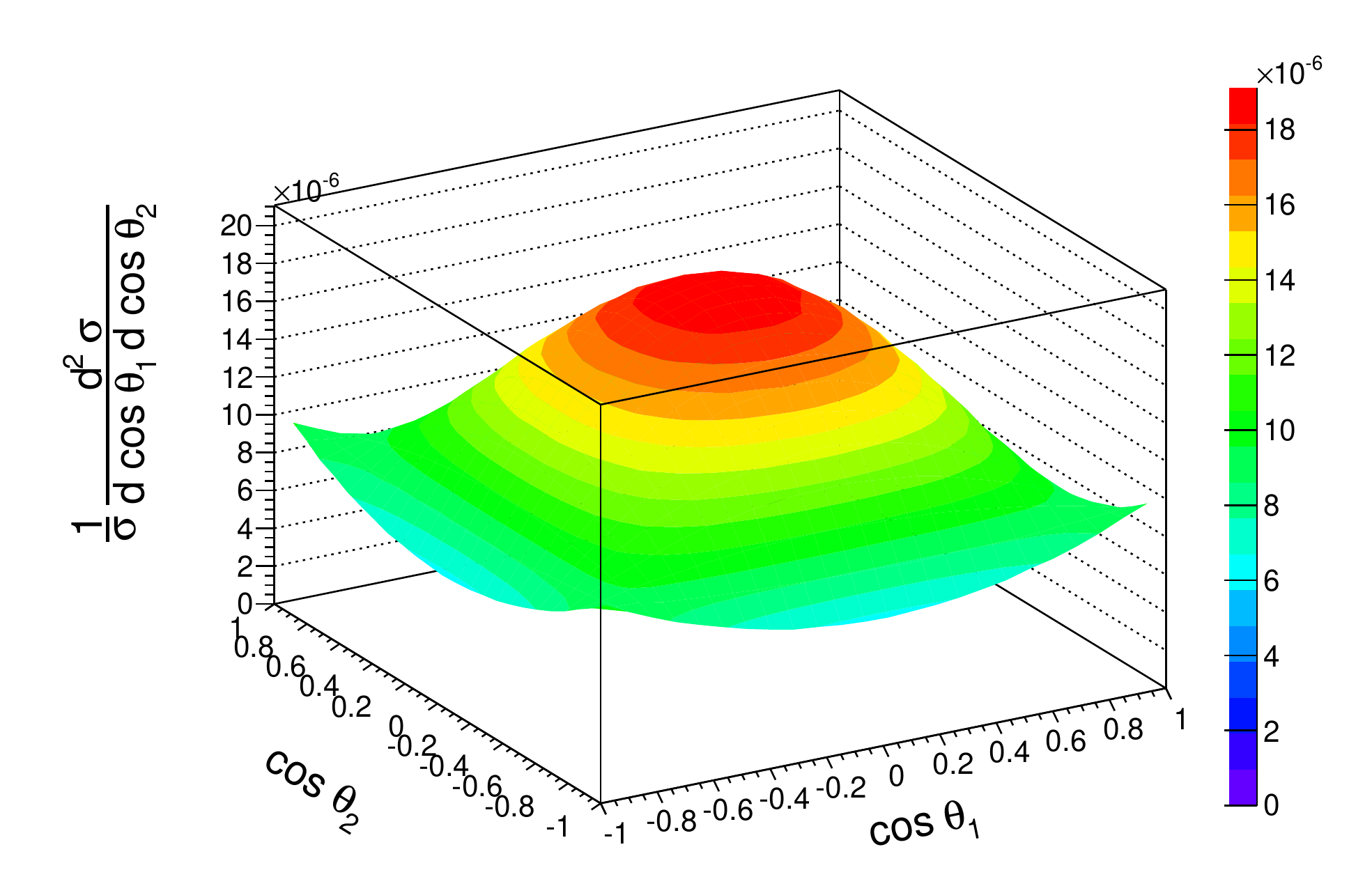}
\includegraphics[width=0.32\textwidth]{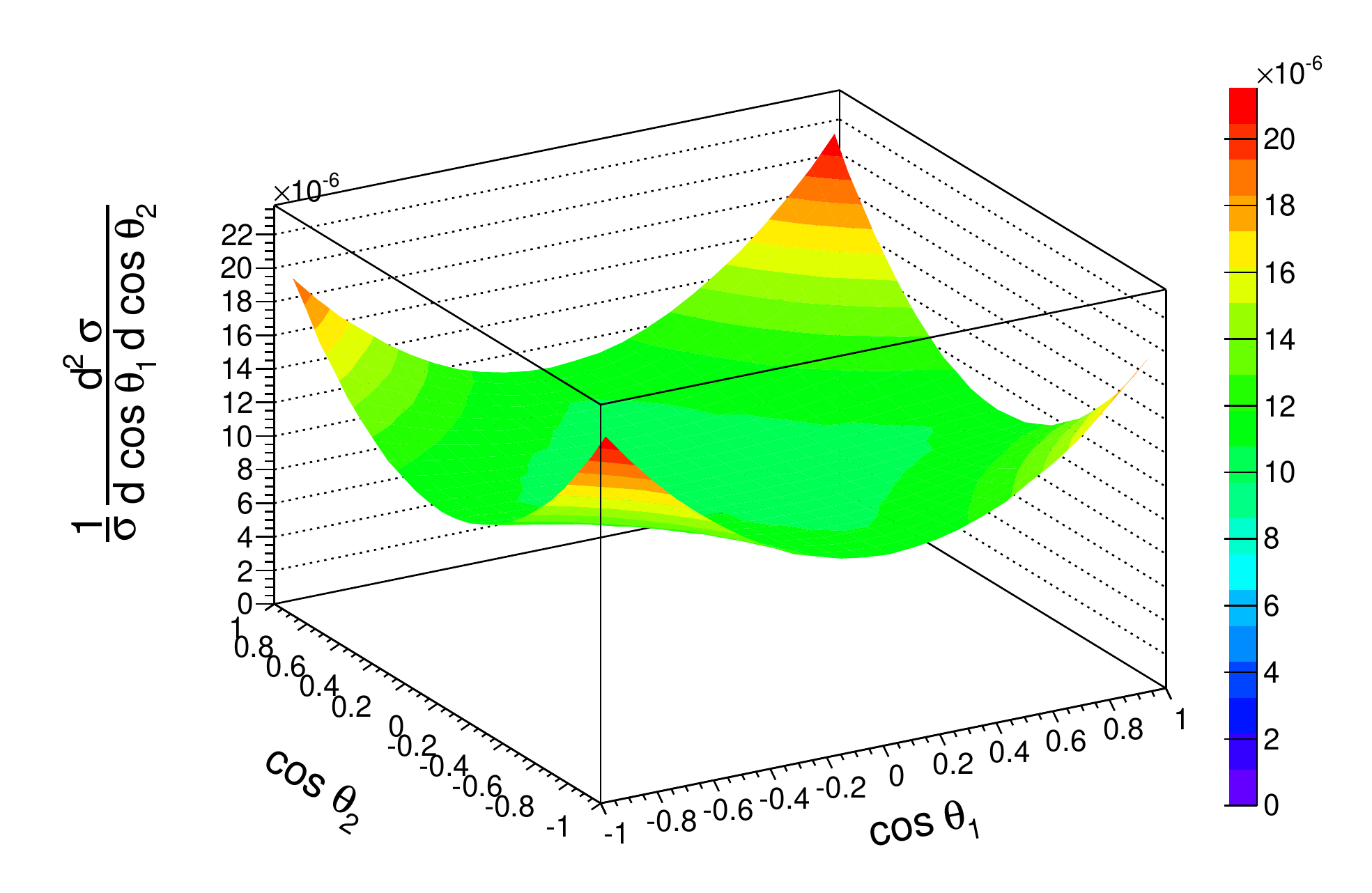}
\includegraphics[width=0.32\textwidth]{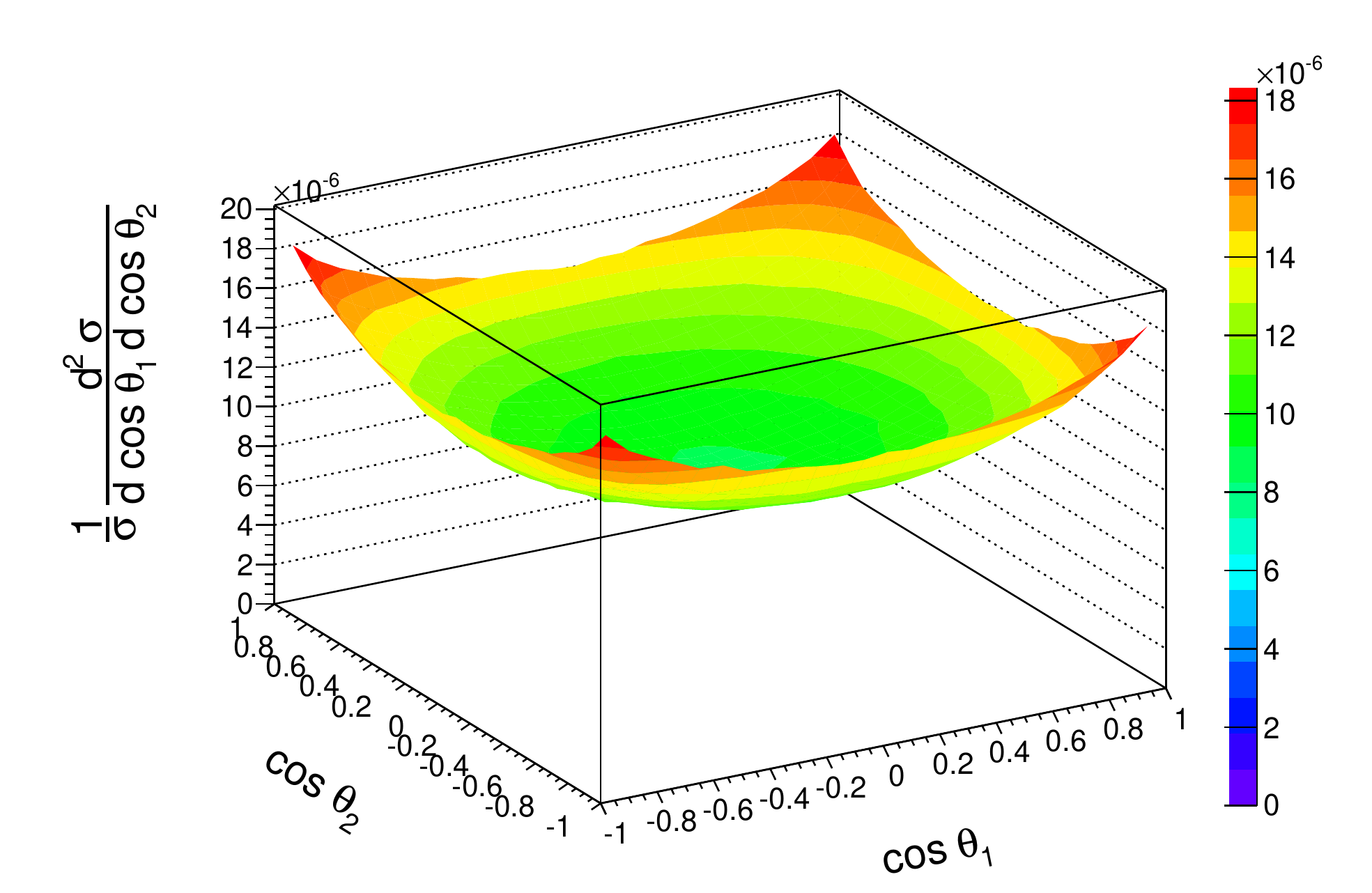}
\caption{The $(\cos{\theta_1}, \cos{\theta_2})$ doubly differential spectrum. The first five distributions are for signal hypotheses 1-5 (hypothesis 1~$\equiv$ SM in top left) defined in Sec.\ref{sec:SigSinglyDoubly} while the bottom right plot is for the full background.}
\label{fig:th1th2diff}
\end{figure*}

\begin{figure*}
\includegraphics[width=0.32\textwidth]{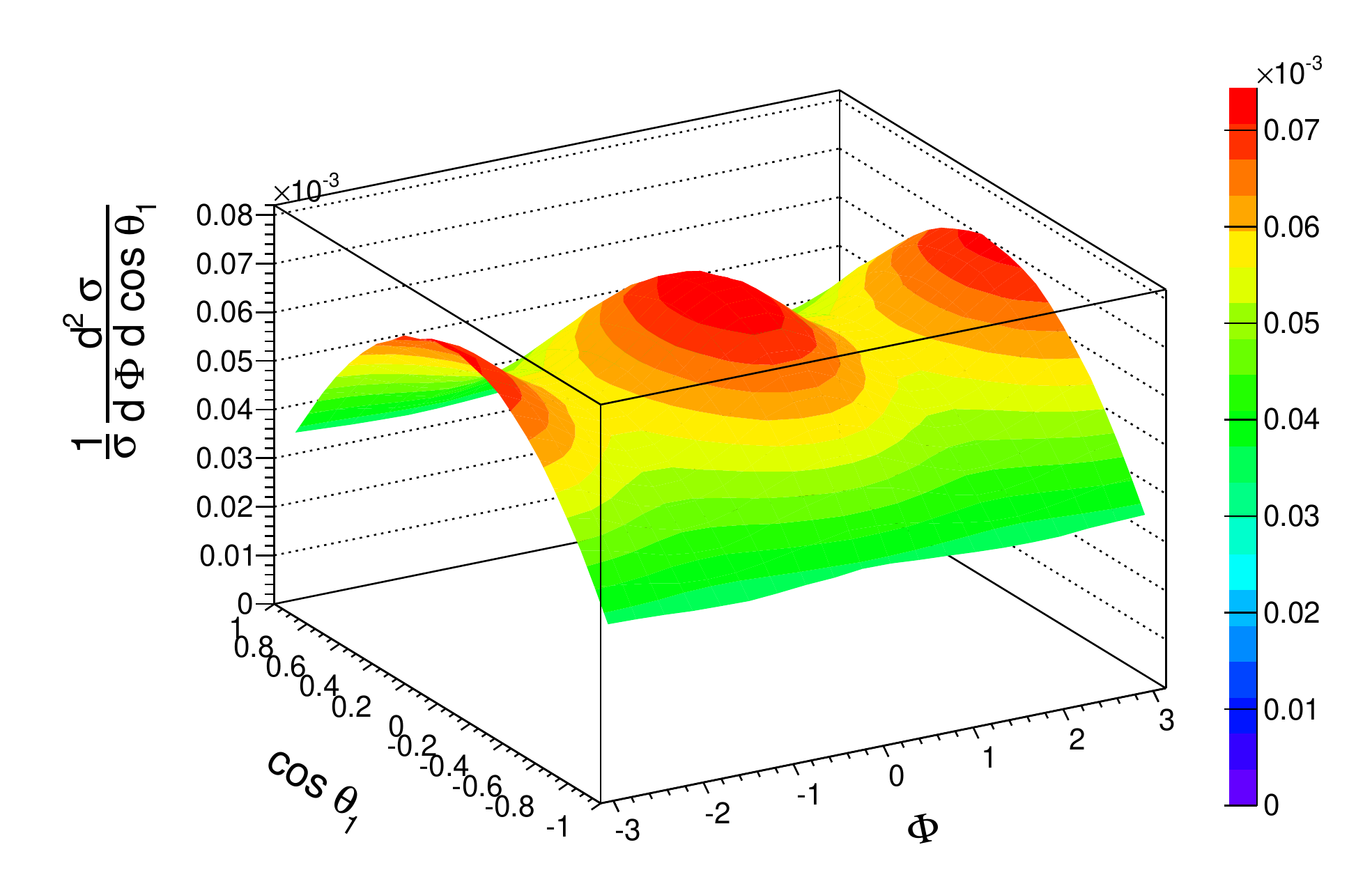}
\includegraphics[width=0.32\textwidth]{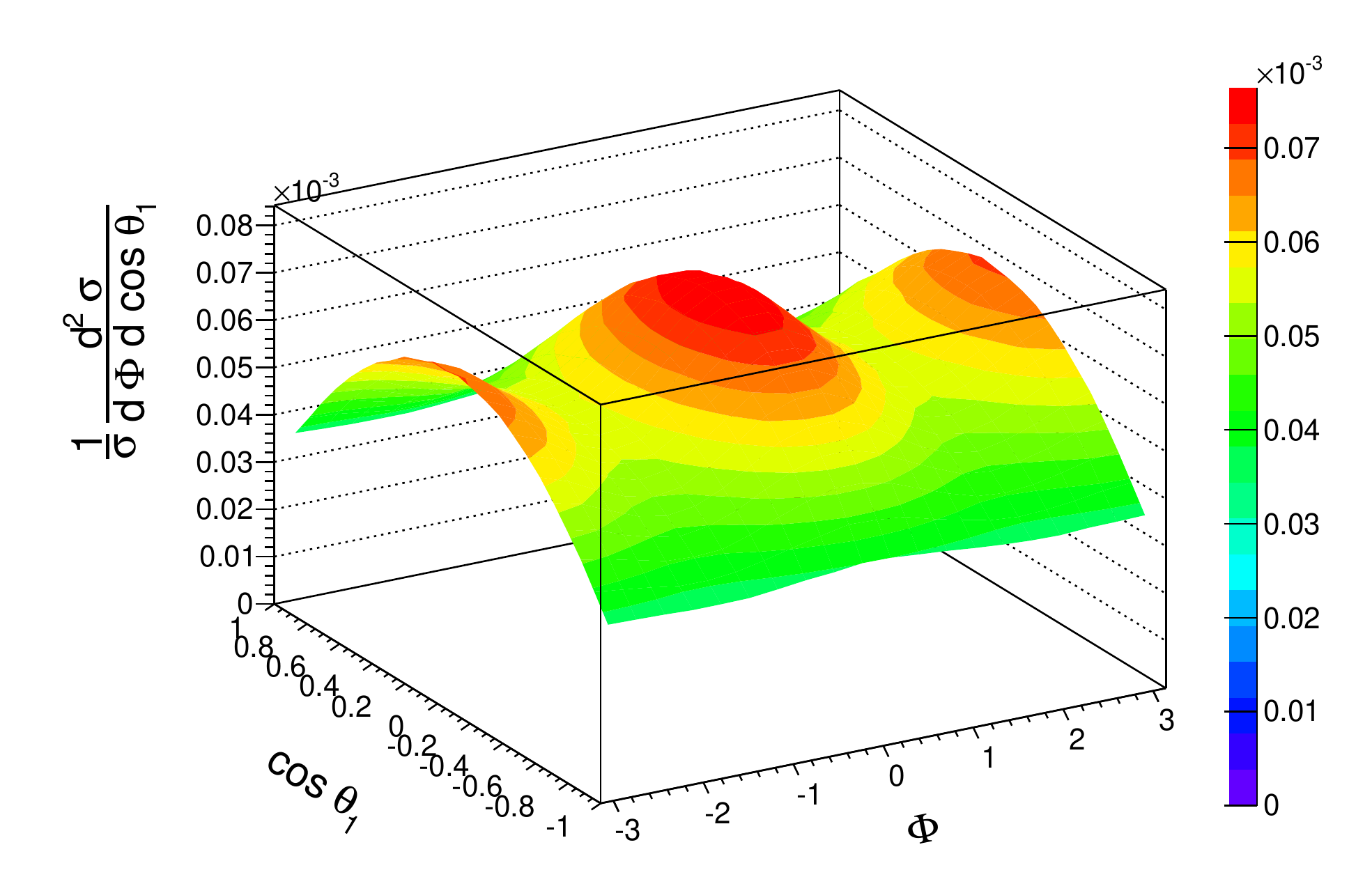}
\includegraphics[width=0.32\textwidth]{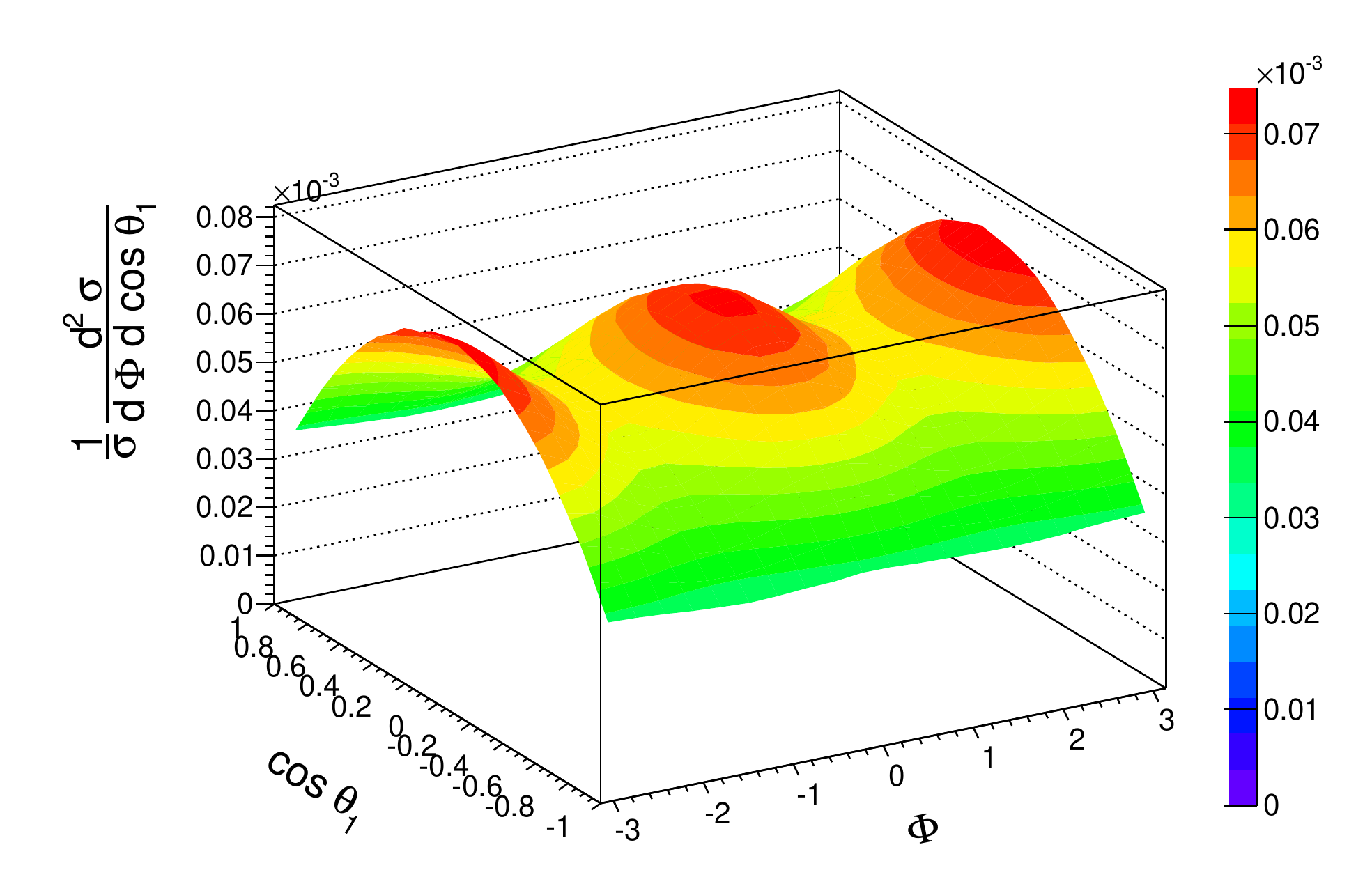}
\includegraphics[width=0.32\textwidth]{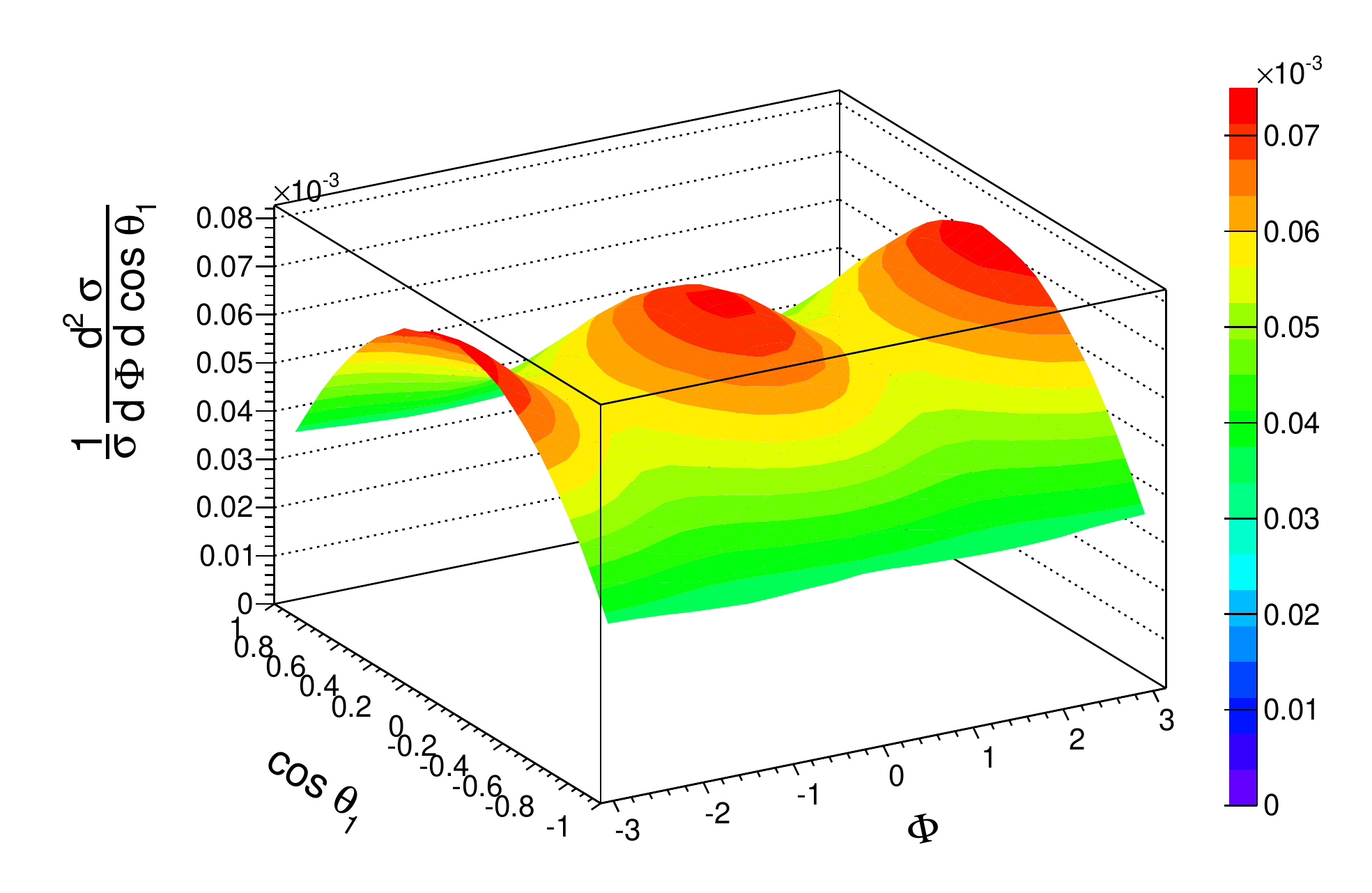}
\includegraphics[width=0.32\textwidth]{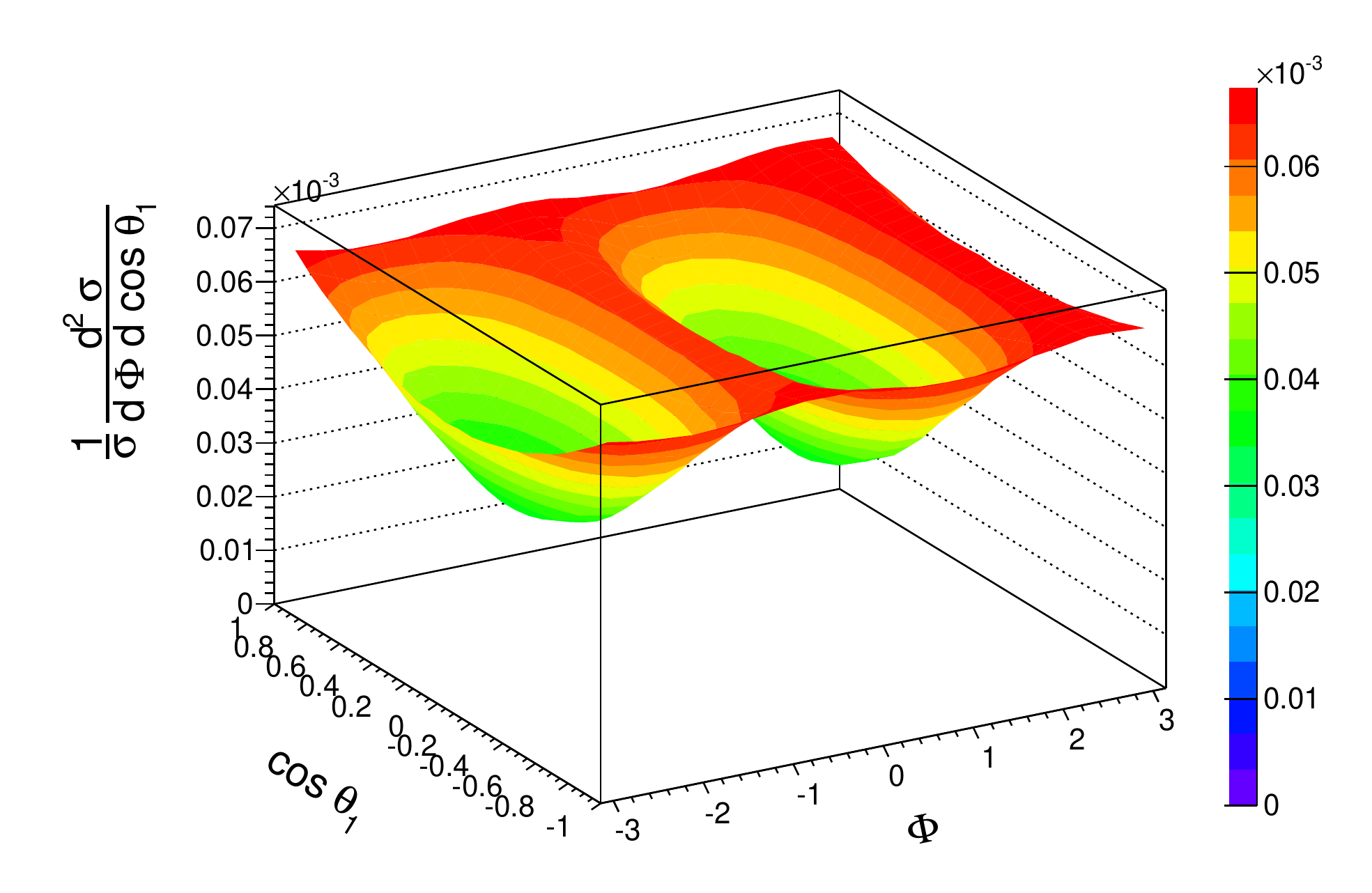}
\includegraphics[width=0.32\textwidth]{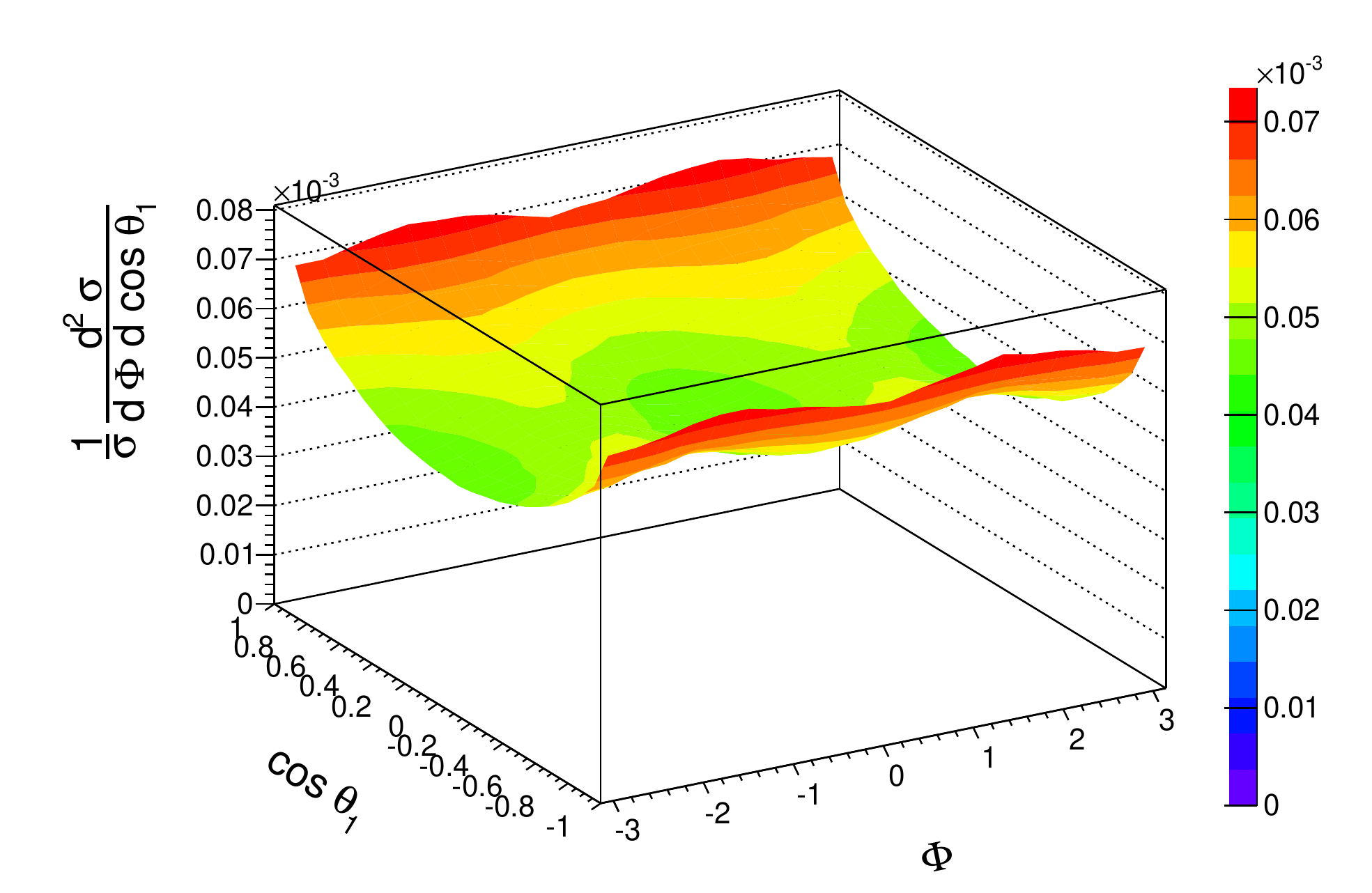}
\caption{The $(\Phi, \cos{\theta_1})$ doubly differential spectrum. The first five distributions are for signal hypotheses 1-5 (hypothesis 1~$\equiv$ SM in top left) defined in Sec.\ref{sec:SigSinglyDoubly} while the bottom right plot is for the full background.}
\label{fig:th1Phdiff}
\end{figure*}

\begin{figure*}
\includegraphics[width=0.32\textwidth]{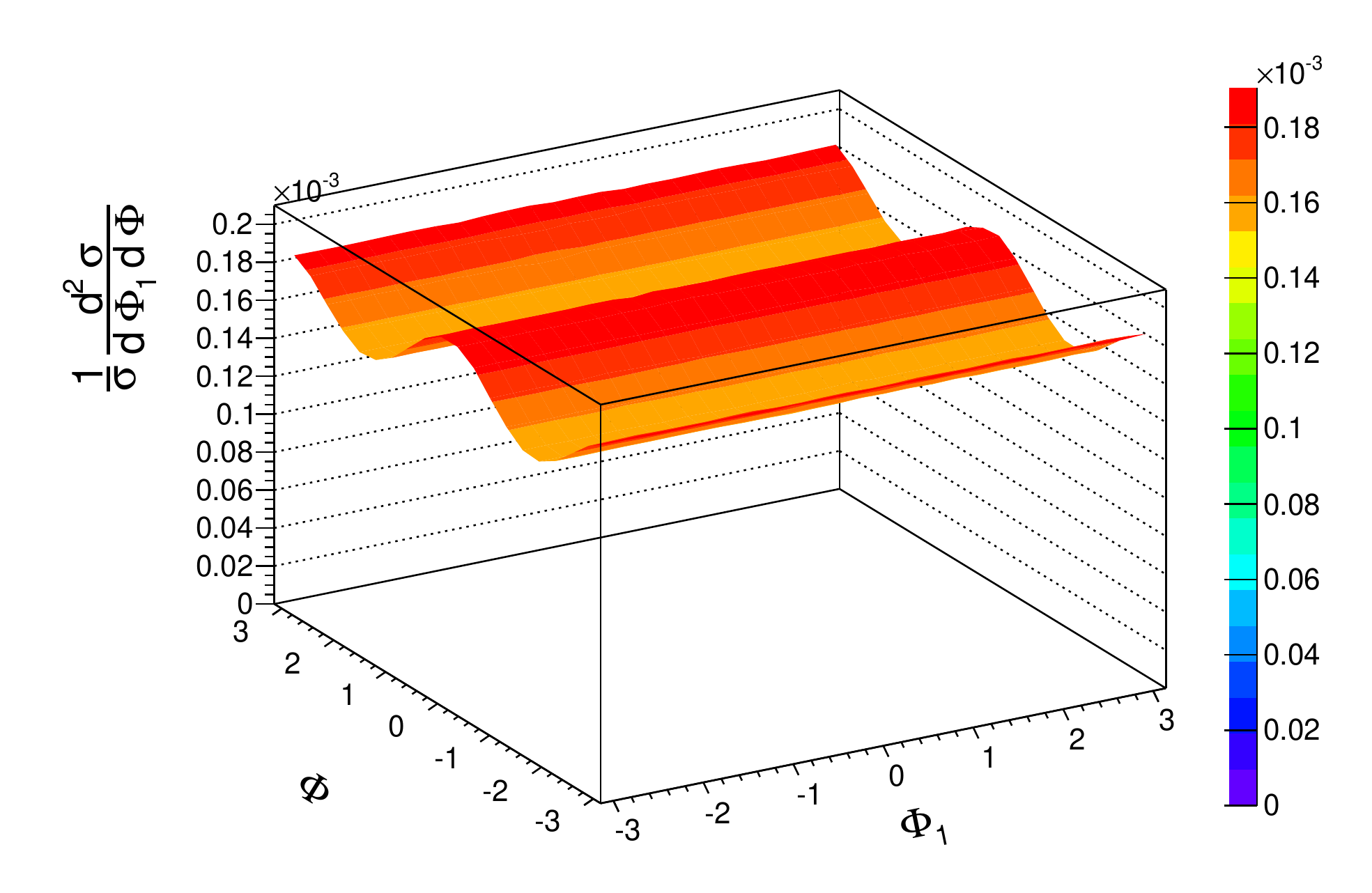}
\includegraphics[width=0.32\textwidth]{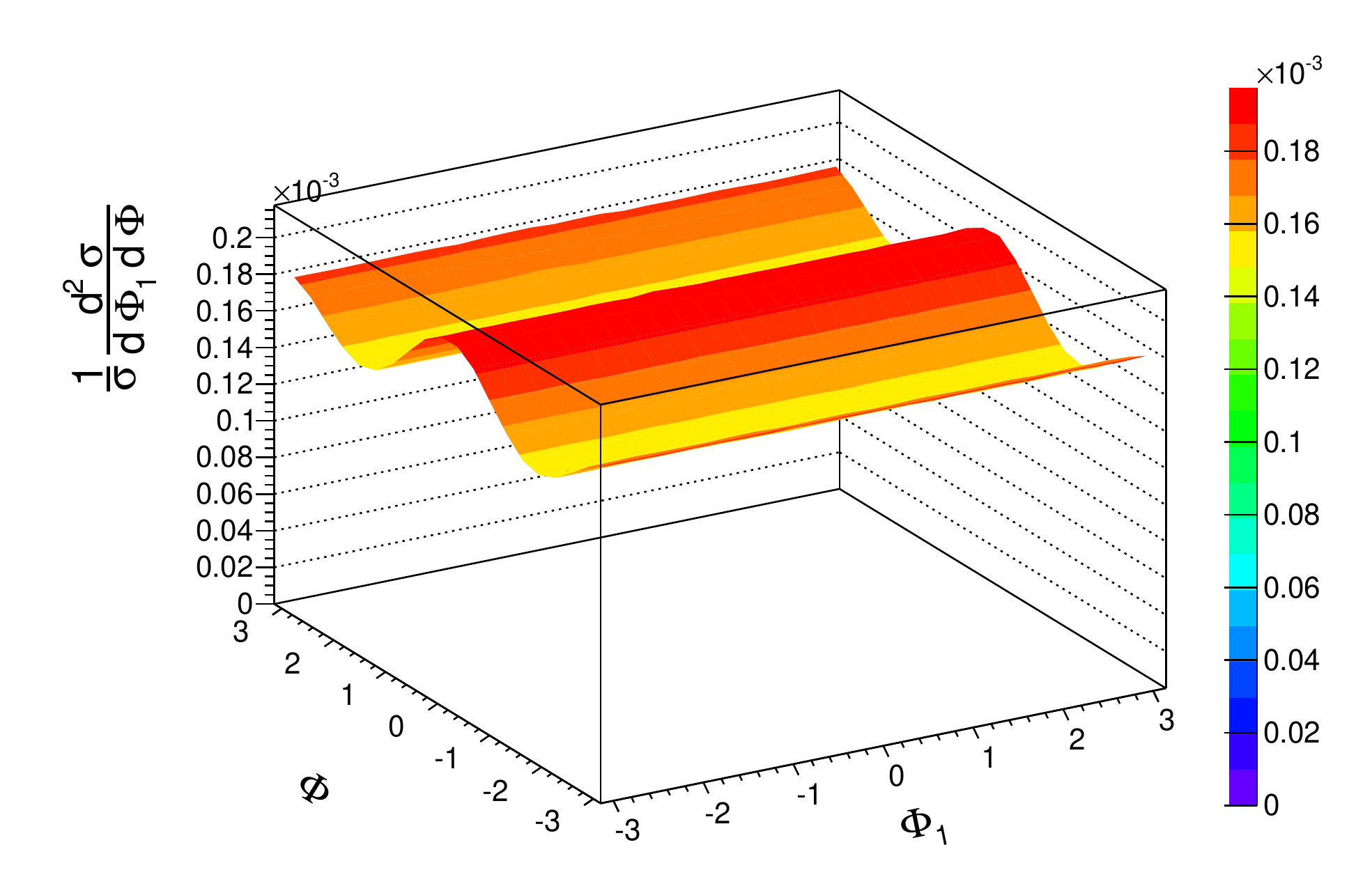}
\includegraphics[width=0.32\textwidth]{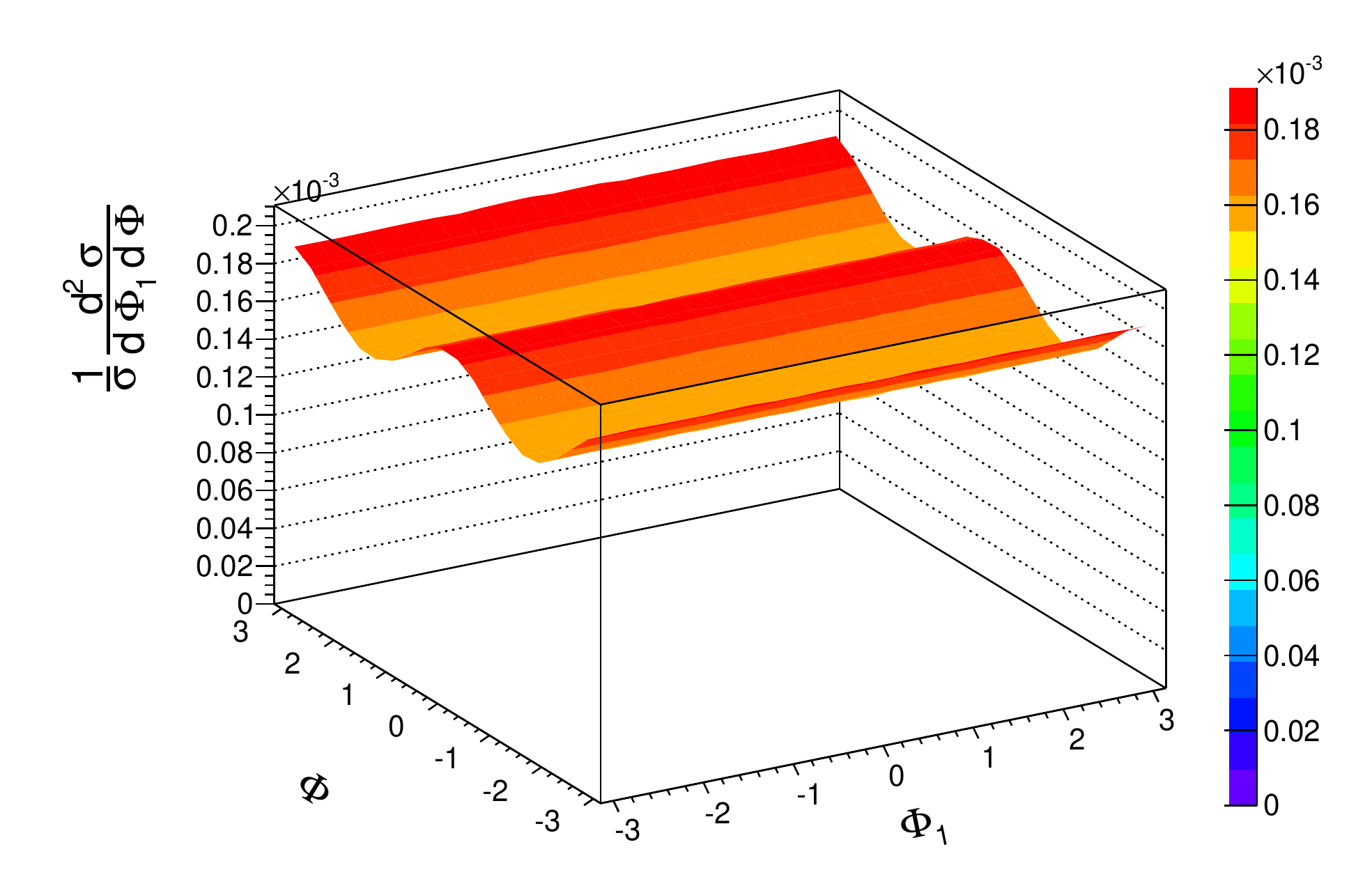}
\includegraphics[width=0.32\textwidth]{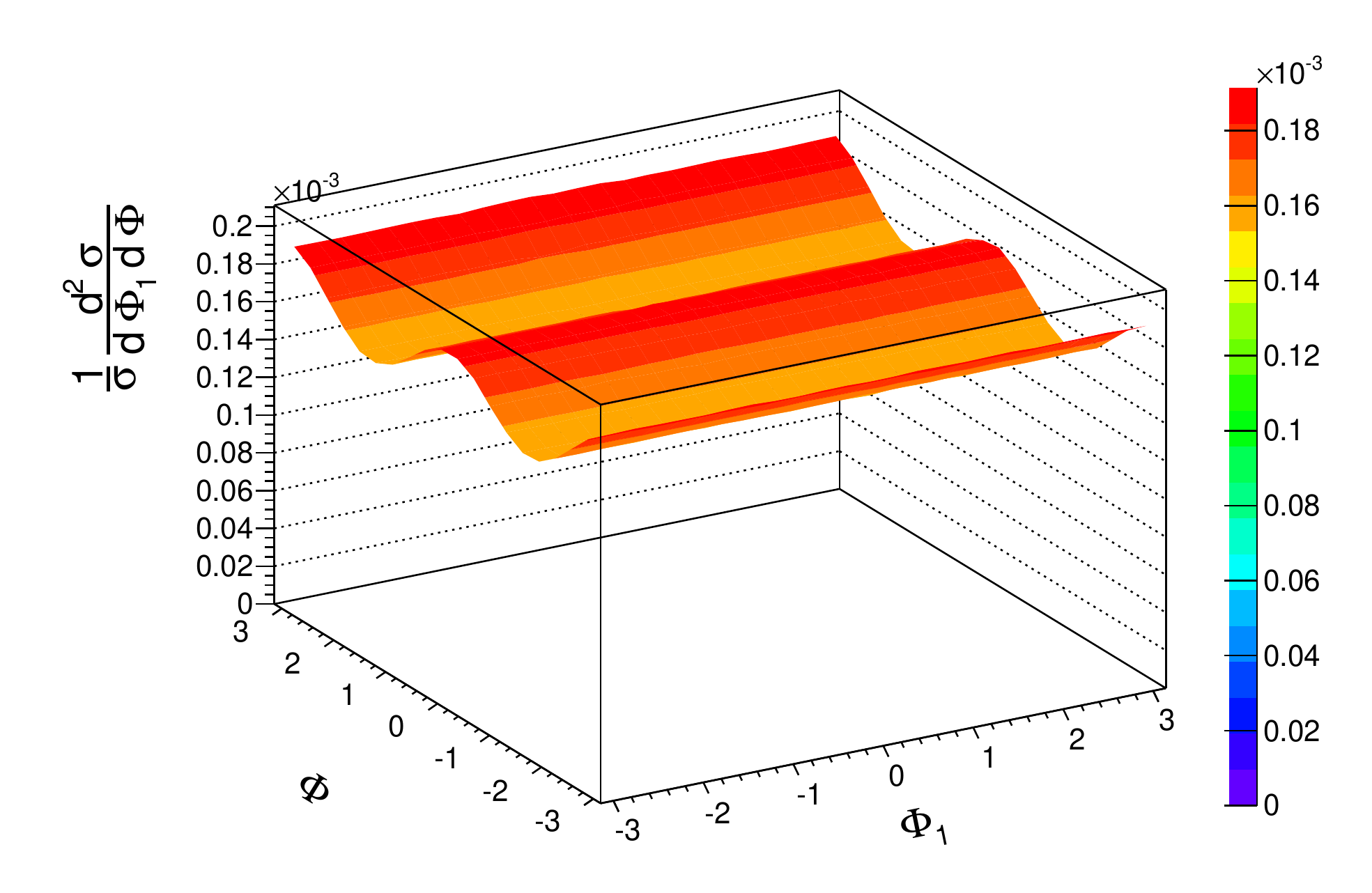}
\includegraphics[width=0.32\textwidth]{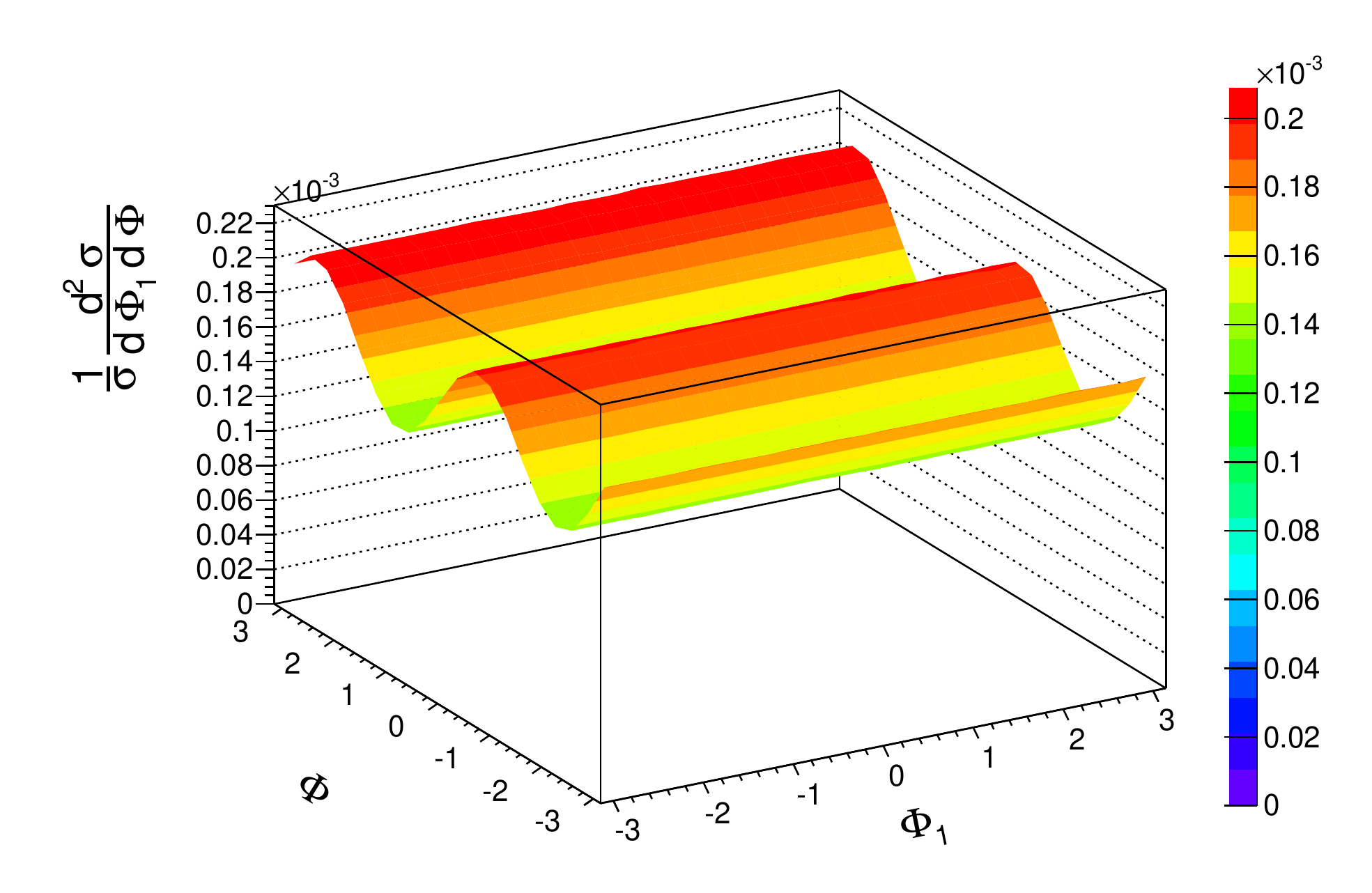}
\includegraphics[width=0.32\textwidth]{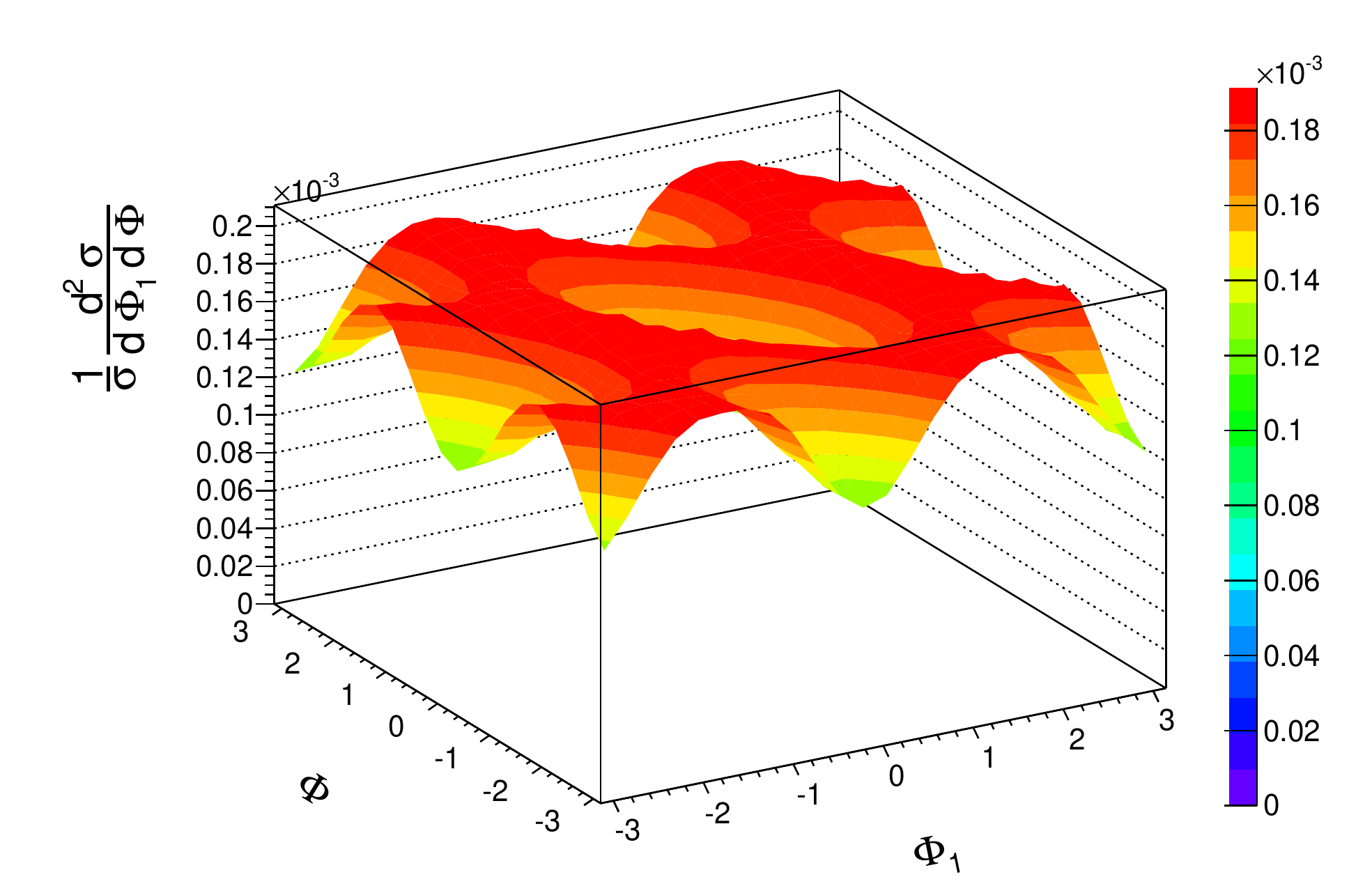}
\caption{The $(\Phi, \Phi_1)$ doubly differential spectrum. The first five distributions are for signal hypotheses 1-5 (hypothesis 1~$\equiv$ SM in top left) defined in Sec.\ref{sec:SigSinglyDoubly} while the bottom right plot is for the full background.}
\label{fig:PhPh1diff}
\end{figure*}

In Fig.~\ref{fig:m2Thdiff}-\ref{fig:PhPh1diff} we show various combinations of the doubly differential spectra for the five signal hypotheses as well as the full background.   These are primarily for illustration purposes, but from these one can get an idea of the correlations between the different kinematic variables\footnote{We do not show all possible combinations, but any not shown here can be obtained by emailing the corresponding author}.  For these plots only the five signal hypotheses and the full result for the background are shown. For all distributions the phase space is defined as 4~GeV $< M_1 < 120$~GeV and 4~GeV $< M_2 <$ 120~GeV with $\sqrt{\hat{s}}=125$~GeV for signal and 110~GeV $< \sqrt{\hat{s}} <$ 140~GeV for background.


\clearpage

\bibliographystyle{apsrev}
\bibliography{GoldenChannelBib}

\end{document}